\documentclass[usenatbib]{mn2e}
\usepackage{wrapfig}
\usepackage{sidecap}
\usepackage{graphicx}
\usepackage{caption}
\usepackage{longtable}
\usepackage{array}
\usepackage{float}
\usepackage{capt-of}
\usepackage{subcaption}
\usepackage[bookmarks, colorlinks, breaklinks, pdftitle={FQXi Essay},pdfauthor={Author}]{hyperref}  
\hypersetup{linkcolor=blue,citecolor=blue,filecolor=black,urlcolor=blue}

\title[Asymmetry in Gas Kinematics]{{The SAMI Galaxy Survey: Asymmetry in Gas Kinematics and its links to Stellar Mass and Star Formation}}

\author[Bloom et al.,]
{J. V. Bloom$^{1,2}$\thanks{jbloom@physics.usyd.edu.au}, L. M. R. Fogarty$^{1,2}$, S. M. Croom$^{1,2}$, A. Schaefer$^{1,2}$, J. J. Bryant$^{1,2,3}$, 
\newauthor L. Cortese$^{4}$, S. Richards$^{1,2,3}$, J. Bland-Hawthorn$^{1}$, I-T. Ho$^{6}$, N. Scott$^{1,2}$, G. Goldstein$^{7}$, 
\newauthor  A. Medling$^{8}$, S. Brough$^{3}$, S.M. Sweet$^{8}$, G. Cecil$^{9}$, A. L\'opez-S\'anchez$^{3,7}$, K. Glazebrook$^{3}$,  \newauthor Q. Parker$^{10,3,7}$, J. T. Allen$^{1,2}$, M. Goodwin$^{3}$, A. W. Green$^{3}$, \newauthor I. S. Konstantopoulos$^{3,11}$,  J. S. Lawrence$^{3}$, N. Lorente$^{3}$,   M. S. Owers$^{3,7}$, R. Sharp$^{8}$
\\
$^{1}$ Sydney Institute for Astronomy, School of Physics, University of Sydney, NSW 2006, Australia.\\ 
$^{2}$ ARC Centre of Excellence for All-Sky Astrophysics (CAASTRO).\\
$^{3}$ Australian Astronomical Observatory, PO Box 915, North Ryde, NSW 1670, Australia.\\
$^{4}$ International Centre for Radio Astronomy Research, The University of Western Australia, 35 Stirling Hwy, Crawley, WA 6009, Australia\\
$^{5}$ Centre for Astrophysics and Supercomputing, Swinburne University of Technology, Hawthorn, VIC 3122, Australia.\\
$^{6}$ Institute for Astronomy, University of Hawaii, 2680 Woodlawn Drive, Honolulu, HI 96822, USA.\\
$^{7}$ Dept. Physics and Astronomy, Macquarie University, NSW 2109,  Australia \\
$^{8}$ Research School for Astronomy \& Astrophysics. Mount Stromlo Observatory, ANU, Cotter Road Weston Creek, ACT 2611 Australia. \\
$^{9}$ Dept. Physics and Astronomy University of North Carolina Chapel Hill, NC 27599 USA \\
$^{10}$ Department of Physics, The University of Hong Kong, Hong Kong \\
$^{11}$  Envizi Group Suite 213, National Innovation Centre, Australian Technology Park, 4 Cornwallis Street, Eveleigh NSW 2015, Australia
}

\begin{document} 

\maketitle
\begin{abstract}

{We study the properties of kinematically disturbed galaxies in the SAMI Galaxy Survey using a quantitative criterion, based on kinemetry (Krajnovi\'{c} et al.). The approach, similar to the application of kinemetry by Shapiro et al. uses ionised gas kinematics, probed by H$\alpha$ emission. By this method 23$\pm$7$\%$ of our 360-galaxy sub-sample of the SAMI Galaxy Survey are kinematically asymmetric. Visual classifications agree with our kinemetric results for 90$\%$ of asymmetric and 95$\%$ of normal galaxies. We find stellar mass and kinematic asymmetry are inversely correlated and that kinematic asymmetry is both more frequent and stronger in low-mass galaxies. {This builds on previous studies that found high fractions of kinematic asymmetry in low mass galaxies using a variety of different methods.} Concentration of star formation and kinematic disturbance are found to be correlated, confirming results found in previous work. This effect is stronger for high mass galaxies (log($M_{*}) > 10$) and indicates that kinematic disturbance is linked to centrally concentrated star formation. Comparison of the inner (within 0.5R$_{e}$) and outer H$\alpha$ equivalent widths of asymmetric and normal galaxies shows a small but significant increase in inner equivalent width for asymmetric galaxies.}
\end{abstract}


\section{Introduction}
Hierarchical galaxy formation, in which dark matter halos form through a series of mergers, is central to $\Lambda$CDM cosmology \citep{peebles1982large,cole2008statistical,neistein2008merger}. It is well-established from N-body simulations [e.g. \citet{mayer2007multi,stewart2009galaxy}] that the rate of both halo and galaxy mergers should vary with redshift, but the precise details are not yet fully understood. There are discrepancies between theory and observation. For example, major mergers are known from theory to transform disky galaxies into thick, flared, bulge-dominated systems, that we would then expect to dominate the local universe. We instead see a large proportion of thin, disky systems at low redshift \citep{mihos1994dense,zucca2005vimos}.

Previous studies of galaxy disturbance have used images to either visually or quantitatively calculate the degree of disturbance within galaxies [e.g. \citet{burkey1994galaxy, lotz2004new, de2007millennium, darg2010galaxy}]. Visual identification of close pairs was used to determine merger rates in the Sloan Digital Sky Survey (SDSS) \citep{ellison2013galaxy} and Galaxy And Mass Assembly (GAMA) Survey \citep{robotham2014galaxy} samples. Structural analysis has also been used \citep{casteels2014galaxy}. These methods have been employed to determine merger rates in the nearby \citep{darg2010galaxy} and high redshift \citep{conselice2003direct} universe. These merger rates are, however, not always consistent. For example, the merger rate calculated by close pair counting in \citet{lin2004deep2} is an order of magnitude lower than that found by \citet{conselice2003direct}. The discrepancy has been noted in the literature, and efforts have been made to reconcile different results from different methods, such as in \citet{lotz2011major}.

Despite the success of these approaches, they have limitations: purely visual analyses are difficult (although not impossible) to quantify \citep{casteels2013galaxy}, and quantitative morphological techniques can be influenced by Hubble type and are restricted in the range of asymmetries to which they are sensitive. This is the case in the Concentration/Asymmetry/Smoothness system \citep{conselice2003evidence}. The Asymmetry parameter can be influenced by the presence of spiral arms, which can mask the effects of minor asymmetries \citep{conselice2003evidence}. Further, at high redshift, galaxies may be photometrically asymmetric but kinematically regular, due to features such as clumpy star formation \citep{glazebrook2012future}. 

Recent technological advances have revolutionised the reach of spectroscopy. Previously, the majority of spectroscopic measurements were taken with a single fibre or slit [e.g. \citet{york2000sloan,percival20012df,driver2009gama}]. For extended sources, such as galaxies, this approach is highly vulnerable to aperture effects, and it is difficult to gather information about spatial variation across the source. Integral field spectroscopy (IFS) solves this problem by taking spectra at various positions across the object, opening up new scientific possibilities.

{Instruments such as KMOS \citep{sharples2013first}, FLAMES \citep{pasquini2002installation} and SINFONI \citep{eisenhauer2003sinfoni} have demonstrated the usefulness of IFS}. Surveys such as SINS \citep{shapiro2008kinemetry} and ATLAS\textsuperscript{3D} \citep{cappellari2011atlas3d} have used IFS technology to spatially resolve and measure disturbances in the kinematics of high and low redshift galaxies, respectively. {The ATLAS\textsuperscript{3D} and SINS surveys differ in redshift and scale, but they have both} demonstrated that the 2D kinematics of galaxies can be used to effectively classify disturbed galaxies at various epochs \citep{shapiro2008kinemetry}. Both SINS and ATLAS\textsuperscript{3D} used monolithic instruments, which involve taking IFS measurements for each object individually. 

The Sydney--AAO Multi-object IFS (SAMI) is a multiplexed spectrograph, able to produce sample sizes into the thousands of galaxies on a much shorter timescale than a single IFS instrument \citep{croom2012sydney}. We here demonstrate how the large sample of IFS data in the SAMI Galaxy Survey can be used to determine an asymmetric fraction {(i.e. fraction of galaxies classified as asymmetric)} from gas kinematics, using a method based on kinemetry. Further, we show that {classification methods based on kinematics} are more robust in distinguishing interacting galaxies within the SAMI Galaxy Survey sample than methods based on quantitative morphology. We finally present results showing links between kinematic asymmetry, concentration of star formation and stellar mass.

In Section \ref{sec:sami}, we give a brief overview of the SAMI instrument and SAMI Galaxy Survey sample, outlining the data reduction pipeline and production of emission line maps, as well as characterising the sub-sample used in this work. In Section \ref{sec:kinemetry} we describe the kinematic classification method used to identify {kinematically} asymmetric galaxies, as well as a visual classification scheme used to calibrate the results from kinemetry. Section \ref{sec:sec4} shows the results of using kinemetry to identify perturbed galaxies, and {compares} them directly with results from quantitative morphology. Section \ref{sec:high_z} contains a comparison of our results to those from high redshift studies. {Sections} \ref{sec:sm} and \ref{sec:sfr} show relationships between {kinematic} asymmetry, stellar mass and star formation rate. Section \ref{sec:agn} briefly discusses the AGN in our sample. We conclude in Section \ref{sec:conclusions}.

\section{The SAMI Galaxy Survey}
\label{sec:sami}

The SAMI Galaxy Survey will {consist of 3400 galaxies across a range of stellar masses and environments, within $0.004<z<0.095$ \citep{croom2012sydney}}. The increased size of the survey sample is possible within a relatively short time frame because the SAMI instrument can take observations of up to 12 galaxies at a time (plus one calibration star), greatly increasing the ease with which large samples of IFS data can be obtained.

\subsection{The Sydney AAO Multi-object Integral Field Spectrograph}
The SAMI instrument takes integral field spectra for multiple objects using innovative imaging fibre bundles, called hexabundles \citep{bryant2011characterization}. The SAMI hexabundles consist of 61 optical fibres, with each core subtending $\sim$1.6 arcsec on sky, so that the total bundle diameter is $\sim$15 arcsec. Each bundle has a physical size $<$1mm, and a filling factor of 75\% \citep{bryant2012sami}. Thirteen bundles are manually plugged into a field plate, installed at the prime focus of the Anglo--Australian Telescope (AAT), and a fibre cable then feeds to  the double-beamed AAOmega spectrograph. AAOmega is configured with a dichroic splitting the light at 5700\AA\, with the 580V blue arm grating having a wavelength range of 3700�-5700\AA\  and the 1000R red arm grating having a wavelength range of 6300�-7400\AA. This gives resolutions of $R\sim1730$ in the blue arm and $R\sim4500$ in the red arm  \citep{croom2012sydney}. {All objects are observed with both arms of the spectrograph. Measurements used in this work are of the H$\alpha$ line, which is observed using the high resolution arm of the spectrograph.}

\subsection{Sample Selection and Data Reduction}
\label{sec:dr}
The 3400 galaxies of the SAMI Galaxy Survey were selected from the GAMA survey \citep{driver2009gama}, supplemented by 8 galaxy clusters. The GAMA galaxies selected consist of both group and field galaxies, and were chosen to reflect a broad range in stellar mass, in accordance with the science drivers of the SAMI Galaxy Survey. The GAMA galaxies lie on the celestial equator at RA $\sim$ 9, 12 and 15 hours, covering a total of 144 square degrees. {The GAMA survey provides a uniform and highly complete ($\sim99\%$), $r$-band flux limited survey of three large representative regions of the sky \citep{liske2015galaxy}.}

From the GAMA survey galaxies, those with unreliable redshifts or magnitudes were rejected. The final sample consists of four stellar mass, volume-limited sub-samples, along with additional dwarf galaxies at low redshift \citep{bryant2015sami}.

In order to provide more complete spatial coverage, each galaxy in the sample is observed in a series of 7 exposures. The individual exposures are combined to produce two datacubes per galaxy, one for each arm of the spectrograph, with 0.5 arcsec spatial pixels (spaxels).

All reduction of data taken using the AAOmega spectrograph at the AAT uses the software \textsc{2dfdr}\footnote{\url{http://www.aao.gov.au/science/software/2dfdr}}. The \textsc{2dfdr} package conducts all steps of the data reduction up to the production of wavelength calibration and sky subtracted spectra \citep{sharp2014sami}.

The first stage of data reduction using \textsc{2dfdr} is the subtraction of bias and dark frames. In order to ensure good extraction of spectra from the 2D data frame, it is necessary to accurately map the positions and profiles of the fibres across the detectors. This is accomplished using a fibre flat field frame, which is taken using an illumination of a white spot on the inside of the AAT dome. Wavelength calibration is then performed, using frames of standard CuAr arc-lamp exposures. Twilight exposures are used to measure the relative throughput between all fibres in the instrument. Finally, there are 26 dedicated sky fibres in the SAMI instrument. These fibres are set to blank sky positions for each field, to measure the sky spectrum, that can then be subtracted from all spectra.

The next steps of the data reduction process, the flux calibration correction and correction for telluric absorption, are independent of \textsc{2dfdr}, and are conducted using an external software suite, written in Python \citep{2014ascl.soft07006A}. For the flux calibration, spectrophotometric standard stars are observed, when possible, on the same night as the galaxy observations. Secondary standard stars are observed simultaneously with the galaxies, and are used to derive a correction in the telluric bands at 6850--6960\AA\   and 7130--7360\AA .  Datacubes are produced with regular 0''.5 square spaxels. Some spatial resolution is lost when convolving 1''.6 fibres with 0''.5 spaxels, because the flux in each output spaxel is taken as the mean of the flux in each input fibre, with weightings given by the fractional overlap of the fibre with the spaxel. To regain some of the lost resolution, a 0''.8 diameter fibre footprint is used to calculate the overlaps, a \textit{drizzle}-like process that was initially developed to resample high-resolution imaging from \textit{HST/}WPFC2 \citet{fruchter2002drizzle}.  The final product is a pair of datacubes per galaxy, one for each arm of the spectrograph, with multiple extensions. The extensions contain, respectively, the flux, variance, weight and covariance datacubes.  A full explanation of the data reduction process can be found in \citet{sharp2014sami}

Due to the re-sampling process used to make the datacubes, the noise in neighbouring spaxels is correlated.  This effect is negligible for spaxels spaced more than $\sim$2.5 spaxels apart. A full explanation of how covariance is handled in SAMI Galaxy Survey data can be found in \citet{sharp2014sami}.

\subsection{{\small LZIFU}{\sc} and Data Products}
\label{sec:LZIFU}
{\small LZIFU}{\sc} is an Interactive Data Language ({\small IDL}{\sc}) spectral fitting pipeline, designed to perform flexible emission line fitting in IFS data cubes. It works by fitting, and then removing, the continuum emission in each spaxel by using simple stellar population (SSP) models (Ho et al., in prep; \citealt{ho2014sami}). {\small LZIFU}{\sc} uses the penalized pixel-fitting routine, {\small pPXF}{\sc} \citep{cappellari2004parametric} to do the continuum fit. We use theoretical SSP models, assuming Padova isochrones, of solar metallicity and 18 ages  \citep{delgado2005evolutionary}. After the continuum emission is removed, {\small LZIFU}{\sc} models user-assigned emission lines as Gaussian profiles, and then performs a bounded value nonlinear least-squares fit. This is done using {\small IDL}{\sc}'s Levenberg-Marquardt least-squares method \citet{markwardt2009non}, with {\small LZIFU}{\sc} automatically establishing reasonable initial guesses for the wavelength of emission lines. Users have the choice to model individual emission line as either 1, 2, or 3-component Gaussians, describing possible different kinematic components. We use 1-component fits in this work.

The key data products from the {\small LZIFU}{\sc} pipeline are 2D emission line strength and kinematic maps (showing the kinematics of gas in the galaxy, both velocity and velocity dispersion), for user-assigned lines, with error maps. For a more detailed explanation of the {\small LZIFU}{\sc} pipeline, see \citet{ho2014sami}.

\subsection{Selecting and Characterising the Sample}
\label{sec:characterising}
This work uses the first 451 galaxies to be processed through the {\small LZIFU}{\sc} pipeline. We took the velocity and velocity dispersion maps generated by the H$\alpha$ fit from {\small LZIFU}{\sc} and applied a S/N cut of 10 to the H$\alpha$ emission map. If, after applying the cut, there were fewer than 200 spaxels remaining in the flux map, out of an original $\sim$1225 spaxels, the galaxy was excluded from the sample (the average being $\sim$450 spaxels remaining). This was to avoid {problems fitting severely discontinuous, low S/N maps}. After the above cuts were performed there remained a sample of 360 galaxies, which forms the basis of our study.

\begin{figure}
\centering
\includegraphics[width=9cm]{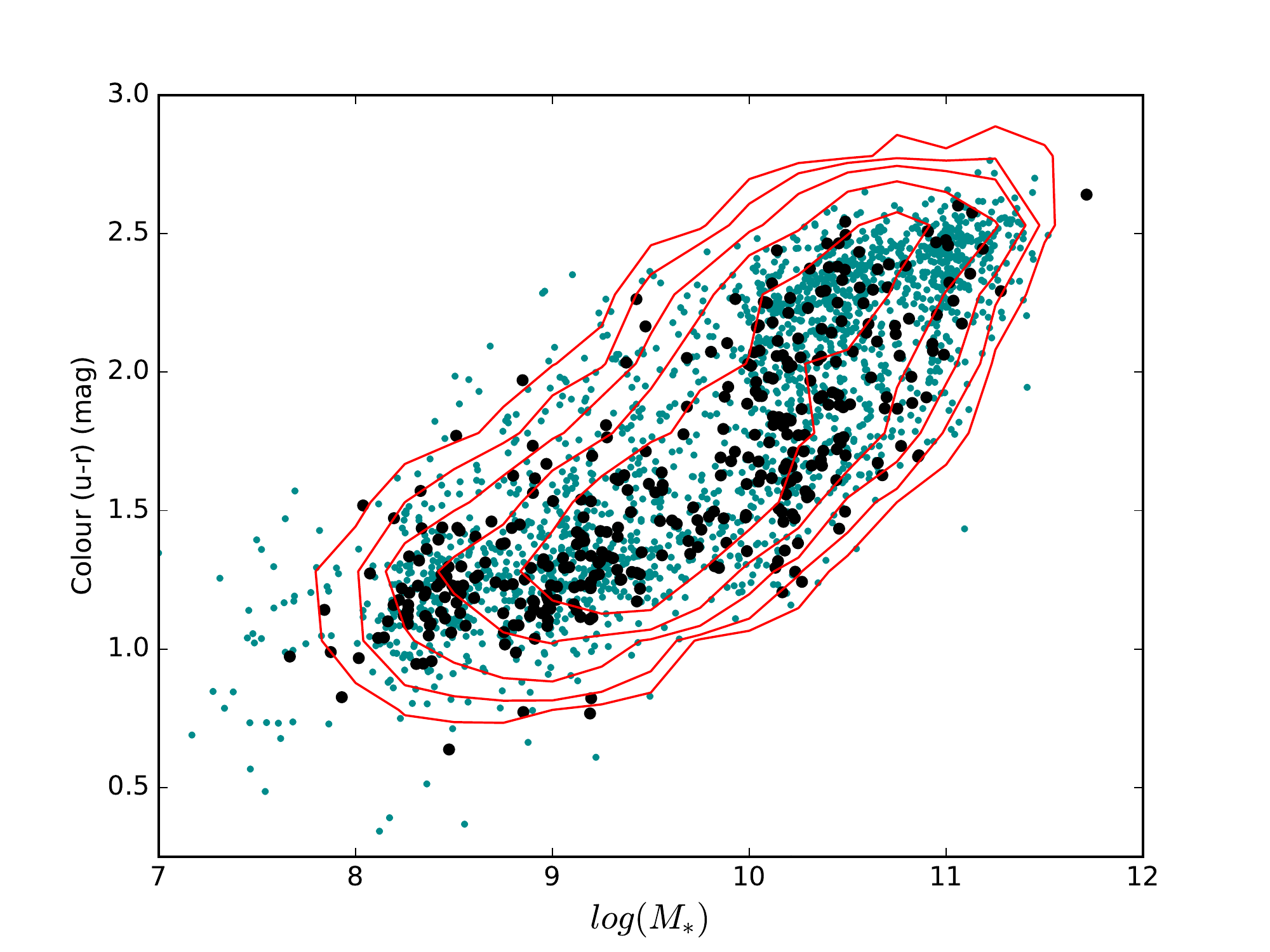}
\caption{{We show colour and stellar mass} for the entire GAMA Survey sample {(log-spaced red contours)} SAMI Galaxy Survey sample (blue points) and the sample used in this work (black dots). Our sample is evenly distributed within the main SAMI Galaxy Survey sample. The data used were taken from the GAMA Data Release 2 catalogues ApMatchedPhotom \citep{hill2011galaxy} and StellarMasses \citep{taylor2011galaxy}.}
\label{fig:colour_mag_1}
\end{figure}

\begin{figure}
\centering
\includegraphics[width=9cm]{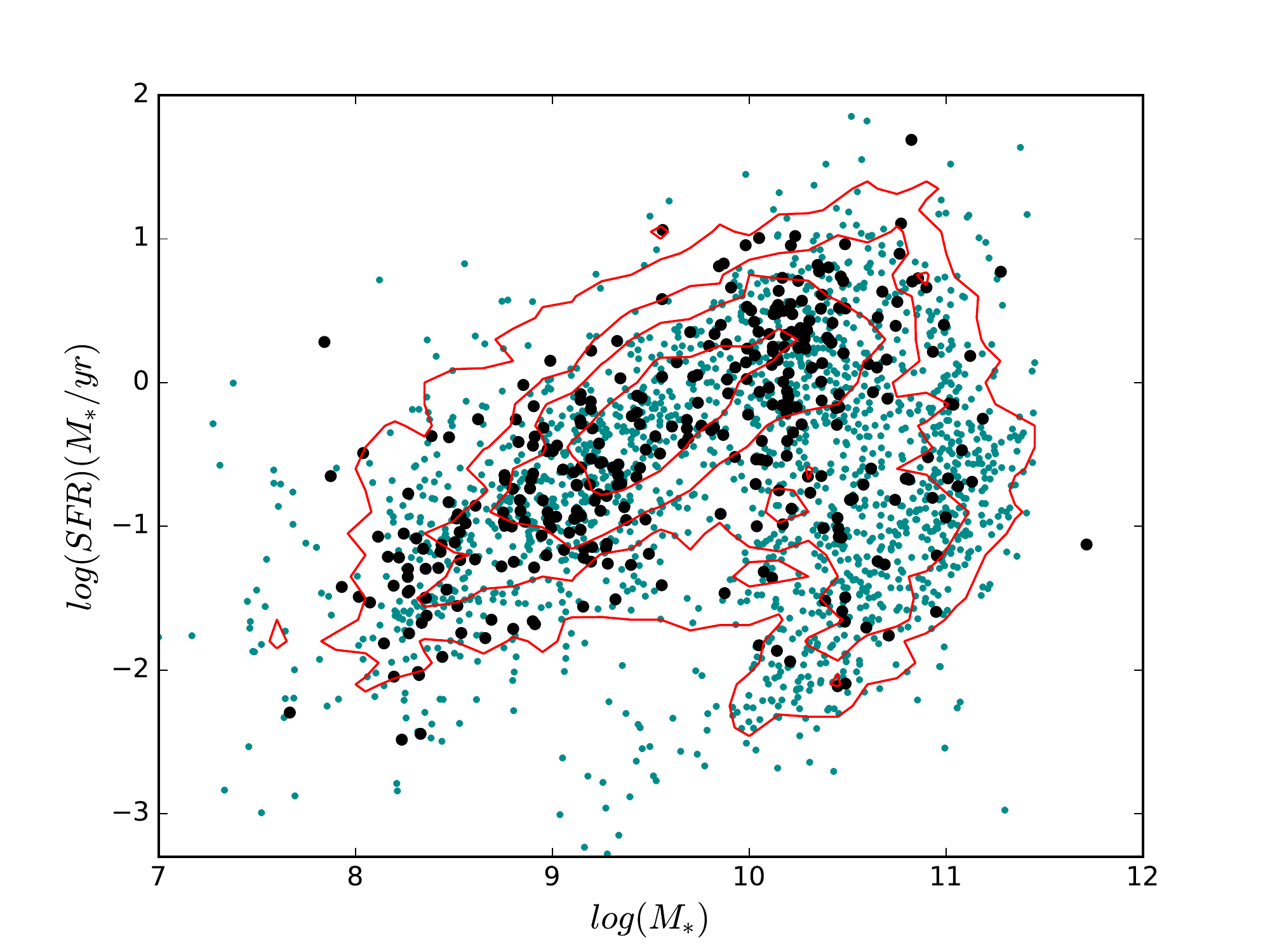}
\caption{{We show} SFR against stellar mass for the GAMA Survey sample {(log-spaced red contours)}, SAMI Galaxy Survey sample (blue points) and our sample (black dots). The H$\alpha$ kinemetry sample has a slight bias against galaxies that fall off the main sequence of star formation, due to the cut in H$\alpha$ S/N. The data used were taken from the GAMA Data Release 2 catalogues StellarMasses \citep{taylor2011galaxy} and SpecLineSFR \citep{hopkins2013galaxy}.}
\label{fig:sfr_1}
\end{figure}

\begin{figure}
\centering
\includegraphics[width=9cm]{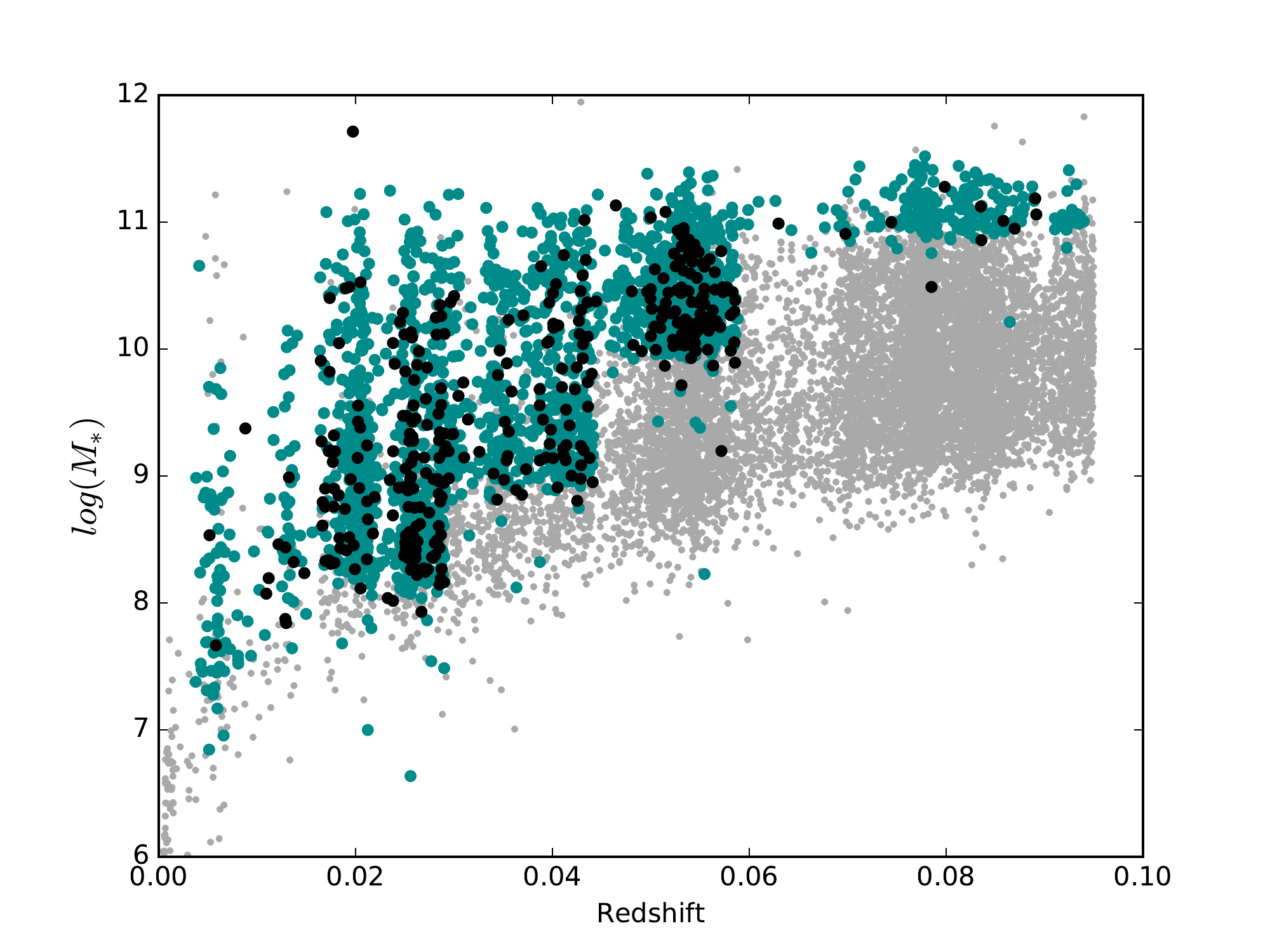}
\caption{{We show} redshift and stellar mass for the GAMA Survey sample (grey points), SAMI Galaxy Survey sample (blue points) and our sample (black dots). Our sample is evenly distributed within the main SAMI Galaxy Survey sample. The data used were taken from the GAMA Data Release 2 catalogues StellarMasses \citep{taylor2011galaxy} and SpecLineSFR \citep{hopkins2013galaxy}.}
\label{fig:redshift_plot}
\end{figure}

\begin{figure}
\centering
\includegraphics[width=9cm]{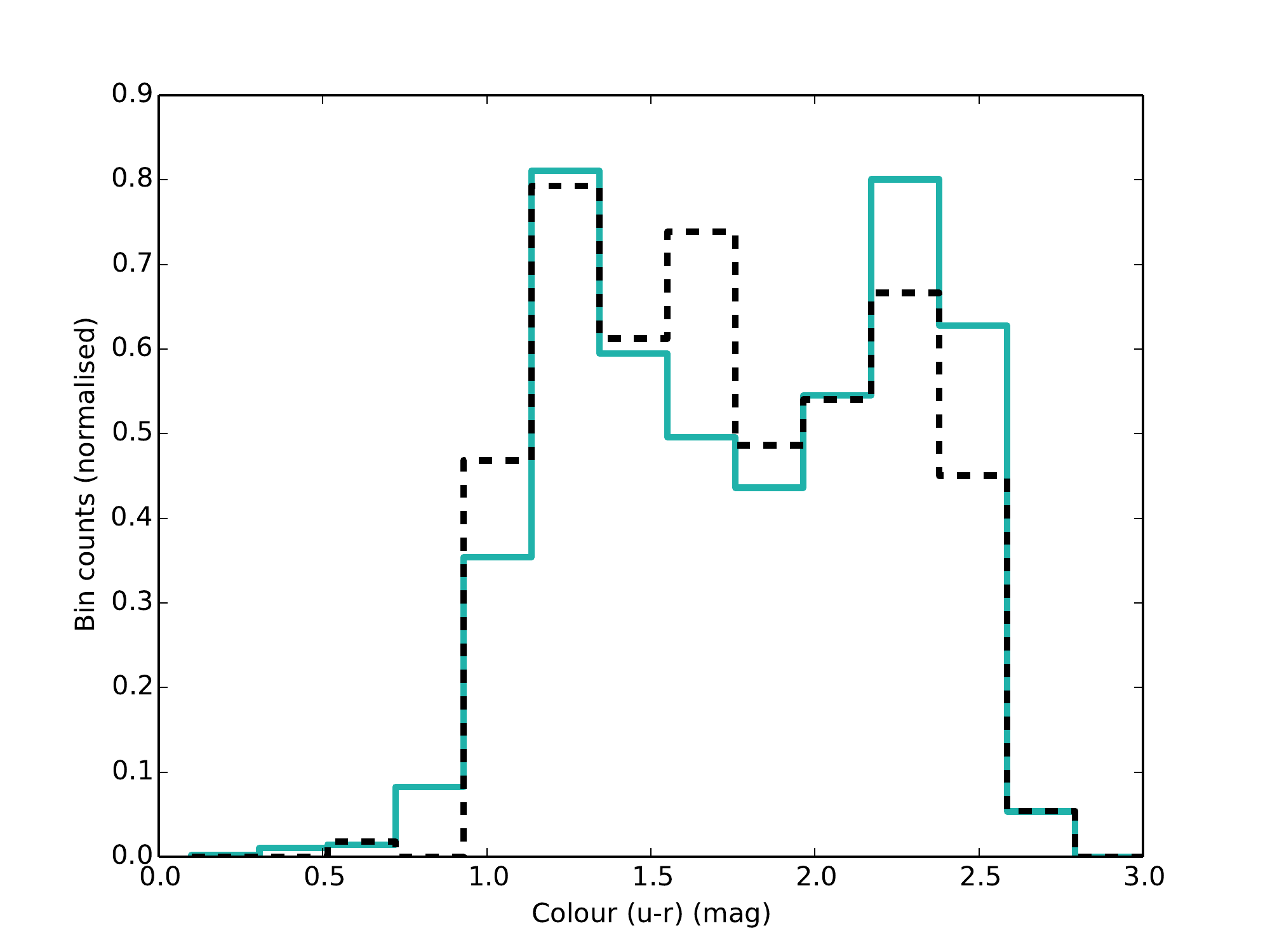}
\caption{{We show a} normalised histogram of the colours in the SAMI Galaxy Survey sample (blue) and the sample used in this work (black, dashed). The data used were taken from the GAMA Data Release 2 catalogue SpecLineSFR \citep{hopkins2013galaxy}.}
\label{fig:colours_hist}
\end{figure}

\begin{figure}
\centering
\includegraphics[width=9cm]{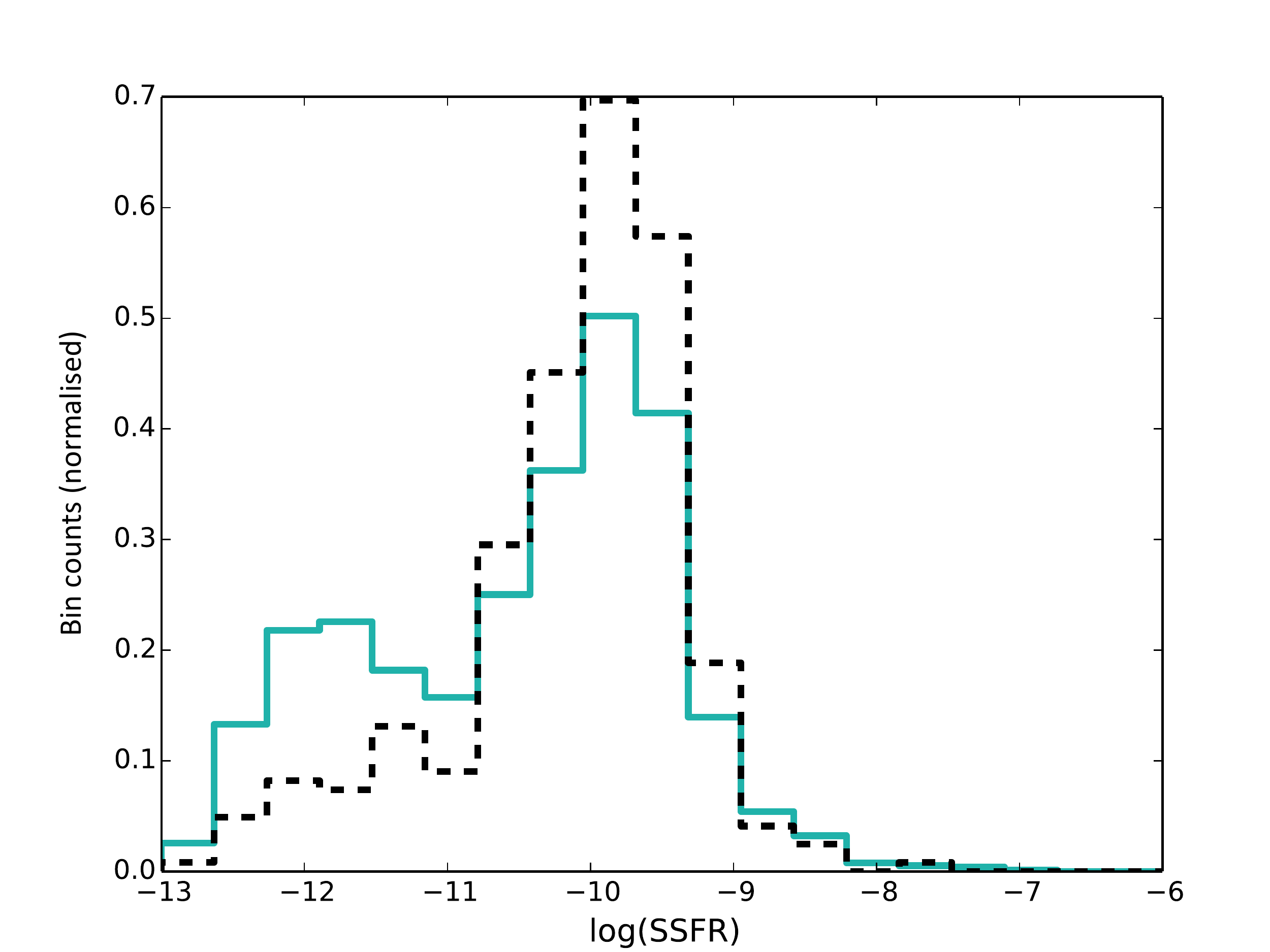}
\caption{{We show a }normalised histogram of the specific SFR in the SAMI Galaxy Survey sample (blue) and the sample used in this work (black, dashed). The data used were taken from the GAMA Data Release 2 catalogues  SpecLineSFR \citep{hill2011galaxy} and StellarMasses \citep{taylor2011galaxy}.}
\label{fig:sfrs_hist}
\end{figure}

Fig.~\ref{fig:colour_mag_1} shows the distribution of the rest frame $(u-r)$ colour-stellar mass for our sample (black points) the full SAMI Galaxy Survey sample (blue points),  and the parent GAMA survey sample (log-spaced red contours). For ease of comparison, the GAMA survey sample in these figures has been restricted to $z\leq 0.095$. Likewise, Fig.~\ref{fig:sfr_1} shows dust obscuration-corrected star formation rate (SFR) against stellar mass, and Fig.~\ref{fig:redshift_plot} shows redshift against stellar mass, with the GAMA Survey sample in grey. Finally, Fig.~\ref{fig:colours_hist} and Fig.~\ref{fig:sfrs_hist} show histograms of the colour and specific star formation rate (SSFR) for both samples. SFRs are calculated from the luminosity of the hydrogen Balmer lines. This is because the Balmer emission-line luminosity is directly proportional to the total ionizing flux of the stars in H\textsc{II} regions of star forming galaxies. Medians for the stellar mass, colour and SFR for the full SAMI Galaxy Survey Sample and our sample,{and the results of two-sample Kolmogorov-Smirnov tests between the samples} are provided in Table~\ref{table:sample_specs}. {Our sample shows a slight bias towards bluer galaxies with low stellar masses and higher SFR {(see Table~\ref{table:sample_specs})}, as is to be expected when applying an H$\alpha$ emission-based S/N cut. Despite the slight bias, we cover the full parameter space (see Fig.~\ref{fig:colour_mag_1},~\ref{fig:sfr_1},~\ref{fig:redshift_plot}).}

\begin{center}
\begin{table}
  \begin{tabular}{ b{2.2cm}  l  b{1.8cm}  l  l }
    \hline
    Sample & log($M_{*}$) & $(u-r)$ Colour & log(SFR)  \\ \hline
    SAMI Galaxy Survey sample & 10.04$\pm 0.024$& 1.77 $\pm0.012$& -0.66$\pm0.023$ \\
    Our sample & 9.70$\pm0.056$ & 1.53$\pm 0.028$ & -0.51$\pm0.053$\\
    \hline
     2KS test $p$-value & 6.15$\times10^{-6}$  & 2.70$\times10^{-10}$ & 0.021  \\
    \hline
  \end{tabular}
\caption[Table caption text]{{This table shows the median stellar mass, colour and SFR values for the SAMI Galaxy Survey sample and the sub-sample used in this work, and the results of two-sample Kolmogorov-Smirnov tests for colour, mass and SFR. There is a slight bias towards blue, high-SFR, low stellar mass galaxies in our sample, as is to be expected after performing an H$\alpha$ S/N cut. We nevertheless cover the full parameter space (see Fig.~\ref{fig:colour_mag_1},~\ref{fig:sfr_1},~\ref{fig:redshift_plot}). {The uncertainty we quote in the table is the statistical error on the median, rather than the uncertainty on SFR or colour for individual galaxies.}}} 
\label{table:sample_specs}
\end{table}
\end{center}

\section{Using kinemetry to quantify {kinematic} asymmetry}
\label{sec:kinemetry}

\subsection{The kinemetry algorithm}
\label{sec:intro_kin}

Kinemetry is an extension of photometry to the higher order moments of the line of sight velocity distribution (LOSVD)\footnote{The kinemetry code is written in IDL, and can be found at \url{http://davor.krajnovic.org/idl/} \citep{krajnovic2006kinemetry}.}. It was developed as a means to quantify asymmetries in stellar velocity (and velocity dispersion) maps. These anomalies may be caused by internal disturbances or by external factors, namely interactions \citep{krajnovic2006kinemetry}.

The method works by modelling kinematic maps as a sequence of concentric ellipses, with parameters defined by the galaxy centre, kinematic position angle (PA) and ellipticity. It is possible to fit the latter two parameters within kinemetry, or to determine them by other means and exclude them from the fitting procedure. For each ellipse, the kinematic moment is extracted and decomposed into the Fourier series:
\begin{equation}
K(a,\psi)=A_{0}(a)+\sum_{n=1}^{N}(A_{n}(a)sin(n\psi)+B_{n}(a)cos(n\psi)),
\end{equation}
where $\psi$ is the azimuthal angle in the galaxy plane, and $a$ is the ellipse semi-major axis length. Points along the ellipse perimeter are sampled uniformly in $\psi$, making them equidistant in circular projection. The series can be expressed more compactly, as  \citep{krajnovic2006kinemetry}:
\begin{equation}
K(a,\psi)=A_{0}(a)+\sum_{n=1}^{N}k_{n}(a)cos[n(\psi-\phi_{n}(a))],
\end{equation}
with the amplitude and phase coefficients ($k_{n},\phi_{n}$) defined as:
\begin{equation}
k_{n}=\sqrt{A^{2}_{n}+B^{2}_{n}}
\end{equation}
and
\begin{equation}\label{eq:phi}
\phi_{n}=arctan\left(\frac{A_{n}}{B_{n}}\right) .
\end{equation}

\begin{figure*}
\centering
\begin{center}
\includegraphics[width=3.5cm]
{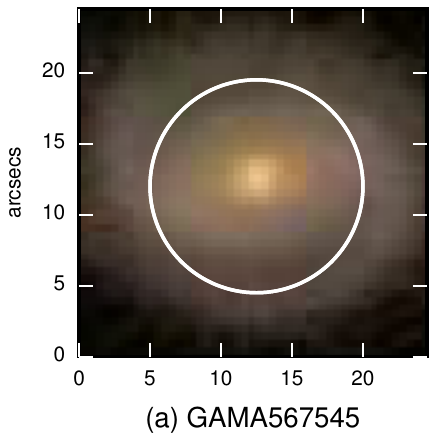}
\includegraphics[width=3.5cm]
{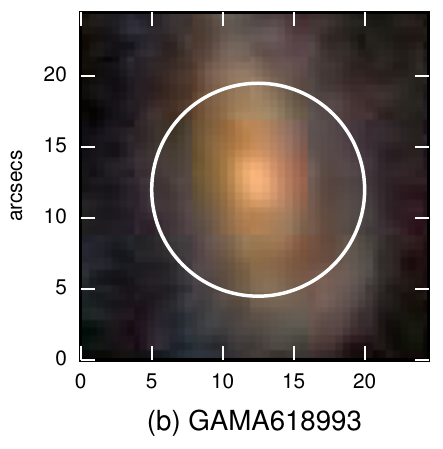}
\includegraphics[width=17.2cm]
{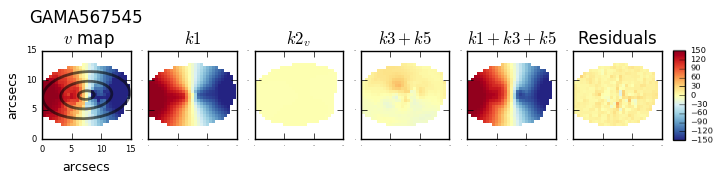}
\includegraphics[width=17cm]
{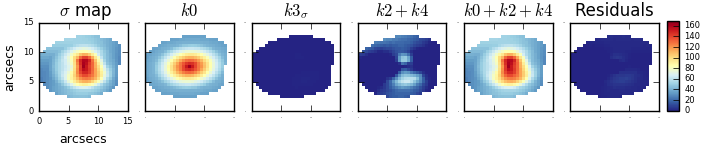}
\includegraphics[width=17.5cm]
{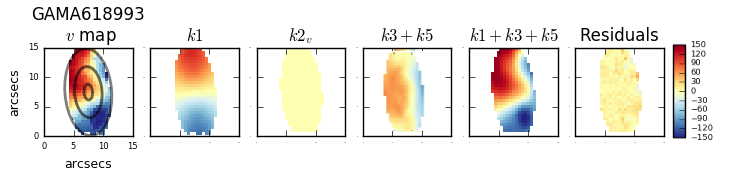}
\includegraphics[width=17cm]
{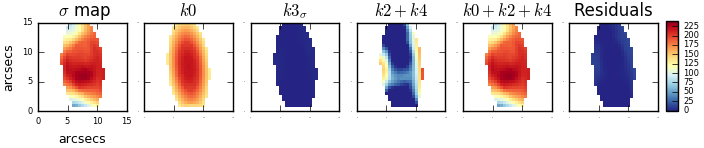}
\caption{{The} {input data and} fitted kinemetric models for the velocity ($v$) and $\sigma$ fields of a morphologically normal galaxy, GAMA567545 (a, top), and a {morphologically} asymmetric galaxy, GAMA618993 (b, bottom), with $gri$ SDSS images. For each galaxy, the top row of plots shows the $v$ models. The leftmost panels are the {observed} SAMI H$\alpha$ $v$ fields, with the ellipses fit by kinemetry overplotted. The PA and ellipticity of the ellipses were fit using photometry from the GAMA Survey DR2 catalogue. The second panels show the first order model, $k1$ i.e. the rotating disk component. The rotating disk model of the normal galaxy is almost identical to the data, indicating that the first component is all that is needed to fit the data for that galaxy, unlike the asymmetric galaxy. The second order moment, designated $k2_{v}$, is shown in the third panel. For both galaxies, the second order moment is close to zero, indicating that there is little `spillover' from the odd into the even moments. The fourth panels explicitly show the  contrast between the galaxies, with the normal galaxy having little power in the higher order modes ($k_{3},k_{5}$), whereas the asymmetric galaxy has significant power. {The fifth plots are of the combined $k_{1}+k_{3}+k_{5}$ models for both galaxies, and the sixth panel shows the residuals of the fit from the data.} The second row of plots for each galaxy shows the $\sigma$ models. The first plots are the $\sigma$ fields, followed by the zeroth order moment model, $k0$. The third order $\sigma$ moment, designated as $k3_{\sigma}$ is shown in the third panel. Analogous to $k2_{v}$, there is minimal power in these modes. The higher order $\sigma$ models are shown fourth, and { the fifth panel shows the full $\sigma$ model, followed by the residuals}. Note that $v$ values in the colourbars are given in km/s.}
\label{fig:kin_output}
\end{center}
\end{figure*}

The moment maps for both velocity and velocity dispersion can thus be described by the geometry of the ellipses and power in the coefficients $k_{n}$ of the Fourier terms as a function of \textit{a} \citep{krajnovic2006kinemetry}.

The velocity field of a completely normal, rotating disk would be entirely contained in the $cos(\psi)$ term {of Equation 2}, with zero power in the higher order modes, since the velocity peaks at the galaxy major axis and goes to zero along the minor axis. As a result, the power in the $B_{1}$ term represents circular motion, with deviations carried in the other coefficients. Fig.~\ref{fig:kin_output} shows results of kinemetric fitting to the H$\alpha$ velocity maps of a normal and an asymmetric galaxy in our sample. For the normal galaxy (top), GAMA567545, all the power is in the first order moment ($k_{1}$), whereas the asymmetric galaxy (bottom), GAMA628993, has significant power in the higher order modes ($k_{3}+k_{5}$). 

Analogously to the velocity field, a map of the velocity dispersion field of a perfectly normal rotating disk would have all power in the $A_{0}$ term (i.e. radial dispersion gradient) \citep{krajnovic2006kinemetry}. The velocity dispersion field is an even moment of the LOSVD, and therefore its kinemetric analysis reduces to traditional surface photometry. 

Kinemetry contains routines to fit the PA and ellipticity of the input kinematic fields. These inbuilt routines were used on the high S/N data fields from the ATLAS\textsuperscript{3D} survey. Similarly to the SINS survey \citep{shapiro2008kinemetry}, we found that the lower S/N of the SAMI Galaxy Survey fields, compared to ATLAS\textsuperscript{3D} data, led to unstable fits to these parameters. We used the PA and ellipticity from the single Sersic fits to the SDSS $r$-band images in the GAMA Survey DR2 catalogue SersicCat \citep{kelvin2012galaxy}. This is a reasonable step for disturbance measures because, for {kinematically} normal {rotating} galaxies, the photometric and kinematic PA {should} agree. {This is not the case for galaxies with misaligned kinematic and photometric PAs. However, we find that galaxies with misaligned PAs are generally classified as visually asymmetric (see Section~\ref{sec:vispert}).}

\subsection{{Using kinemetry to measure kinematic asymmetry}}
\label{sec:classpert}

We perform kinemetry on the H$\alpha$ velocity and velocity dispersion maps for all 360 galaxies in our sample. Following work on the SINS survey \citep{shapiro2008kinemetry}, we calculate radial {kinematic} asymmetry values for the velocity and velocity dispersion fields of each galaxy, respectively:
\begin{equation}
v_{asym}=\frac{k_{3,v}+k_{5,v}}{2k_{1,v}}
\\ \sigma_{asym}=\frac{k_{2,\sigma}+k_{4,\sigma}}{2k_{1,v}},
\end{equation}
We have slightly modified the method of \citet{shapiro2008kinemetry} that fit all moments (odd and even) when calculating both velocity and velocity dispersion asymmetry:
\begin{equation}
v_{asym,SINS}=\frac{k_{2,v}+k_{3,v}+k_{4,v}+k_{5,v}}{4k_{1,v}} \\
\end{equation}
\begin{equation}
\sigma_{asym,SINS}=\frac{k_{1,\sigma}+k_{2,\sigma}+k_{3,\sigma}+k_{4,\sigma}+k_{5,\sigma}}{5k_{1,v}},
\end{equation}
They did this because they were specifically looking for the signatures of major mergers, which produce extremely disturbed velocity fields, with power in all higher order moments \citep{shapiro2008kinemetry}. Due to the comparatively small amplitude of the {kinematic} asymmetries in our sample, we found the asymmetry contributions of even moments to velocity asymmetry to be negligible (see velocity fields of GAMA618993 in Fig.~\ref{fig:kin_output} for an example), and similarly for the contributions of odd moments to the velocity dispersion asymmetry. Accordingly, we used only odd moments for the velocity fields and the even moments for the velocity dispersion, as was done for ATLAS\textsuperscript{3D} data \citep{krajnovic2011atlas3d}. 

\citet{krajnovic2006kinemetry} note that, when studying the velocity dispersion of the stellar component of the galaxy, it is appropriate to normalise over the zeroth moment of the velocity dispersion. We, however,  follow \citet{shapiro2008kinemetry} in normalising both quantities over the velocity ($k_{1}$), rather than the velocity dispersion, because the velocity dispersion of the gas component of a galaxy is extremely sensitive to shocks and other features. This makes normalising to the rotation curve (the first moment of the velocity) a more reliable choice, as it is not as sensitive to these features, but is sensitive to the potential. 

To determine the centre of the kinematic maps used in this analysis, we fit a 2-dimensional Gaussian to the  SAMI Galaxy Survey $r$-band continuum flux maps and took the centroid of the location of the 25 brightest spaxels in a 6x6 pixel area around the centre of the fitted Gaussian. We did not use the the H$\alpha$ emission maps because they contain clumps of star formation and other features, which make determining the centre from these maps potentially unreliable.

\begin{figure} 
\begin{subfigure}{.25\textwidth}
\centering
\includegraphics[width=4.5cm]{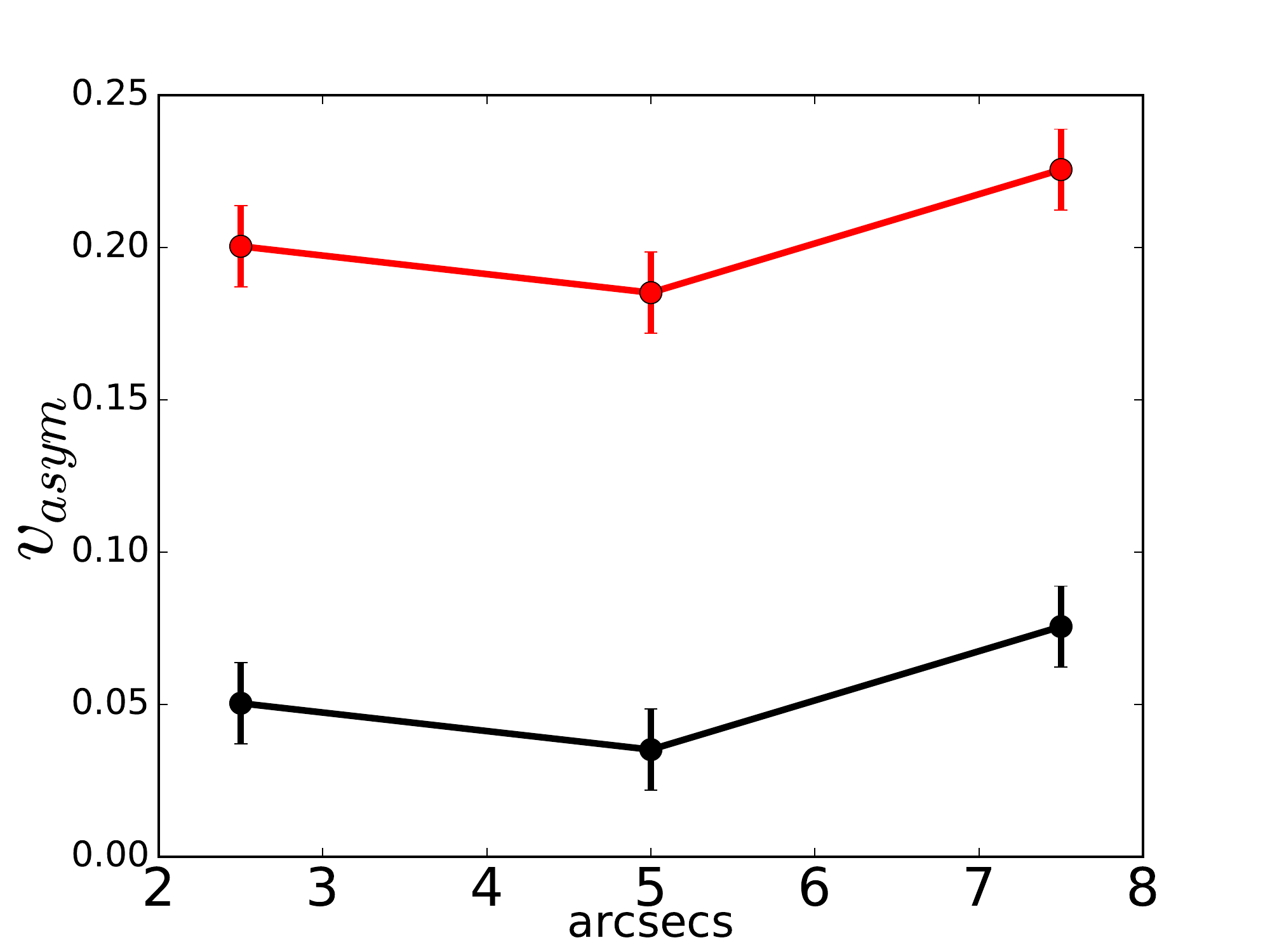}

\label{fig:vel_curves}
\end{subfigure}%
\begin{subfigure}{.25\textwidth}
\includegraphics[width=4.5cm]{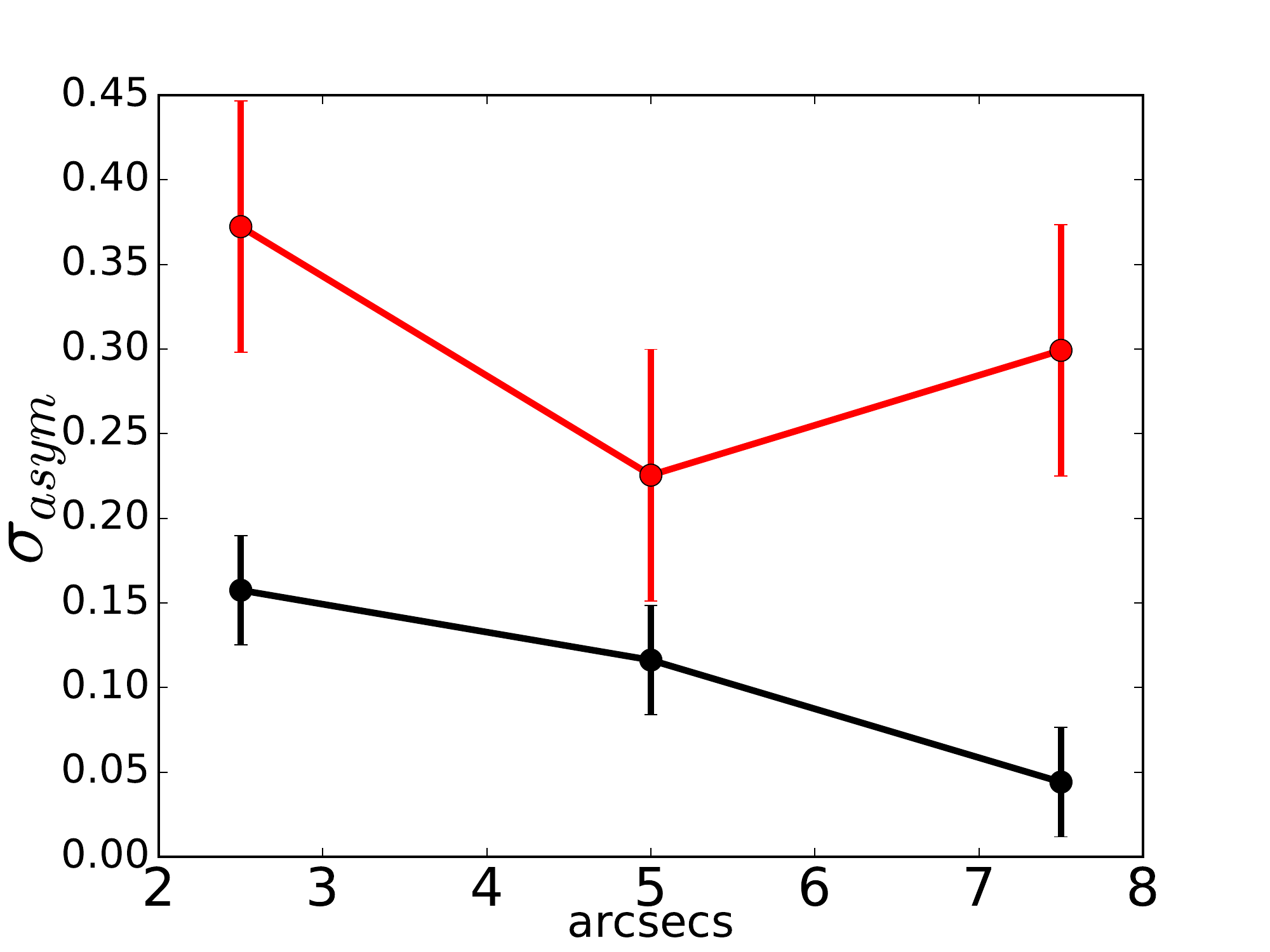}
\end{subfigure}%
\caption{{We show} the values of  $v_{asym}$ (left) and $\sigma_{asym}$ (right), with 1-$\sigma$ errors, at the positions of the semi-major radii of the fitted ellipses for an asymmetric (red, GAMA618993) and non-asymmetric (black, GAMA567545) galaxy. It can clearly be seen that the velocity and velocity dispersion asymmetry values for the asymmetric galaxy are significantly higher than those of the normal galaxy. These are the same two galaxies shown in Fig.\ref{fig:kin_output}.}
\label{fig:sig_curves}
\end{figure}

In order to accommodate the effects of covariance outlined in Section~\ref{sec:dr}, we altered the kinemetry fitting routine, such that the spacing between ellipses was constrained to be $\geq$2.5 arcsecs. The effective covariance diameter is $\sim$ 2.5 arcsecs, so this ensures that data points used by adjacent ellipses are independent of each other. The 1-$\sigma$ errors on $v_{asym}$ and $\sigma_{asym}$ were then bootstrapped to produce the errors on $\overline{v_{asym}}$ and $\overline{\sigma_{asym}}$. The bootstrapping involved 100 iterations of adding random noise to the velocity and velocity dispersion fields, and then recalculating the median of $v_{asym}$ and $\sigma_{asym}$. The final values for the errors on $\overline{v_{asym}}$ and $\overline{\sigma_{asym}}$ are taken from the scatter in the results of the iterations. Because of the enforced separation between ellipses, we only fit across three kinemetric ellipses for each galaxy (see Fig.~\ref{fig:kin_output}). This does entail the possibility of losing detail, as small {kinematic} asymmetries may be lost `between' successive ellipses. {It is also possible that small features identified in the morphological classification in Section ~\ref{sec:vispert} may not be detected using this method. However, due to covariance, a more finely sampled kinematic fit would not increase the accuracy of our results. The qualitative nature of the visual classification means that we cannot quantitatively define a lower limit for the size of features that caused a galaxy to be classified as `definitely asymmetric'. However, the process focussed on large-scale features, such as tidal tails.}

In Fig. ~\ref{fig:sig_curves}, $v_{asym}$ and $\sigma_{asym}$ are plotted at the semi-major radii of each of the fitted ellipses for a visually normal and an asymmetric galaxy (the same galaxies as shown in Fig. \ref{fig:kin_output}). We see that the visually asymmetric galaxy (in red) has consistently higher kinematic asymmetry than the visually normal galaxy.

\begin{figure*}
\centering
\includegraphics[width=13cm]{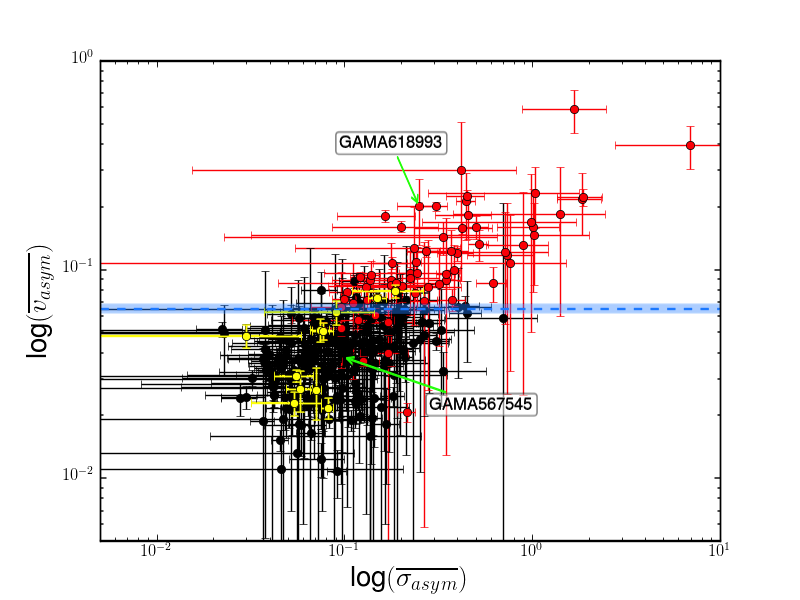}

\includegraphics[width=18cm]{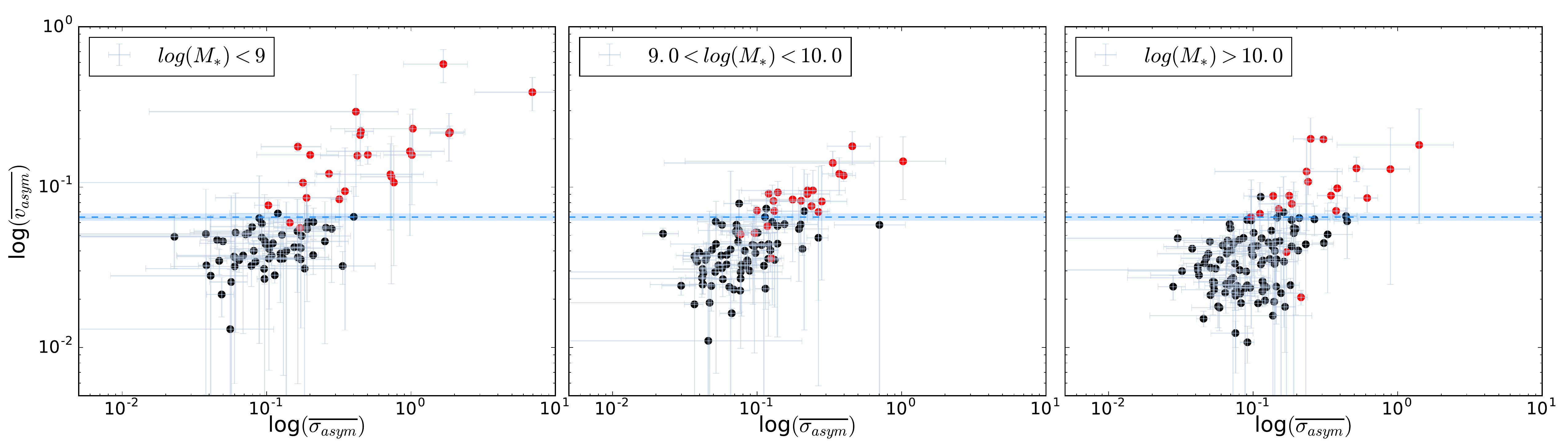}
\caption{{These plots summarise the results of kinemetry for the 360 galaxies in our sample (top) and three stellar mass bins (bottom row).} The median kinemetric asymmetries for both velocity and velocity dispersion across fitted ellipse radii are normalised over the rotational velocity and plotted against each other. The sample was visually classified morphologically (using the SDSS DR10 images), with visually asymmetric galaxies shown here in red, and `possibly asymmetric' galaxies in yellow. We show the $\overline{v_{asym}}$ cutoff for {kinematic} asymmetry derived in the text (blue, dashed line). The two example galaxies from Fig.~\ref{fig:kin_output}, GAMA567545 (the normal galaxy) and GAMA618993 (the asymmetric galaxy) are indicated by the green arrows, for reference.}
\label{fig:sami_asyms}
\end{figure*}

We take the median of $v_{asym}$ and $\sigma_{asym}$ over all ellipses (relative to the centre of the continuum emission), so that each galaxy's total kinemetric asymmetry can be expressed as the combination of $\overline{v_{asym}}$ and $\overline{\sigma_{asym}}$. {Fig.~\ref{fig:sami_asyms} shows $\overline{\sigma_{asym}}$ against $\overline{v_{asym}}$ for the whole sample and for three mass bins. A discussion of the relationship between stellar mass and $\overline{v_{asym}}$ can be found in Section \ref{sec:sm}.} As in \citet{shapiro2008kinemetry}, it will be useful to define a cutoff on this plane, above which galaxies may be considered kinematically asymmetric. Given that both $\overline{\sigma_{asym}}$ against $\overline{v_{asym}}$ form continuous distributions, it is not immediately obvious, based purely on kinematics, where to draw such a cutoff. Accordingly, we use visual classification to provide a guide.

\subsection{Visually classifying asymmetric galaxies}
\label{sec:vispert}
The 360 galaxies in our sample were morphologically classified visually, by members of the team, using $gri$ composite images from the SDSS DR10 catalogue \citep{ahn2014tenth}. The SDSS DR10 catalogue has sufficient depth and spatial resolution to show the large-scale features relevant to our work. The median seeing, defined as the FWHM of the point spread function, in the SDSS sample is 1.43'' in the $r$-band. This is much smaller than the median effective radius of galaxies in the SAMI Galaxy Survey sample, 4.4''  \citep{bryant2015sami}. {The features that define morphological asymmetry for this paper (e.g. double cores and large tidal tails) are larger than the seeing in the SDSS DR10 images.} The SDSS DR10 images have been previously successfully used to identify large-scale features such as mergers in Galaxy Zoo \citep{darg2010galaxy,casteels2013galaxy}.

Of course, visual classification is a qualitative approach, and vulnerable to bias or error. We compensated for this by having multiple people classify each galaxy, with an average of 5 individuals classifying each galaxy. 

\begin{figure}
\centering
\includegraphics[width=9cm]{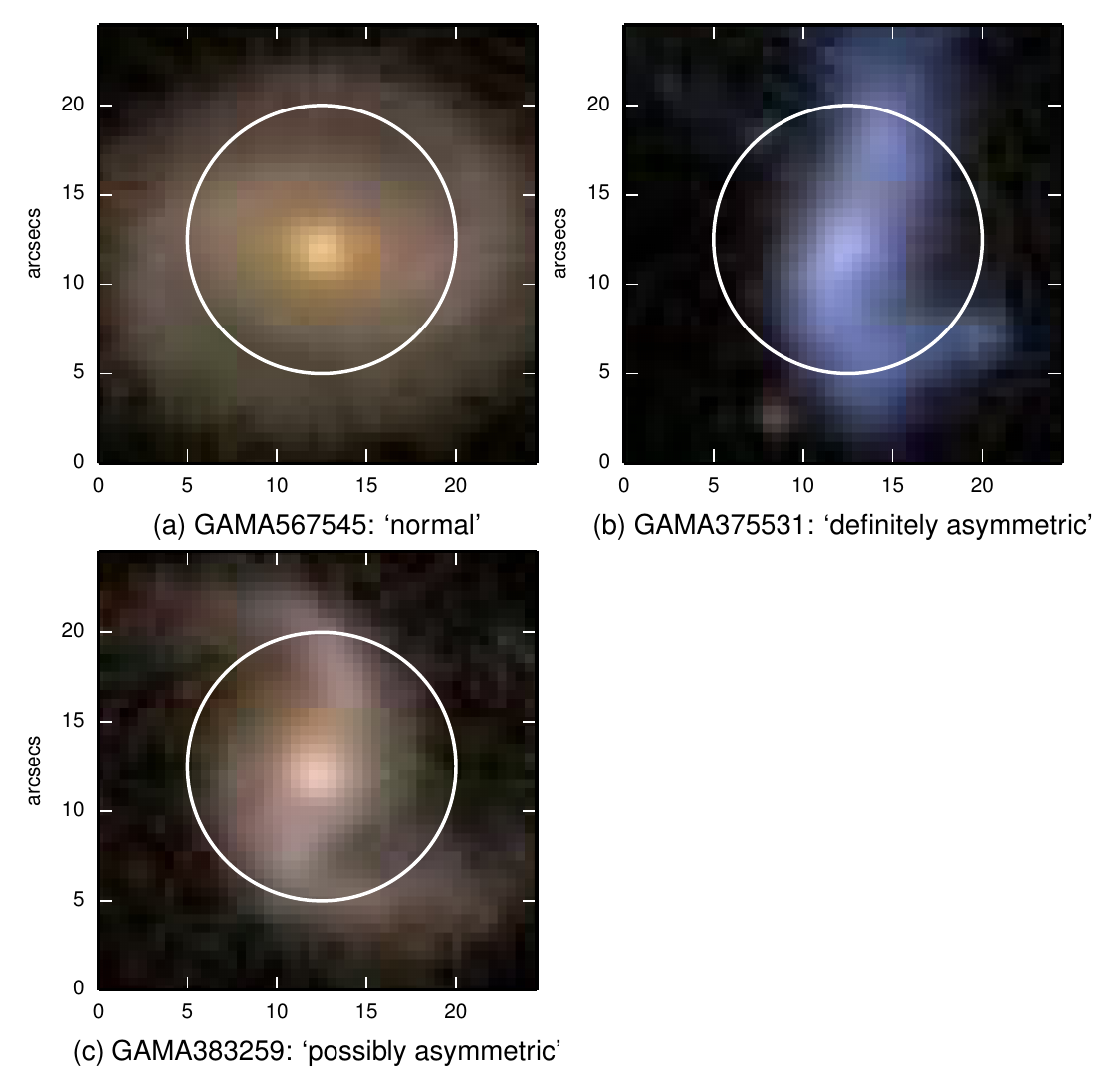}
\caption{From top left {we give}: examples of `normal', `definitely {visually} asymmetric', and `possibly {visually} asymmetric' galaxies from the morphological classification. The SAMI instrument field of view is shown as a white circle. Images are from the SDSS DR10 catalogue used to classify the galaxies.}
\label{fig:examples}
\end{figure}

The categories for classification were (with quoted instructions for classification into each group {and number of galaxies sorted into each category}):
\begin{itemize}
\item normal: ``No evidence of interaction. These galaxies are almost completely smooth and symmetrical." {246 galaxies}
\item definitely {visually} asymmetric: ``These galaxies do not have to be major merger remnants, but are nevertheless clearly distinguishable from normal/slightly/possibly asymmetric galaxies. They might have evidence of a double core, extremely distorted spiral arms, or long tidal tails, indicating a major interaction." {81 galaxies}
\item possibly {visually} asymmetric:``Possible/mild asymmetry. For example, a galaxy might be slightly `bent', indicating some interaction with a passing galaxy, or might show evidence of a very minor merger." {19 galaxies}
\item uncertain: ``Galaxies that are small or unclear enough that a reliable classification is impossible." {15 galaxies}
\end{itemize}
 Fig.~\ref{fig:examples} shows examples of galaxies in the normal, definitely {visually} asymmetric and uncertain categories. The features included in the `definitely {visually} asymmetric' category were: tidal tails, warps, and evidence of double cores or in-progress mergers. Importantly, we were not simply identifying major mergers, but rather {a range of visual} asymmetries. If there was at least 66$\%$ agreement that a galaxy was `definitely {visually} asymmetric', it was placed in the visually asymmetric sample used throughout this work. Individual classification results are given in the appendix to this paper.

We note the similarity of our method to that used by the IMAGES survey \citep{yang2008images} to classify kinematic maps. However, unlike the IMAGES survey work, we used a qualitative approach to identify visual asymmetry and a quantitative standard for kinematic asymmetry.

We chose to be conservative in excluding galaxies that were only possibly {visually} asymmetric from the list of {visually} asymmetric galaxies. This was because we did not want to err on the side of misclassifying normal galaxies as {visually} asymmetric. For our purposes, it was better to have a cleaner sample of {visually} asymmetric galaxies, even at the potential cost of losing a small fraction. Similarly, galaxies with apparent near companions, but no visual evidence of asymmetry, were classified as normal. 

{We find no difference in the errors on the SDSS photometric PAs for galaxies classified as visually asymmetric and normal. Galaxies with offset kinematic and photometric PAs were generally visually classified as asymmetric. The relationship between offset between kinematic and photometric PAs and kinematic asymmetry will be the subject of future work by the SAMI Galaxy Survey team (Bryant et al., in prep, Bloom et al., in prep). An offset between the kinematic and photometric PA would likely qualify a galaxy as `asymmetric' for the purposes of this work. }

\section{Distinguishing between kinematically normal and asymmetric galaxies in the SAMI Galaxy Survey}
\label{sec:sec4}

\begin{figure}
\centering
\includegraphics[width=9cm]{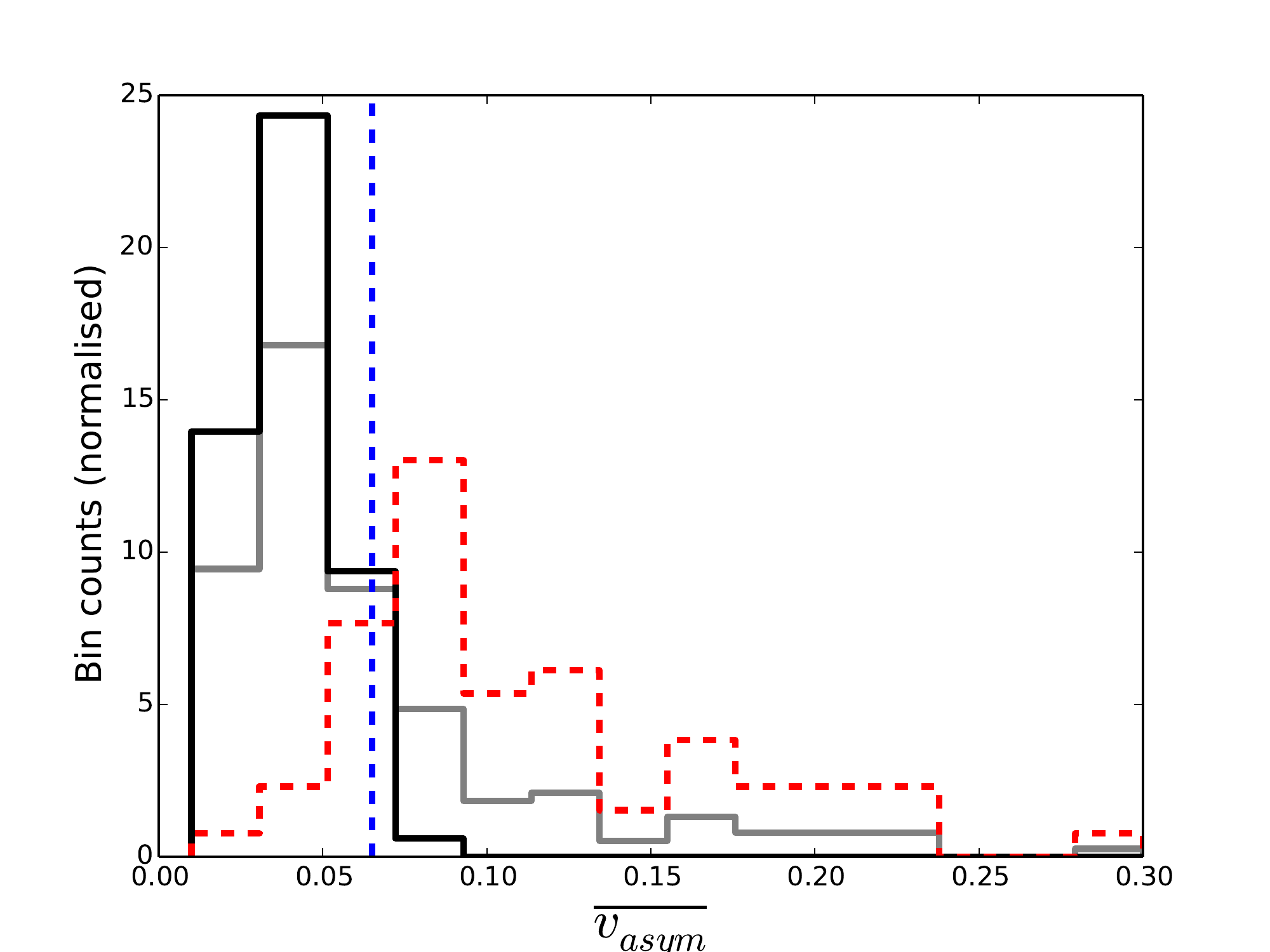}
\includegraphics[width=9cm]{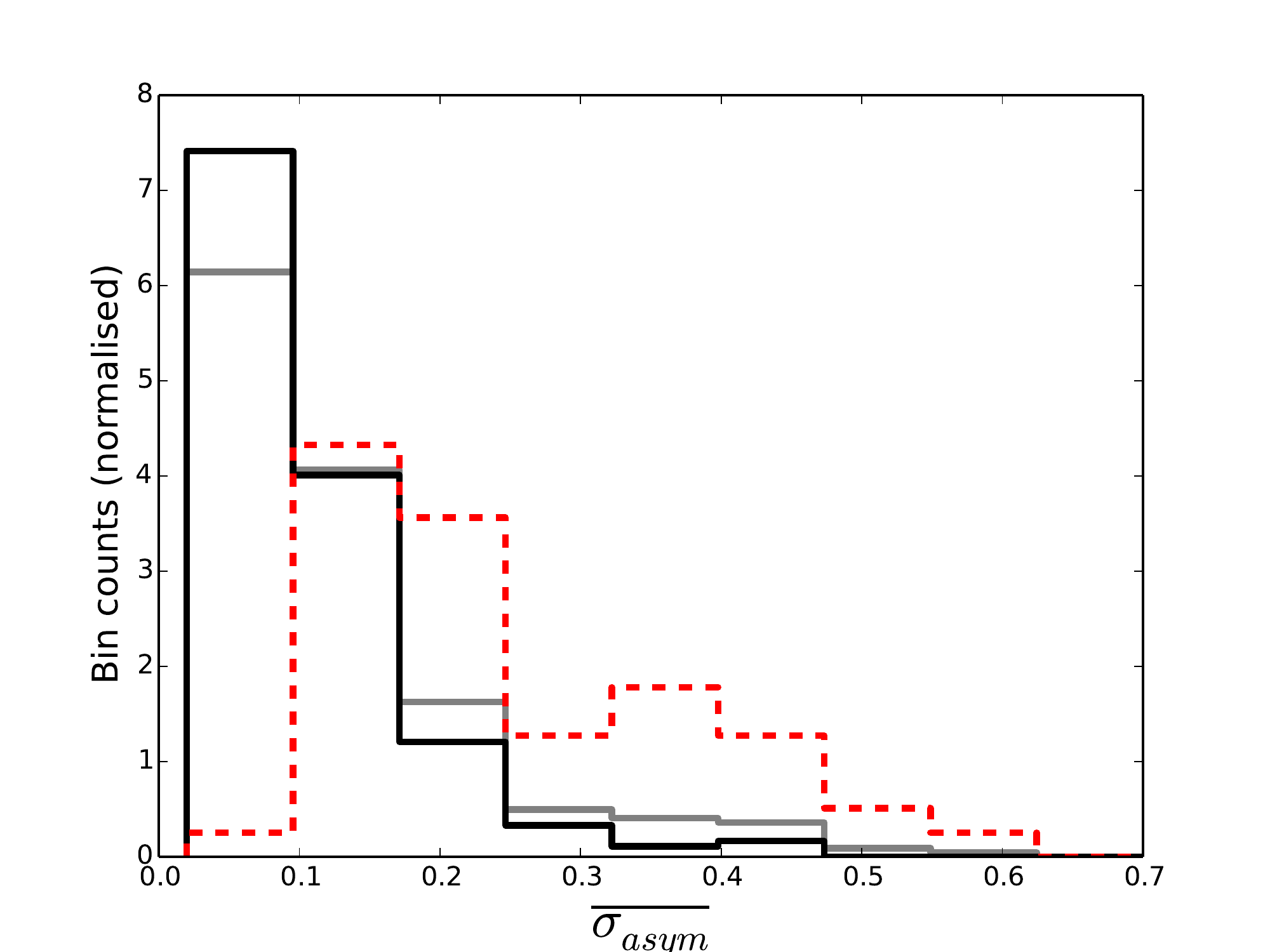}
\caption{{We show} normalised histograms of $\overline{v_{asym}}$ (top panel) and $\overline{\sigma_{asym}}$ (bottom panel) for the whole sample (grey), visually classified normal (black) and asymmetric (red, dashed) galaxies. The cutoff value, $\overline{v_{asym}}=0.065$, is shown in blue.}
\label{fig:asym_histograms}
\end{figure}

Fig~\ref{fig:asym_histograms} shows histograms of the distributions of median velocity and velocity dispersion asymmetries, $\overline{v_{asym}}$ and $\overline{\sigma_{asym}}$.  The complete sample is shown in grey. The distribution in both cases is smooth, and peaked at low values, with a tail. It is not immediately obvious where to place a cutoff for galaxies to be classified as kinematically asymmetric or normal. 

Fig~\ref{fig:asym_histograms} also shows galaxies visually identified as asymmetric in red, and normal galaxies in black. Visually asymmetric galaxies tend to have higher $\overline{v_{asym}}$ and $\overline{\sigma_{asym}}$, which provides a reasonable basis for the cutoff.

{The good agreement between the visual classification and measured kinematic asymmetry suggests that, if there is kinematic disturbance, the image will almost always show a signature of that disturbance. While kinematics provide a more physical measure of disturbance, the images show the effects of that disturbance.  Galaxies `incorrectly' visually classified cannot be quantitatively distinguished, due the qualitative nature of the visual classification.}

\begin{figure}
\centering
\includegraphics[width=9cm]{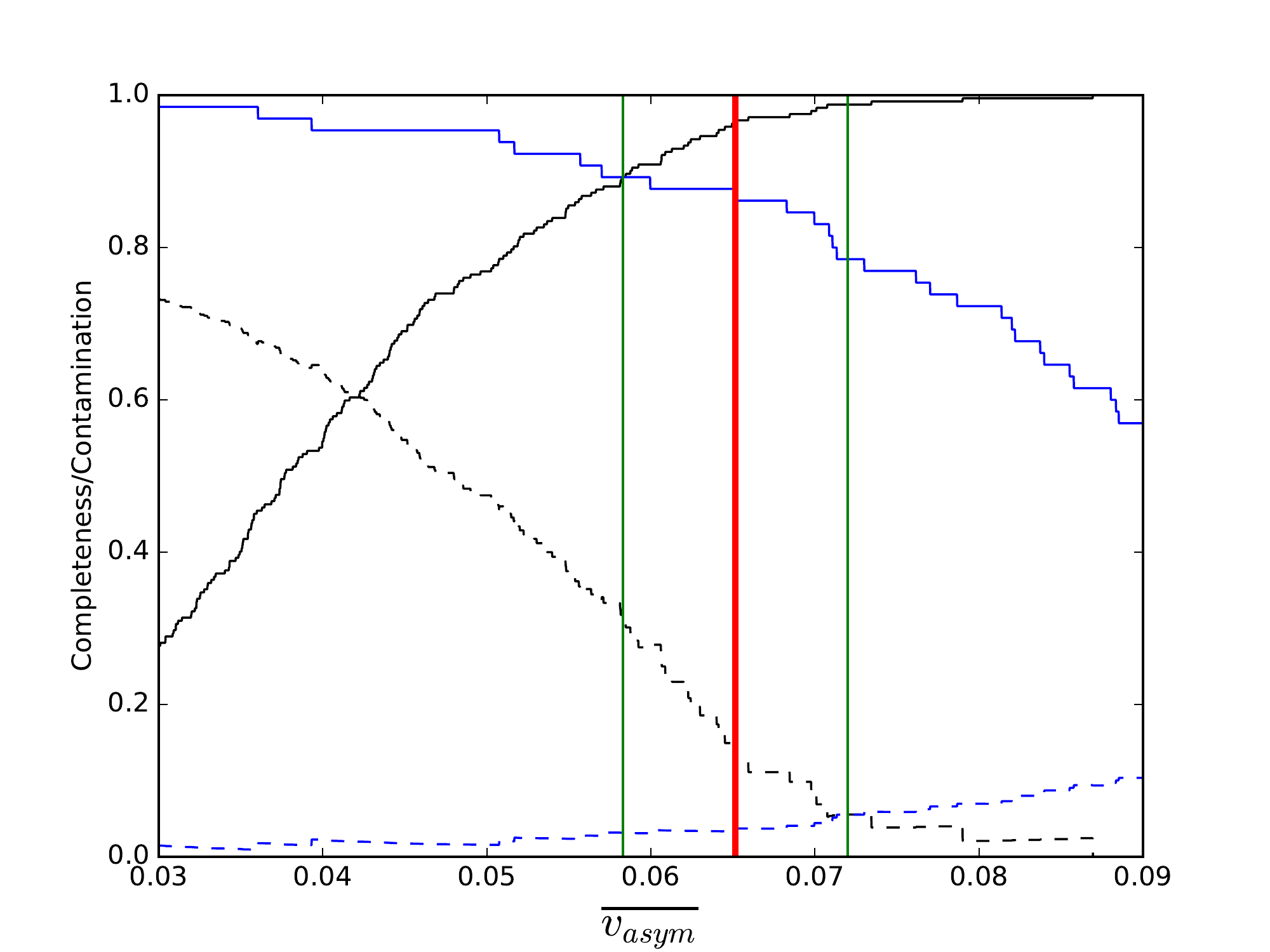}
\caption{{These curves show} contamination (dashed) and completeness {solid} for visually asymmetric (blue) and visually normal (black) galaxies, at a range of values of $\overline{v_{asym}}$. We choose our cutoff point (shown in red) as the midpoint between the {points of intersection of} the curves for contamination (number of visually normal galaxies classified as kinematically asymmetric and \emph{vice versa}) and completeness (number of visually asymmetric galaxies classified as kinematically asymmetric, similarly for visually normal galaxies) (shown in green). The {points of intersection} for contamination and completeness are 0.071 and 0.058, respectively, so we choose a cutoff value of $\overline{v_{asym}}=$0.065. {See Section~\ref{sec:sec4} for a complete explanation.}}
\label{fig:contamination_completeness}
\end{figure}

Fig.~\ref{fig:contamination_completeness} shows the {fractional} completeness and contamination {from 0 to 1.0} of the samples of visually asymmetric and normal galaxies across a range of values on either side of the crossover points of the  visual classification histograms in Fig.~\ref{fig:asym_histograms}, for $\overline{v_{asym}}$. Ideally, we would like to maximise completeness {(which we define as fraction of galaxies with the same kinematic classifications as visual)} and minimise contamination {(defined as fraction in each kinematic bin with opposite visual classifications)} for both {visually} normal and asymmetric galaxies. The distance between the {intersection points of the curves for}  {visually} normal and asymmetric galaxies indicates how viable it is to choose a single {kinematic} cutoff point for that purpose. We found that the distance between {intersection} points was much greater for $\overline{\sigma_{asym}}$ than for $\overline{v_{asym}}$ (0.032 against 0.013). This means that it is significantly harder to choose a cutoff based on $\overline{\sigma_{asym}}$ that will give both high completeness and low contamination for both normal and asymmetric galaxy populations. {This is the case whether we try a cut based on $\overline{\sigma_{asym}}$ or a combination with $\overline{v_{asym}}$.}

Accordingly, we cut by median velocity asymmetry, $\overline{v_{asym}}$. The {intersection points of the curves representing} contamination and completeness are 0.071 and 0.058, respectively. We thus choose the midpoint of these values, giving a cutoff value of $\overline{v_{asym}}>0.065$ for a galaxy to be considered asymmetric (shown in blue in Fig~\ref{fig:asym_histograms}). This yields a contamination of 3$\%$ and completeness of 90$\%$ for {visually} asymmetric galaxies, and contamination of 10$\%$ and completeness of 95$\%$ for {visually} normal galaxies. Of course, the by-eye classification we take as our guide is imperfect and qualitative, so there is some uncertainty in these measures. Nevertheless, the degree to which it is possible to map the visual classification onto the kinemetric one does point to an underlying physical similarity between the features selected for by each.

The red points in Fig. \ref{fig:sami_asyms} are galaxies classified as visually asymmetric, and the black points are visually normal galaxies. The larger scatter between the visually asymmetric and normal galaxies in $\overline{\sigma_{asym}}$ shows again that more visually asymmetric galaxies would be misclassified as asymmetric by applying a cutoff in both $\overline{v_{asym}}$ and $\overline{\sigma_{asym}}$. We note that the SINS survey team also used the mean of the higher order mode values, whereas we took the median. We found that taking the mean increased the contamination between the distributions of {visually} asymmetric and normal galaxies. A similar analysis to that performed above yielded a crossover contamination of 8$\%$ for {visually} asymmetric galaxies. 

We also show galaxies classified as visually `possibly {visually} asymmetric' in yellow in Fig. \ref{fig:sami_asyms}. We note that these galaxies are almost all (90$\%$) in the lower section, or `normal area' of the plot, with only two galaxies falling above our $\overline{v_{asym}}>0.065$ line. This supports our decision to exclude them from the sample of visually asymmetry galaxies.

{Galaxies classified as visually normal but kinematically asymmetric are clustered around the boundary value of $\overline{v_{asym}}\sim0.065$. The qualitative visual classification does not allow us to further distinguish these galaxies. The advantage of kinemetry is that it removes this qualitative component. Of the galaxies classified as morphologically perturbed but kinematically normal, some have images which appear to include foreground objects, which may contribute to the `misclassification'.}

We note that galaxies in the sample have different effective radii, so are covered to different extents by the SAMI instrument field of view. We do not, however, find that coverage influences the outcome of fits by kinemetry, so we disregard this as a possible bias in our results.

Using our kinemetry results, and cutting the sample at $\overline{v_{asym}}>0.065$, we calculate a {kinematic} asymmetric fraction of 23$\% \pm 7\%$. This is comparable to the complex kinematics fraction of $26\% \pm 7$ calculated by the IMAGES survey \citep{yang2008images} for galaxies with $z\sim1$, using a visual classification method similar to ours described in Section ~\ref{sec:vispert}.

\subsection{Comparison of the kinemetry technique with quantitative morphology}
\label{sec:q_m}

{Given that major mergers are known to cause kinematic asymmetry \citep{shapiro2008kinemetry}, we determine whether quantitatively morphologically asymmetric galaxies are kinematically asymmetric, and whether kinematically asymmetric galaxies are quantitatively morphologically distinct from kinematically normal galaxies. Quantitative morphology techniques, such as the Gini \citep{gini1912variabilita} and $M_{20}$ \citep{lotz2004new} coefficients, and the Concentration/Asymmetry/Smoothness (CAS) categorisation method \citep{conselice2003relationship}  have been used in previous studies to identify major mergers, as has kinemetry \citep{shapiro2008kinemetry}, providing a useful basis for comparison.} {\citet{simons2015transition} compare $\frac{\sigma}{V}$ and quantitative morphology, concluding that kinematically disturbed galaxies, falling off the Tully-Fisher line, also tend to be asymmetric by the standards of quantitative morphology.}

 To perform this analysis, we use the $r$-band SDSS DR10 images \citep{ahn2014tenth}. We follow methods in \citet{lotz2004new} and \citet{conselice2003relationship}. To determine the centre of the images used in this analysis, we replicate the method described in Section \ref{sec:kinemetry} on the $r$-band SDSS DR10 images for each galaxy.

In this work, errors on $G$, $M_{20}$ and the CAS Asymmetry coefficient are found by bootstrapping the calculations of the coefficient for each galaxy, adding random noise to the data and calculating the uncertainties from the distribution of these iterations.

\subsubsection{The Gini coefficient}

The Gini coefficient, $G$, was developed by economists \citep{gini1912variabilita} as a descriptor of the distribution of resources amongst a population. $G=1$ indicates that the entire wealth of a population is concentrated with one individual, whereas $G=0$ signifies complete equality. It was first used in an astronomical context by \citet{abraham2003new}. Applied to galaxy flux, $G$ becomes a measure of concentration of light, similar to $C$ in the CAS system (see Section~\ref{sec:cas}). In high-$G$ galaxies, the light is locally concentrated (for instance, there could be a highly dominant bulge). A low-$G$ galaxy would have a smooth, uniform light distribution, without a significant bulge. Alternatively, it could be composed of many small clumps, including regions of star formation that would `balance out' a central bulge.

$G$ is mathematically defined as \citep{glasser1962variance}:
\begin{equation}
G=\frac{1}{\overline{X}n(n-1)}\sum_{i}^{n}(2i-n-1)X_{i} ,
\end{equation}
where $\overline{X}$ is (in this case) the mean of the galaxy flux in $n$ total pixels, with $X_{i}$ being the flux in the $i$th pixel.

Previous work has used $G$ in combination with $M_{20}$ (\citet{lotz2004new}, see Section~\ref{sec:M20}) and CAS (\citet{conselice2003relationship}, see Section~\ref{sec:cas}) to identify major mergers.

\subsubsection{The $M_{20}$ coefficient}
\label{sec:M20}
The $M_{20}$ coefficient is similar to $G$, in that it measures the concentration of the spatial distribution of light within a galaxy, but can be used to distinguish galaxies with different Hubble types \citep{lotz2004new}. The total second-order moment of galaxy flux, $M_{tot}$, is defined as the flux $f_{i}$ in each pixel, multiplied by the squared distance between pixel $i$ and the centre of the galaxy $(x_{c},y_{c})$, summed over all galaxy pixels:
\begin{equation}
M_{tot}=\sum_{i}^{n}M_{i}=\sum_{i}^{n}f_{i}[(x_{i}-x_{c})^{2}+(y_{i}-y_{c})^{2}] .
\end{equation}

$M_{20}$ is then defined as the normalised second-order moment of the brightest 20\% of the galaxy's flux. The galaxy pixels are rank-ordered by flux, and then $M_{i}$ (the second-order moment of light for each pixel $i$) is summed over the brightest pixels until the sum of the brightest pixels is equal to 20\% of the total flux:
\begin{equation}
M_{20}=log_{10}\left(\frac{\sum_{i}M_{i}}{M_{tot}}\right)
\end{equation}
for $\sum_{i}f_{i}<0.2f_{tot}$, where $f_{tot}$ is the total flux, $\sum_{i}^{n}f_{i}$.

\subsubsection{CAS Asymmetry}
\label{sec:cas}
The CAS (Concentration, Asymmetry, Smoothness) system was developed as a means to distinguish galaxies in different stages of evolution, based on where they fall within a volume derived from the CAS structural parameters \citep{conselice2003relationship}. The $C$ and $A$ parameters were first developed by \citet{abraham1996galaxy}. We will discuss only the application of Asymmetry, henceforth $A$, as $C$ and $S$ are not intended to describe asymmetries.

To find $A$, a galaxy image is rotated 180$^{\circ}$ around its central point and the result is subtracted from the original image. $A$ is then computed as the sum of the absolute values of the residuals, normalised over the sum of the flux in the original image. $A$ is clearly sensitive to any feature that is not rotationally symmetric. These may include spiral arms, areas of intense star formation, and merger signatures \citep{conselice2003relationship}.

It is important to note that $A$ was specifically developed to identify the middle stage of major mergers, in which case the {visual} asymmetric features resulting from the merger would far outweigh those from ordinary morphology, such as spiral arms. However, in the case of more minor asymmetries, other morphological features may dominate. For example, a relatively small and dim tidal tail would be overshadowed by the presence of bright spiral arms.

\subsubsection{Comparison of classifications using kinematics and quantitative morphology}
\begin{figure}
\centering
\includegraphics[width=9cm]
{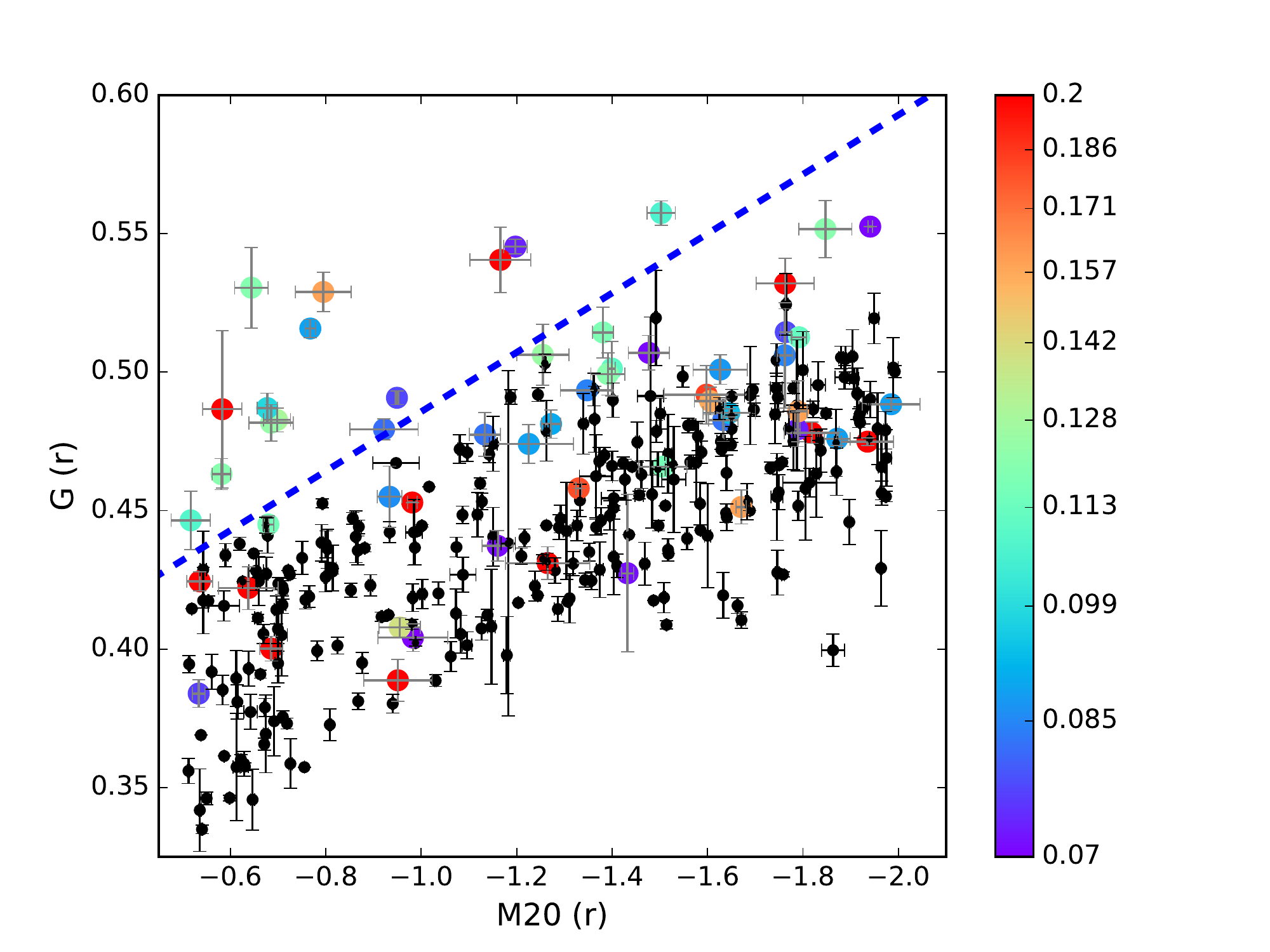}
\caption{{We show} $M_{20}$ against $G$, with kinemetric classifications (normal galaxies in black, {kinematically} asymmetric galaxies in colour). We see that, even though the majority of galaxies falling above the line defined in \citet{lotz2004new} are {kinematically} asymmetric, there are many that would not be identified using this system. The colourbar shows the $\overline{v_{asym}}$ values on a log scale for the kinematically asymmetric points. We see that many of the most kinematically asymmetric points fall below the line.}
\label{fig:gini_m20_2}
\end{figure}

Following previous work \citep{lotz2004new}, Fig.~\ref{fig:gini_m20_2} shows $M_{20}$ against $G$ for the galaxies in our sample. The {kinematically} asymmetric galaxies (coloured) are those with $\overline{v_{asym}}>0.065$. \citet{lotz2004new} drew a line on the plane to distinguish ULIRGS from `normal' galaxies, shown in Fig.~\ref{fig:gini_m20_2} in dashed blue. Whilst all of the galaxies above the line in our sample are kinematically asymmetric, there are also many kinematically asymmetric galaxies which would not be identified using this method.  Quantitatively, 79$\% \pm 3\%$ of kinemetrically asymmetric galaxies would be {classified as normal} by this method. Further, many of the most kinematically asymmetric galaxies, indicated by the colour bar, fall below the line. This is because $G$ and $M_{20}$ measure the spread of light. A galaxy with multiple, bright clumps has a different light distribution profile from a normal galaxy (with one bright, central bulge), but a galaxy with a strong {kinematic} twist may not, as the ratio of light in the bulge and body of the galaxy may be the same, regardless of the twist.

\begin{figure}
\centering
\includegraphics[width=9cm]{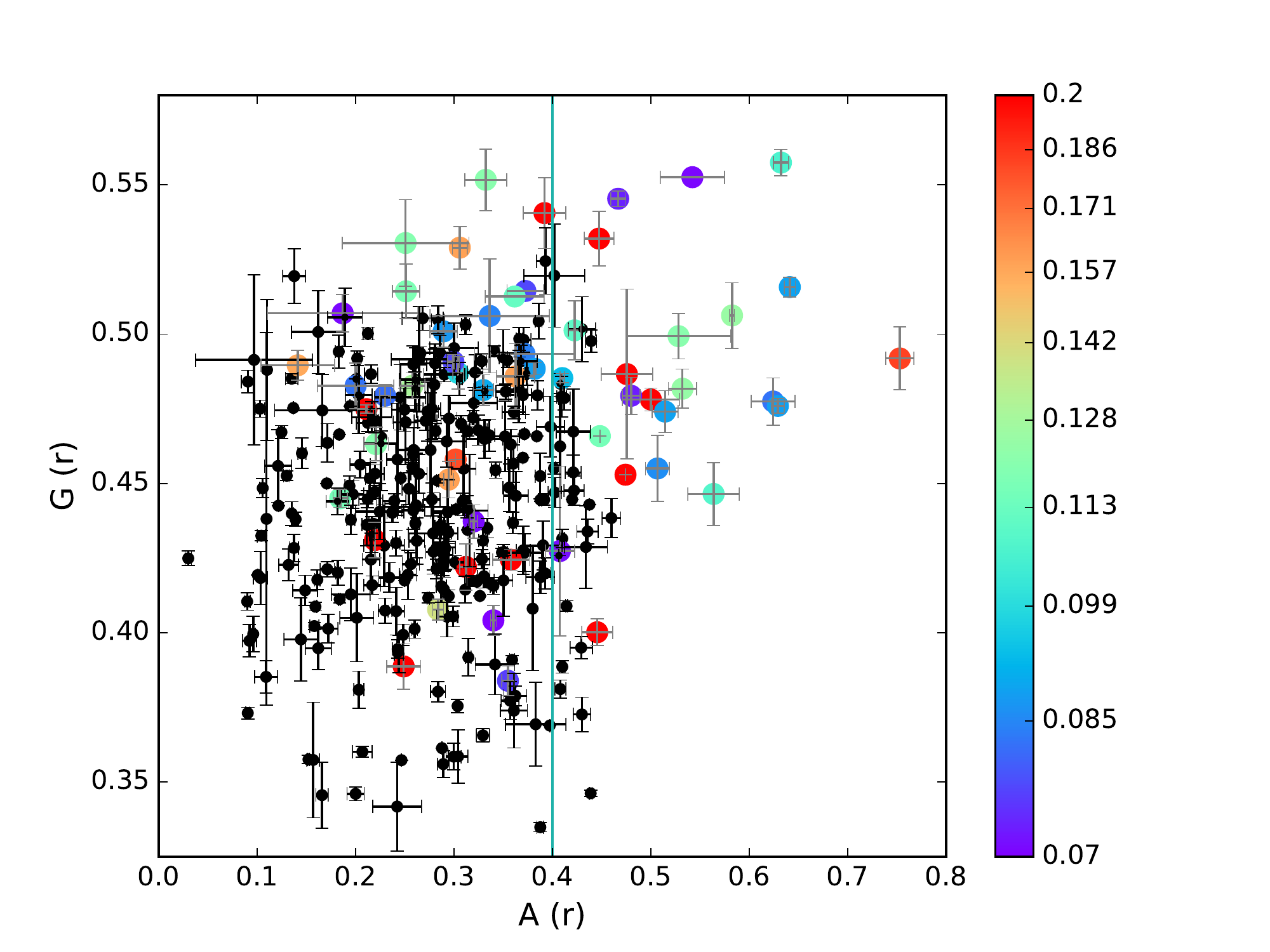}
\caption{{We show} the $A-G$ plane for {kinematically} normal (black) and  {kinematically} asymmetric (coloured) galaxies. Errors are 1-$\sigma$ errors on both axes. As anticipated, only a weak relation emerges between asymmetry and plane location. The blue line indicates $A\	sim 0.4$, below which the relationship between $A$ and asymmetry becomes significantly weaker.}
\label{fig:gini_cas}
\end{figure}

In Fig.~\ref{fig:gini_cas}, following \citet{conselice2008structures}, we show $G$ against $A$. As in Fig.~\ref{fig:gini_m20_2}, coloured points are those with $\overline{v_{asym}}>0.065$. We once again see only a weak relationship between kinematic asymmetry and placement on the plane. Galaxies with $A>0.4$ are mostly kinematically asymmetric, but below this threshold there is no significant relation. Indeed, $59\%\pm4\%$ of {kinematically} asymmetric galaxies in our sample lie below $A=0.4$. As in Fig.~\ref{fig:gini_m20_2}, some of the most kinematically asymmetric galaxies {are not classified as morphologically disturbed} using this method. This is because the relationship between $A$ and {photometric (also, by extension, kinematic)} asymmetry becomes significantly weaker below $A=0.4$ \citep{conselice2003relationship, conselice2008structures}. Whilst galaxies with $A>0.4$ are very likely to be disturbed, below this threshold normal morphological features (particularly spiral arms) dominate, so true asymmetries are lost. Given that most of our kinematically asymmetric galaxies fall below this threshold, our result is not unexpected. 

{We conclude that almost all quantitatively morphologically asymmetric galaxies are kinematically asymmetric. Further, kinematics can identify asymmetry in galaxies that appear normal when using these quantitative morphology methods.}

\section{Comparison of our results with high redshift studies}
\label{sec:high_z}

\begin{figure}
\centering
\includegraphics[width=9cm]{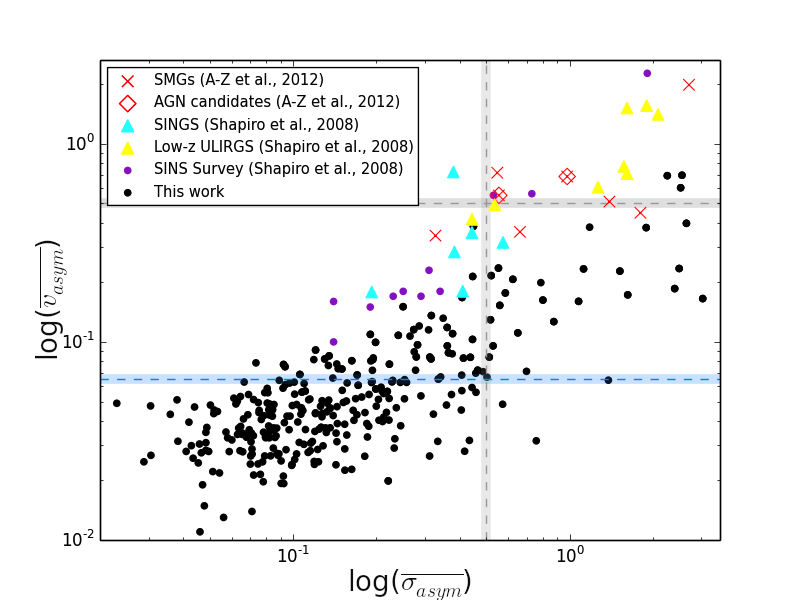}
\caption{{This figure shows the} mean velocity and velocity dispersion asymmetries for galaxies in our sample and those from several other works using kinemetry (as in legend).  Error bars have been removed for clarity. The grey lines indicate the {kinematic} asymmetry cutoffs from the SINS Survey work \citep{shapiro2008kinemetry}, and the blue line is the cutoff we developed for use on our data.}
\label{fig:sins_points}
\end{figure}

Although our method is based on that used by the SINS Survey team, there are some differences. For example, they used the mean of radially calculated {kinematic} asymmetry values, whereas we used the median, and they included the even modes of the velocity fields and odd modes of velocity dispersion fields in the calculation of velocity and velocity dispersion asymmetry. In order to more directly compare our results with the SINS Survey results, we recalculated our {kinematic} asymmetry values, applying their method. Fig~\ref{fig:sins_points} shows our results (black points), along with a sample of those from the SINS Survey, including artificially redshifted local spiral galaxies from the SINGS Survey (turquoise triangles), artificially redshifted local ULIRGS (yellow triangles) and galaxies observed by the SINS Survey team at high redshift (purple points) \citep{shapiro2008kinemetry}. We also show a sample of high redshift ($2.0<z<2.7$) sub-millimetre galaxies (SMGs) from \citet{alaghband2012integral} (red crosses). Of the SMGs, those which are AGN candidates are also noted (red diamonds superimposed on the red crosses). We see that the spread of our values does not change significantly, despite the recalculation.

Applying their version of kinemetry to the SINS Survey sample led to an estimated major merger fraction of $\sim 25\%$ \citep{shapiro2008kinemetry}. Due to the high redshift of their sample ($z\sim 2$), which led to correspondingly low H$\alpha$ S/N in their data, even their normally-classified galaxies have higher {kinematic} asymmetry values than the majority of the galaxies in our sample (see Fig~\ref{fig:sins_points}). \citet{gonccalves2010kinematics} have argued that, in fact, degrading the resolution of data (as a proxy for higher redshift) leads to $decreased$ kinematic asymmetry values, rather than the increased values found by the SINS Survey. However, it is not clear how a corresponding increase in noise for high-redshift data will affect the results. This is a still open question and, accordingly, our result of a {kinematic} asymmetric fraction of 23$\%\pm7\%$ should not be directly compared with the result in \citet{shapiro2008kinemetry}. The SINS Survey also required their galaxies to have continuum S/N$\geq 3$ that may have biased their sample towards galaxies with high SFR, influencing the distribution of {kinematic} asymmetries.

All high-redshift samples in Fig~\ref{fig:sins_points} (both observed and artificially redshifted) fall entirely above our cutoff for {kinematic} asymmetry, with the exception of two redshifted SINGS spiral galaxy. Given the results from the SINS Survey, that showed how decreased S/N (either artificially or from increased redshift) can increase kinemetric coefficients, this is not surprising. Nevertheless, the higher kinemetric coefficients may also be reflective of intrinsic {kinematic} asymmetry.


If we were to naively apply the cutoff from the SINS Survey work (mean velocity and velocity dispersion asymmetry $=0.5$), we would have a major merger rate of $1\pm4\%$. {There have been a number of other local merger rates, including those calculated using simply close pairs  (e.g. \citealt{de2014galaxy}), quantitative morphology (e.g. \citealt{lotz2008evolution,lotz2011major}) and a combination of the two methods \citep{casteels2014galaxy}. These local major merger rates range from $\sim$1-4\%, and so are broadly consistent with the value from this work combined with the SINS Survey cutoff. However, given the difference in redshift between the SINS Survey and SAMI Galaxy Survey, this should not be interpreted as necessarily signifying scientific agreement. At $z\sim1$, \citet{stott2014relationship} use kinemetry (approximately following the \citealt{shapiro2008kinemetry}) method, using a cutoff of $K_{tot}>0.5$) to derive a major merger rate of $10\%$. }

{We further note that, of the SAMI Galaxy Survey sample galaxies that fall above the SINS Survey cutoff, 66\% fall above the major merger lines in Fig.~\ref{fig:gini_m20_2} and Fig.~\ref{fig:gini_cas}. }

\section{The relationship between kinematic asymmetry, colour and stellar mass}
\label{sec:sm}

\begin{figure}
\centering
\includegraphics[width=9cm]{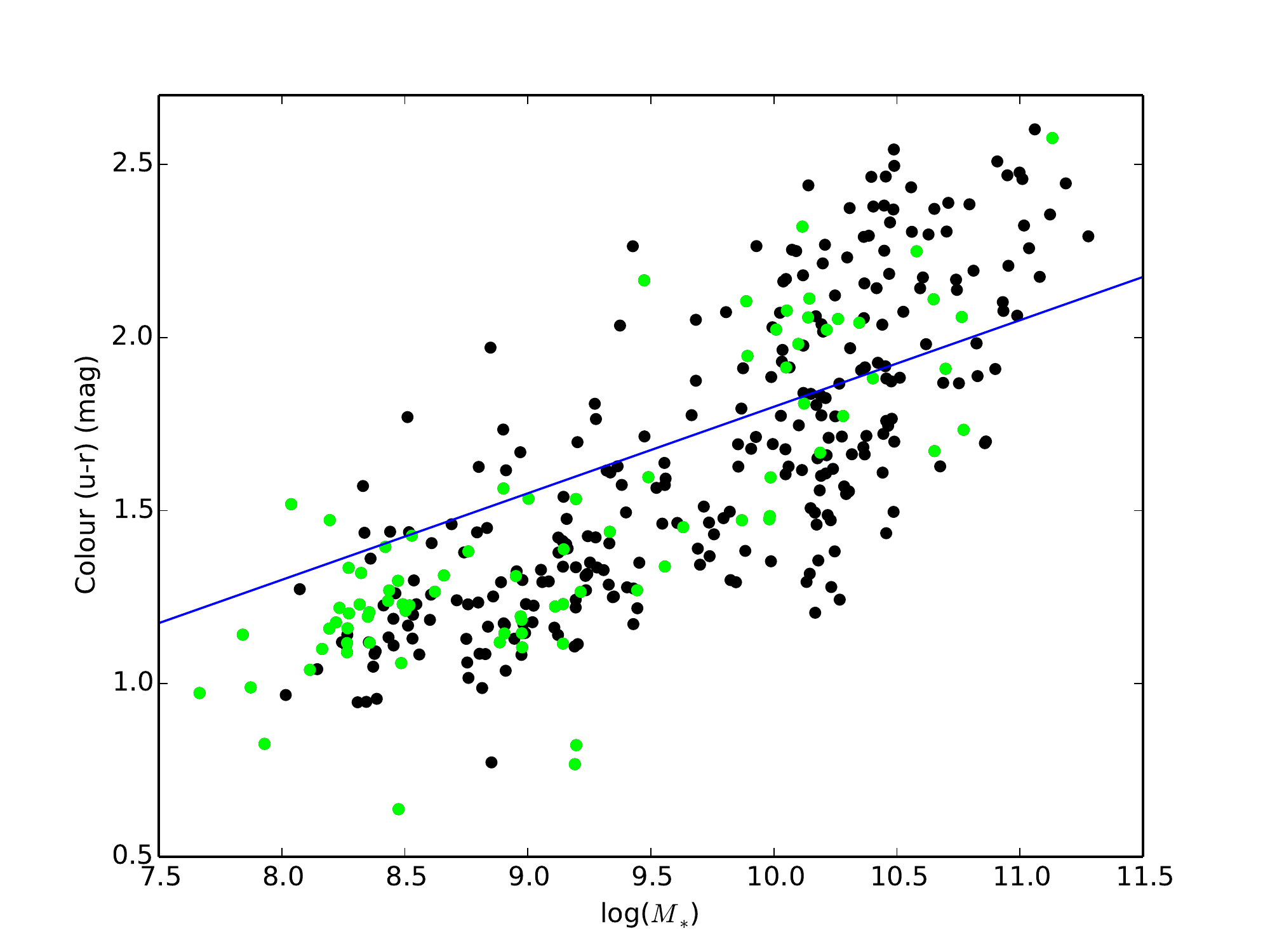}
\caption{{This plot gives colour and stellar mass} for the sample used in this work of kinematically identified normal (black) and {kinematically} asymmetric galaxies (green). The dark blue line separates the blue cloud and red sequence, and is taken from \citet{schawinski2014green}. The data used were taken from the GAMA Data Release 2 catalogues StellarMasses and ApMatchedPhotom \citep{hill2011galaxy,taylor2011galaxy}.}
\label{fig:cm_2}
\end{figure}

{Colour is directly linked to star formation history. We investigate whether processes causing kinematic asymmetry influence star formation history, as revealed by a change in colour. Several studies have established links between colour and kinematic abnormality, e.g. \citet{conselice2000physical,kannappan2002physical,neichel2008images}, finding that asymmetric galaxies are bluer than their normal counterparts. We here determine the position of our sample of kinematically asymmetric galaxies on the colour-stellar mass plane.} Fig. \ref{fig:cm_2} shows the $u-r$ rest frame colour-stellar mass distribution for normal (black) and {kinematically} asymmetric (green) galaxies. We used a cut from \citet{schawinski2014green} (dark blue in Fig. \ref{fig:cm_2}) to separate the blue cloud and red sequence. We see that 75$\pm 10\%$ of {kinematically} asymmetric galaxies fall in the blue cloud, and only 25$\pm 11\%$ fall on the red sequence.  For comparison, 65$\pm 7\%$ of normal galaxies are in the blue cloud, and 34$\pm 8\%$ of normal galaxies are in the red sequence.

\begin{figure}
\centering
\includegraphics[width=9cm]{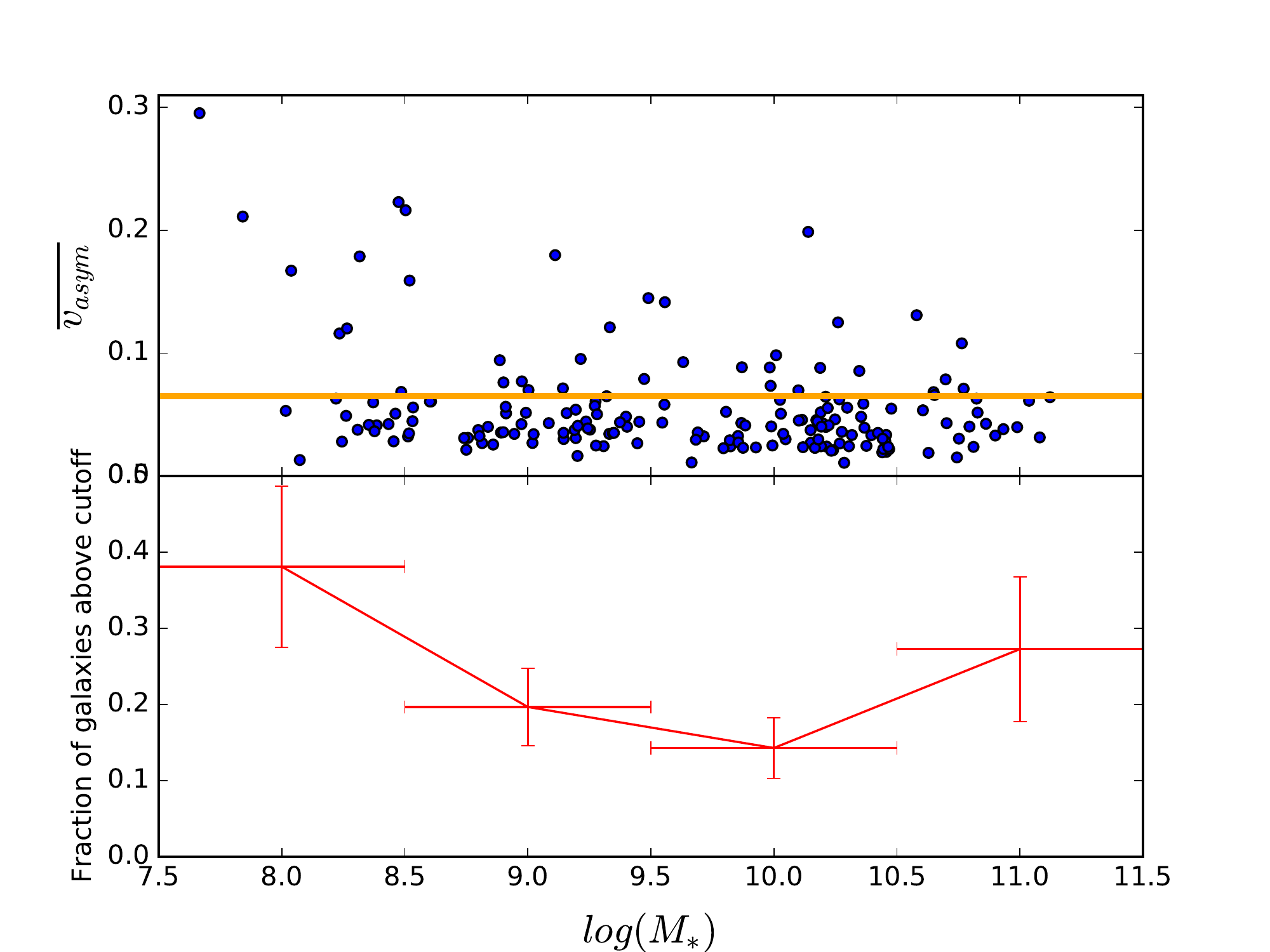}
\caption{$\overline{v_{asym}}$ {is plotted against} against stellar mass for the galaxies in our sample. The orange line shows our {kinematic} asymmetry cutoff, such that all galaxies above it are considered asymmetric. The bottom panel shows the fraction of galaxies, in bins of stellar mass (shown as horizontal error bars), above the asymmetry cutoff. This fraction decreases as a function of stellar mass.}
\label{fig:mass_asym}
\end{figure}

{Fig. \ref{fig:mass_asym} shows an inverse relationship between stellar mass and kinematic asymmetry, and a Spearman rank correlation test for $\overline{v_{asym}}$ and stellar mass gives $\rho=-0.30$, p-value$=8.40\times10^{-9}$. We further show the fraction of {kinematically} asymmetric galaxies in bins of stellar mass that also dramatically declines with increasing stellar mass.} Given that colour and stellar mass are strongly correlated, the apparent tendency of {kinematically} asymmetric galaxies to be bluer, compared to their normal counterparts, is not surprising. 

To test whether $\overline{v_{asym}}$ and colour are related independently of mass, we performed a series of partial Pearson correlation tests between $\overline{v_{asym}}$ and colour (accounting for the correlation of colour with stellar mass) on the individual branches and the whole sample. The results of the correlation test between colour and $\overline{v_{asym}}$ for the red and blue branches were $\rho=$0.054, p-value$=0.57$ and $\rho=$0.096, p-value$=0.13$, respectively. For the whole sample the result was $\rho=$0.11, p-value$=0.028$. This indicates that there is at best marginal correlation between colour and {kinematic} asymmetry, either in the separate branches or the full sample. 

The small positive correlation between $\overline{v_{asym}}$ and colour over the full sample is explained by a slight deficit of high mass, blue branch {kinematically} asymmetric galaxies in Fig. \ref{fig:cm_2}. Examining the proportion of {kinematically} asymmetric galaxies in each branch above $log(M_{*})>10.0$, 15 of 86 ($17\pm3\%$) of red galaxies are {kinematically} asymmetric, compared to 7 of 66 (10$\pm3\%$) of blue galaxies. Although the difference is marginally significant, a possible physical explanation for this comparative deficit of high-mass, blue branch {kinematically} asymmetric galaxies may be related to the lower gas fraction in red galaxies. Incoming gas from an interaction is likely to settle to rotate with the main galaxy mass in a high-gas fraction (blue) galaxy. However, in a low-gas fraction, red galaxy, it may be easier to preserve any difference in angle of rotation between newly accreted gas and the main disk. Given that, for kinemetry, we fit the PA of the disk from $r$-band photometry that traces the stellar component, any offset in the PA of the gas would register in the higher order moments [Bryant et al., 2016 (in prep)].

\begin{figure}
\centering
\includegraphics[width=9cm]{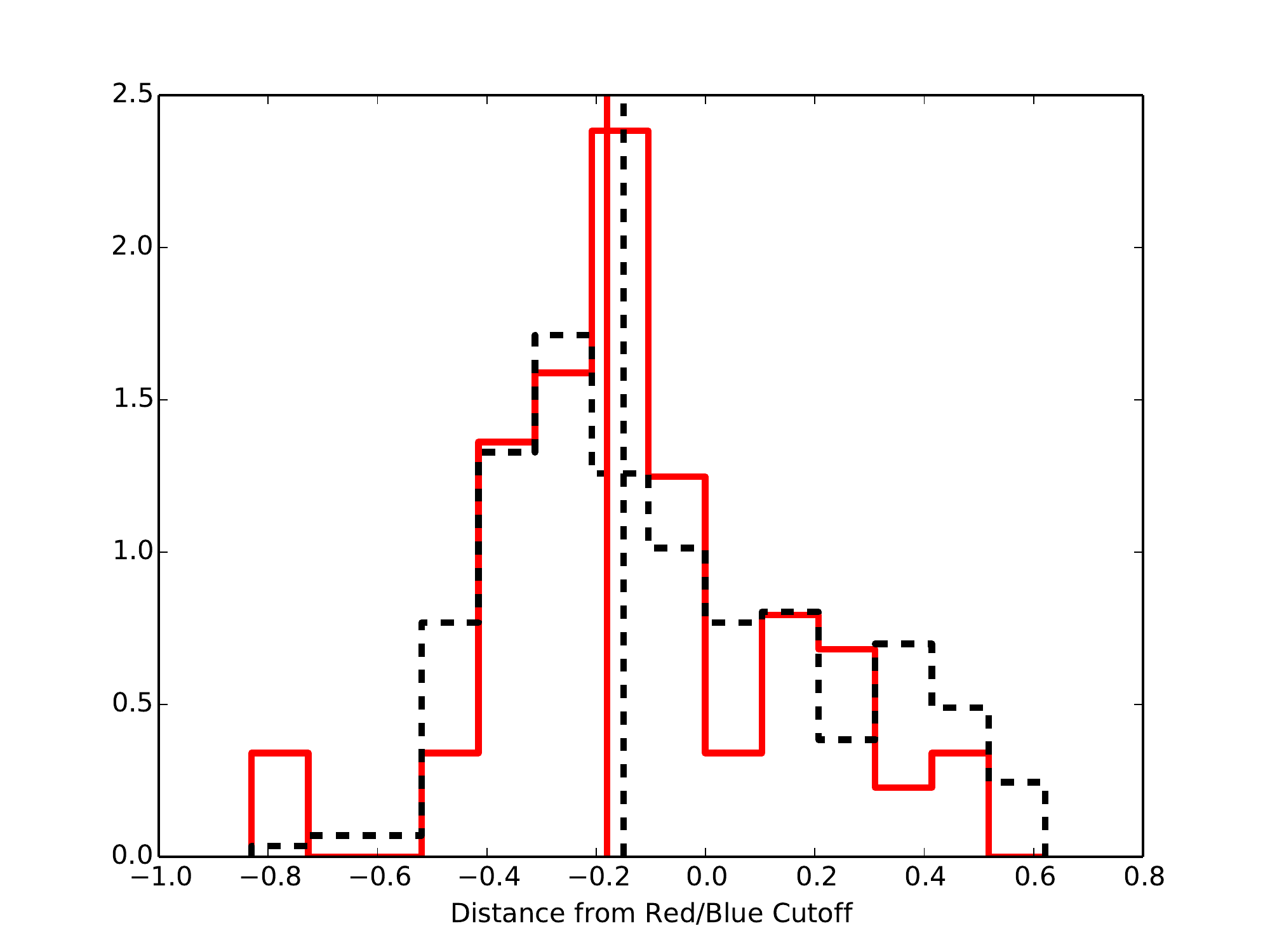}
\caption{{We show a} normalised histogram of distance (in colour space) from red/blue cutoff for {kinematically} asymmetric (red) and normal (black, dashed) galaxies. The median offset for each distribution is shown by the vertical lines. There is no significant offset between the {kinematically} asymmetric and normal medians, indicating that, independent of mass, there is no relationship between colour and $\overline{v_{asym}}$.}
\label{fig:dist_cutoff}
\end{figure}

Fig. \ref{fig:dist_cutoff} shows a histogram of the distance of each galaxy from the red/blue cutoff in Fig. \ref{fig:cm_2}, in terms of colour, as well as the medians of the distributions. A negative distance indicates that the galaxy is in the blue branch, whereas red galaxies have positive distances. A negative shift in the median distance would indicate that {kinematically} asymmetric galaxies are bluer than normal galaxies, independent of the negative correlation between mass and $\overline{v_{asym}}$. The medians are -0.14$\pm0.014$ (normal galaxies) and -0.18$\pm$0.027 ({kinematically} asymmetric galaxies). There is an offset of $0.03\pm0.03$, which is not significant. From this result, we conclude that the dominant relationship is between $\overline{v_{asym}}$ and stellar mass.

\subsection{Explaining the inverse correlation between stellar mass and $\overline{v_{asym}}$}

There are several possible reasons for the {strong} observed relationship between mass and {kinematic} asymmetry. We rule out that the increased $\overline{v_{asym}}$ is a result of poorer S/N in the low-mass objects, as there is no significant trend between the H$\alpha$ S/N and galaxy mass. Further, there is no statistical relationship between the number of effective radii covered in the analysis and {kinematic} asymmetry, which excludes any boost in $\overline{v_{asym}}$ in low-mass galaxies purely due to their size. This is because the low-mass galaxies in the SAMI Galaxy Survey sample are at lower redshift and have proportionately more H$\alpha$ gas \citep{allen2014sami}.

The Tully-Fisher relation \citep{tully1977new} links the rotational velocity of galaxies to their absolute magnitude, which is, in turn, proportional to stellar mass. The dwarf galaxies in our sample have lower rotational velocities than high-mass galaxies. Given that, in this work, higher order kinemetry terms are normalised relative to the first order velocity field, it is plausible that {kinematic} asymmetries of the same intrinsic amplitude would yield greater kinemetric signatures in low-mass galaxies, as low mass galaxies have disproportionately low velocities. Low mass galaxies have also previously been found to have complex kinematics by a variety of metrics [e.g. \citet{van1998neutral,cannon2004complex,lelli2014dynamics}]. Further, alternative measures of kinematic asymmetry, such as that in the rotation curve \citep{barton2001tully,garrido2005ghasp}, have found dwarf galaxies to have both high {kinematic} asymmetry and low rotational support . It has additionally been long established that low-mass galaxies have irregular morphologies \citep{roberts1994physical,mahajan2015galaxy, simons2015transition}, that may be linked to their irregular kinematics.  

\citet{escala2008stability} link the stochastic formation of massive instabilities to rotational velocity. They find that the mass of gaseous instabilities is inversely proportional to  angular rotational velocity:
\begin{equation}
M^{max}_{cl}=\frac{\pi^{4}G^{2}\Sigma^{3}_{gas}}{4\Omega^{4}} ,
\end{equation} 
where $M^{max}_{cl}$ is the maximum mass of unstable regions, $\Sigma$ is the gas surface density and $\Omega$ is the angular rotation speed, $\frac{V_{c}(R)}{R}$, of the disk. Given that, in order to meet the condition for rotational support inside a given radius $R$,
\begin{equation}
\Omega^{2}=\frac{\pi G\Sigma_{gas}}{\nu R},
\end{equation}
where $\nu=M_{gas}/M_{tot}$, the maximum instability mass can be expressed as a function of gas fraction:
\begin{equation}
M^{max}_{cl}=\frac{\pi^{2}\nu^{2}R^{2}\Sigma_{gas}}{4} .
\end{equation}

Given their typically high gas fractions \citep{geha2006baryon,huang2012gas}, i.e. high $\nu$, low-mass galaxies are likely to have comparatively large maximal sizes of these instabilities. Assuming that large instabilities leave kinematic signatures corresponding to increased kinemetric coefficients, this effect may contribute to the trend observed here. This theory is supported by work such as \citet{amorin2012complex}, in which complex H$\alpha$ kinematics in dwarf galaxies are linked to the presence of multiple star-forming clumps. Further, \citet{green2013dynamo} find that, for star-forming galaxies, high gas fraction (inferred from SFR density) is linked to high  $\frac{\sigma_{m}}{V}$, where $\sigma_{m}$ is the mean velocity dispersion, and $V$ is the circular velocity. This measure can be used as a proxy for turbulent support, relative to rotational support. An excess in turbulent support may contribute to the higher {kinematic} asymmetry in low-mass, high-gas fraction galaxies in our sample. \citet{simons2015transition} find that, below $log(M_{*})\sim9.5$, there is increased scatter off the Tully-Fisher relation to lower velocity, due to low mass galaxies having high $\frac{\sigma_{m}}{V}$. This scatter, { [e.g. \citet{kannappan2002physical,cortese2014sami}] } would amplify the effect of low rotation on the normalisation of higher-order kinemetry terms. We note that kinemetry and $\frac{\sigma_{m}}{V}$ probe kinematic disturbance on different scales. Whereas kinemetry is used to identify features dominating the optical galaxy that are larger than a resolution element, $\frac{\sigma_{m}}{V}$ measures turbulence on scales of less than a pixel. Further exploration of the relationships between $\overline{v_{asym}}$, $\frac{\sigma_{m}}{V}$ and stellar mass will be the subject of future work.

It is also possible that the observed relationship is due to environmental effects. For example, dwarf galaxies are more likely to be satellites of high-mass galaxies, which would perturb their kinematics. They are also more likely to undergo interactions with more massive partner galaxies, which would cause them to experience greater {kinematic perturbation} than a higher-mass galaxy under the same circumstances. In contrast, however, \citet{kirby2014dynamics} find that the disturbance in the kinematics of isolated and satellite low mass galaxies are similar. The degree to which environment influences kinematic asymmetry will be the subject of future study.

\section{Star formation in kinematically asymmetric galaxies}
\label{sec:sfr}

Several processes known to cause kinematic asymmetry have also been suggested to influence star formation, such as major mergers \citep{ellison2013galaxy}, minor mergers \citep{kaviraj2014importance,kaviraj2009role} and tidal interactions \citep{bekki2011transformation}. {We here quantitatively determine this relationship by comparing different measures of SFR and distribution of star formation.}

\subsection{Comparison of SAMI Galaxy Survey, SDSS and GAMA SFRs}
The SAMI Galaxy Survey SFRs, derived from the H$\alpha$ flux, were calculated from annular Voronoi binned data cubes [in annular Voronoi binning, the adaptive bins are constrained to an annulus of a specific radius, see Schaefer et al., (submitted)]. These cubes were binned to a target continuum emission S/N of 10 in a 200\AA -wide window around the wavelength of H$\beta$. The S/N calculation includes covariance between spaxels when calculating the variance.  Each binned spectrum is then fit by {\small LZIFU}{\sc} (Ho et al., in prep). The H$\alpha$ flux is then dust-corrected in each bin, using the \citet{calzetti2001dust} dust correction, which models the dust as a foreground screen. The H$\alpha$ fluxes in each bin were converted to luminosities and from that to SFRs using the \citet{kennicutt1998global} relation assuming a Salpeter \citep{salpeter1955luminosity} IMF. For a complete description of the SAMI Galaxy Survey SFRs, see Schaefer et al. (submitted.), from which the data used here were taken.

\begin{figure}
\centering
\includegraphics[width=9cm]{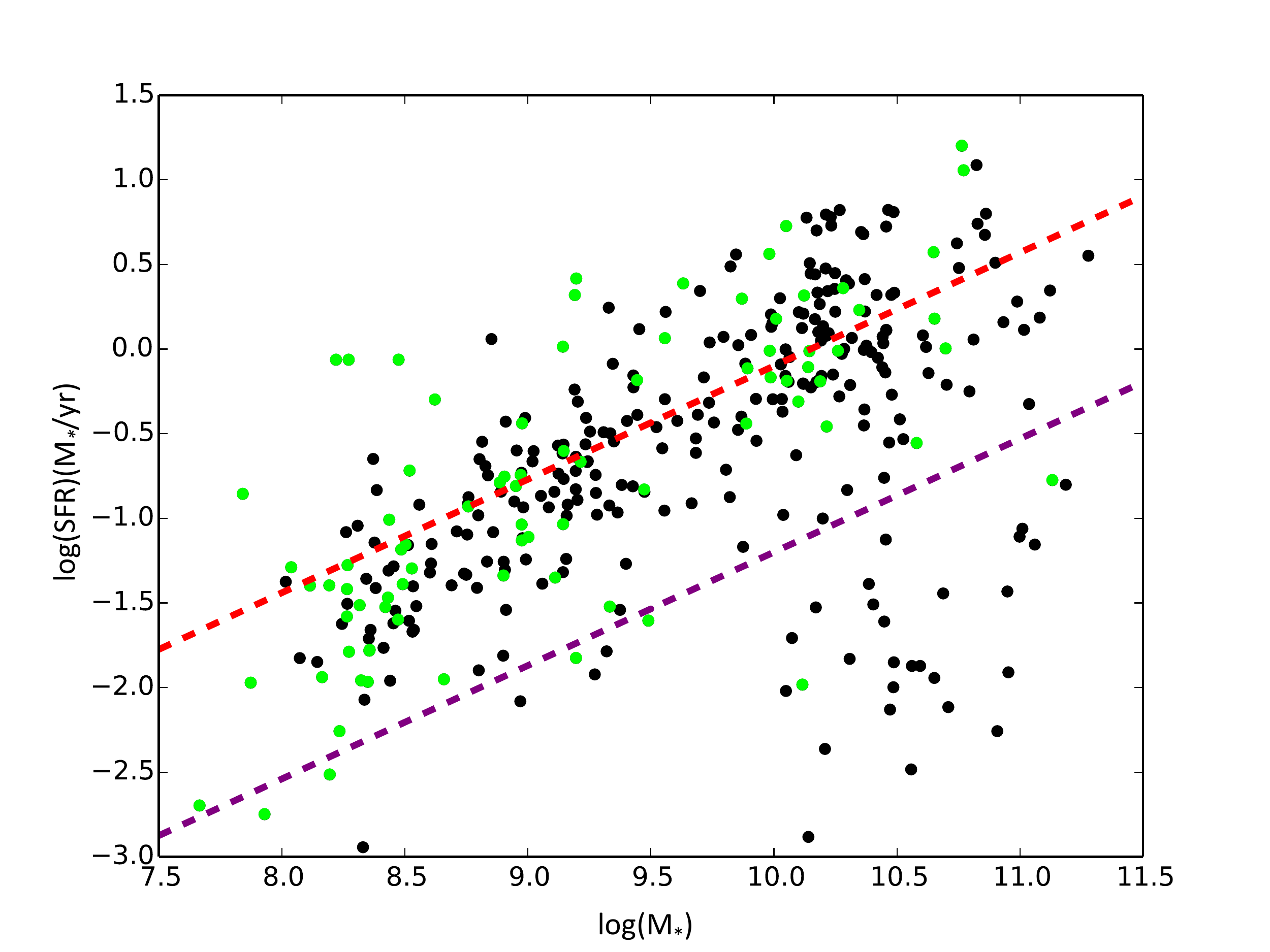}
\caption{{This figure shows} SFR and stellar mass for normal galaxies (black) and kinematically identified asymmetric galaxies (green). We show the SFR main sequence from \citet{noeske2007star} (red dashed line), and the main sequence cutoff (purple dashed line). We see that asymmetric galaxies lie almost exclusively above the purple line, whereas normal galaxies are more evenly distributed.}
\label{fig:sfr_2}
\end{figure}

{The main sequence of star formation is a relation defined by \citet{noeske2007star} (for $z<1$) that describes the SFR of typical galaxies, at a given stellar mass. The main sequence is defined as:
\begin{equation}
\log(SFR) = (0.67\pm 0.08)\log(M_{*})-(6.19\pm 0.78) .
\end{equation} 
 Fig. \ref{fig:sfr_2} shows the SFR-mass plane for our galaxies, using the SAMI Galaxy Survey SFRs, as well as the main sequence of star formation (in red).}  {Kinematically asymmetric galaxies are shown in green, and we see that there are few asymmetric galaxies below the main sequence of star formation. This means that almost all kinematically asymmetric galaxies are star-forming, rather than quiescent.} The lower bound for galaxies to be considered part of the main sequence (shown in purple in Fig. \ref{fig:sfr_2}) was derived using the same method as \citet{noeske2007star}, i.e. that there should be an equal number of galaxies above and below the line of the main sequence itself. We find a lower bound of:
\begin{equation}
\log(SFR) = (0.67\pm 0.08)\log(M_{*})-(7.34\pm 0.78) .
\end{equation} 

\begin{figure}
\centering
\includegraphics[width=9cm]{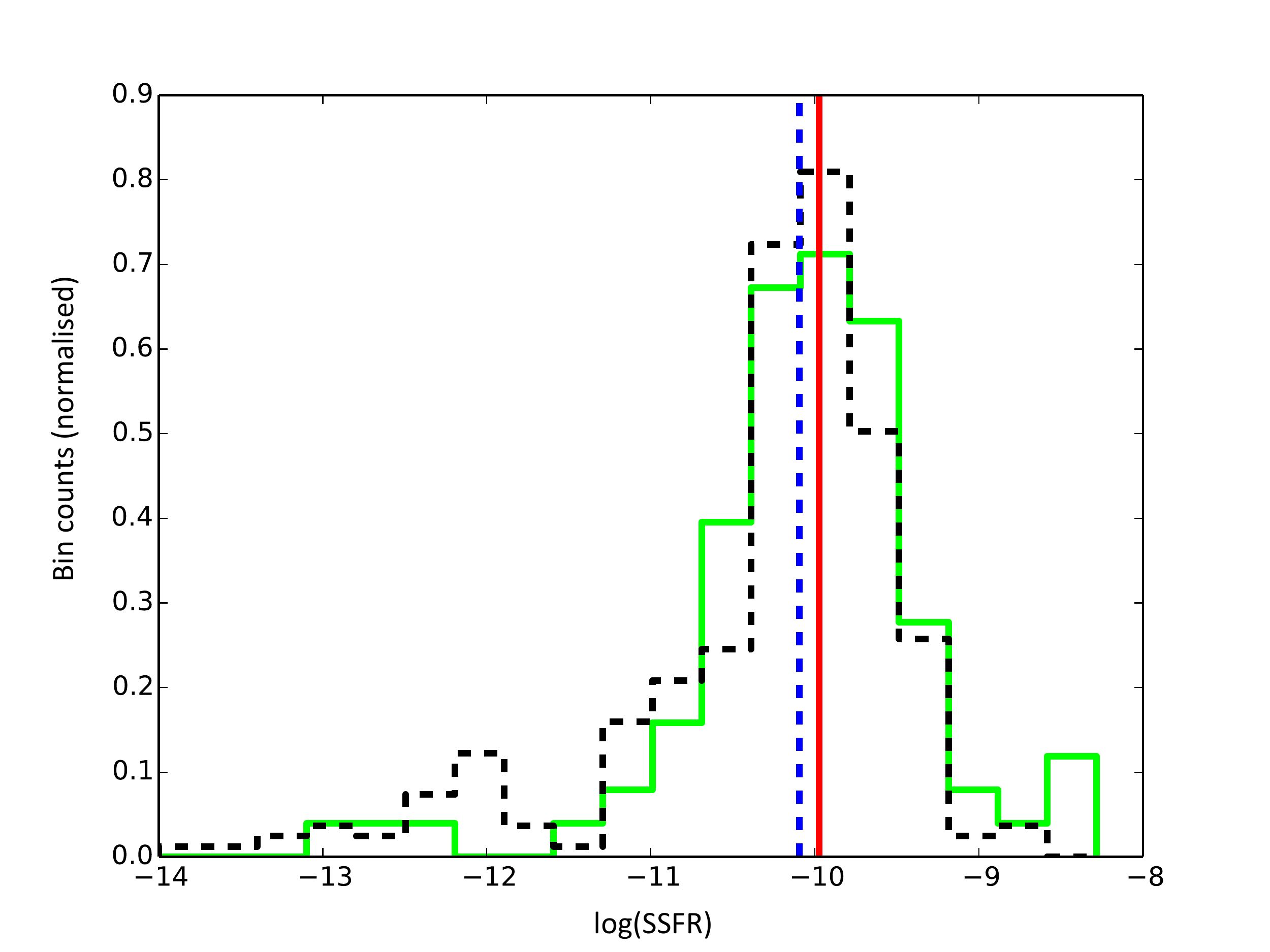}
\caption{{This figure is a} histogram of the SSFR for normal (black, dashed) and {kinematically} asymmetric (green) galaxies in the sample. The medians SSFRs for {kinematically} asymmetric and normal galaxies are red and blue (dashed), respectively. The offset between the medians is within the error, so is not considered significant.}
\label{fig:ssfr_hist_2}
\end{figure}

{Examining the SSFRs of kinematically asymmetric galaxies compared to normal galaxies, they are only marginally different (with a two-sample Kolmogorov-Smirnov test giving $p=0.06$). Fig. \ref{fig:ssfr_hist_2} shows histograms of the SSFR for kinematically asymmetric and normal galaxies, again using the SAMI Galaxy Survey SFRs. We find that the offset in the medians for the {kinematically} asymmetric and normal galaxies is not significant,} indicating that there is no significant increase in SSFR in {kinematically} asymmetric galaxies (see Table \ref{table:ssfr_offsets}). The error is calculated, here and henceforth, using the median correction to the standard error on the mean (where $\sigma$ is the standard deviation of the population, and $N$ is the population size):
\begin{equation}
error_{median}=1.253\times\frac{\sigma}{\sqrt{N}}
\end{equation}

To further analyse the relationship between star formation rate and {kinematic} asymmetry, we used two more measurements of SFRs to compare with the SAMI Galaxy Survey values. These were the GAMA Survey SFRs used in \ref{sec:characterising} and SDSS DR7 SFRs\footnote{SDSS DR7 SFRs are taken from `gal\_totsfr\_dr7\_v5\_2.fits.gz', obtained at \url{http://www.mpa-garching.mpg.de/SDSS/DR7/sfrs.html}},  both of which were also obtained from the H$\alpha$ flux.

The GAMA Survey SFRs were calculated by taking a measurement of the H$\alpha$ emission at the centre of the galaxy using a 2'' fibre and applying an aperture correction. The correction is calculated based on the proportion of the $r$-band continuum light of the galaxy captured within the size of the fibre, using a method from \citet{hopkins2003star}. There is an assumption that the H$\alpha$ emission scales directly with the $r$-band stellar continuum. A dust correction is applied from the Balmer decrement.

The SDSS DR7 SFRs are constructed using a method based on that in \citet{brinchmann2004physical} and a model from \citet{charlot2001nebular}, in which a measurement of H$\alpha$ emission is taken at the centre of the galaxy and then an aperture correction is applied, but the colour of the galaxy  is considered, yielding a more accurate total measurement. They do this by calculating the light outside the fibre, and then fit stochastic models, similar to those used in \citet{salim2007uv}, in which $ugriz$ photometry is fit to a variety of dust-attenuated population synthesis models. We note that only $\sim$250 of our galaxies had corresponding SDSS DR 7 SFRs. However, this reduction in sample size was spread equally across {kinematically} asymmetric and normal galaxies, so did not introduce bias.

Table \ref{table:ssfr_offsets} shows median SSFRs for {kinematically} asymmetric and normal galaxies, and the offsets between the two. The GAMA Survey SFRs yield a marginally significant offset in the medians, whereas there is no offset from the other two methods. Whilst the GAMA Survey results are still only marginally significant, they do represent a significant difference from the spatially resolved SAMI Galaxy Survey results. This is a result of the different methods of calculating SFR. The GAMA Survey SFRs are predicated on the assumption that the global SFR is directly proportional to the SFR of the region of the galaxy contained within a fibre placed at the centre. This leads to an over-emphasis on central SFR. By contrast, the SDSS DR7 SFRs, whilst still measuring SFR within a central fibre, modify their aperture correction by considering the global colour of the galaxy, which is linked to global SFR. Finally, the SAMI Galaxy Survey SFRs are intrinsically global, as they measure SFR in individual spaxels across the galaxy, although there is a small aperture effect outside the SAMI instrument bundle [Richards et al., (submitted)].

If we calculate the SAMI Galaxy Survey SFR within 2'' (the size of the GAMA  Survey fibre measurements), and then apply the GAMA Survey aperture corrections, we find the offset between the SSFRs of {kinematically} asymmetric and normal galaxies is $0.36\pm0.14$ dex. When comparing the aperture corrected values inside 2'' to the standard SAMI Galaxy Survey values, the median SSFRs for {kinematically} asymmetric and normal galaxies increase by 0.27$\pm0.12$ dex and $0.04\pm0.06$ dex, respectively. Although the effect is only $\sim 2\sigma$ significant, the GAMA Survey method of extrapolating central SFR throughout the disk appears to lead to an overestimation of extended SFR in some galaxies. The more centrally peaked the SFR, the greater the resulting overestimation. That this effect is strongest for {kinematically} asymmetric galaxies indicates that SFR in asymmetric galaxies is more centrally concentrated than it is in normal galaxies.

\subsection{Kinematic asymmetry and concentration of star formation}

\begin{figure}
\centering
\includegraphics[width=9cm]{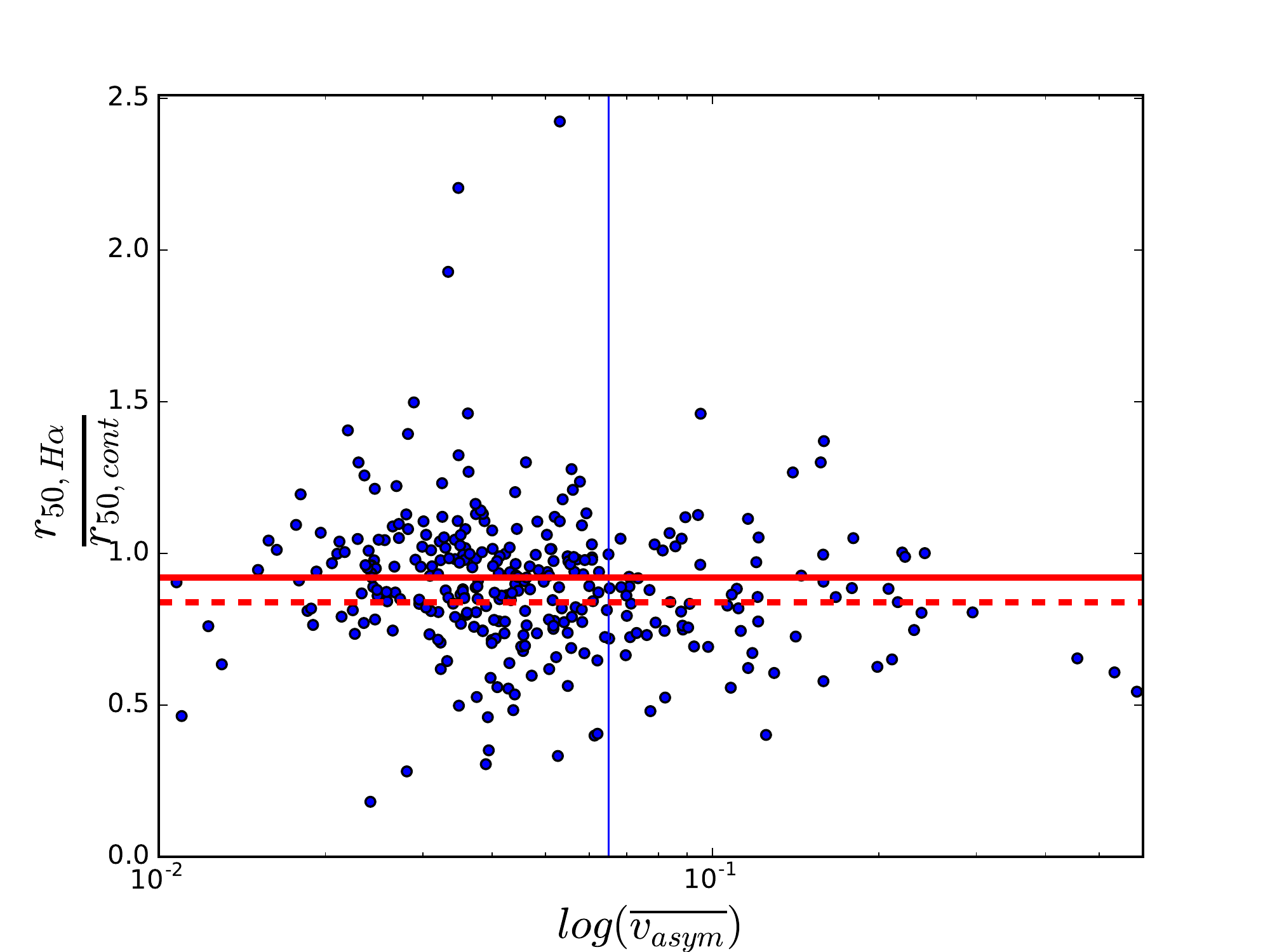}
\includegraphics[width=9cm]{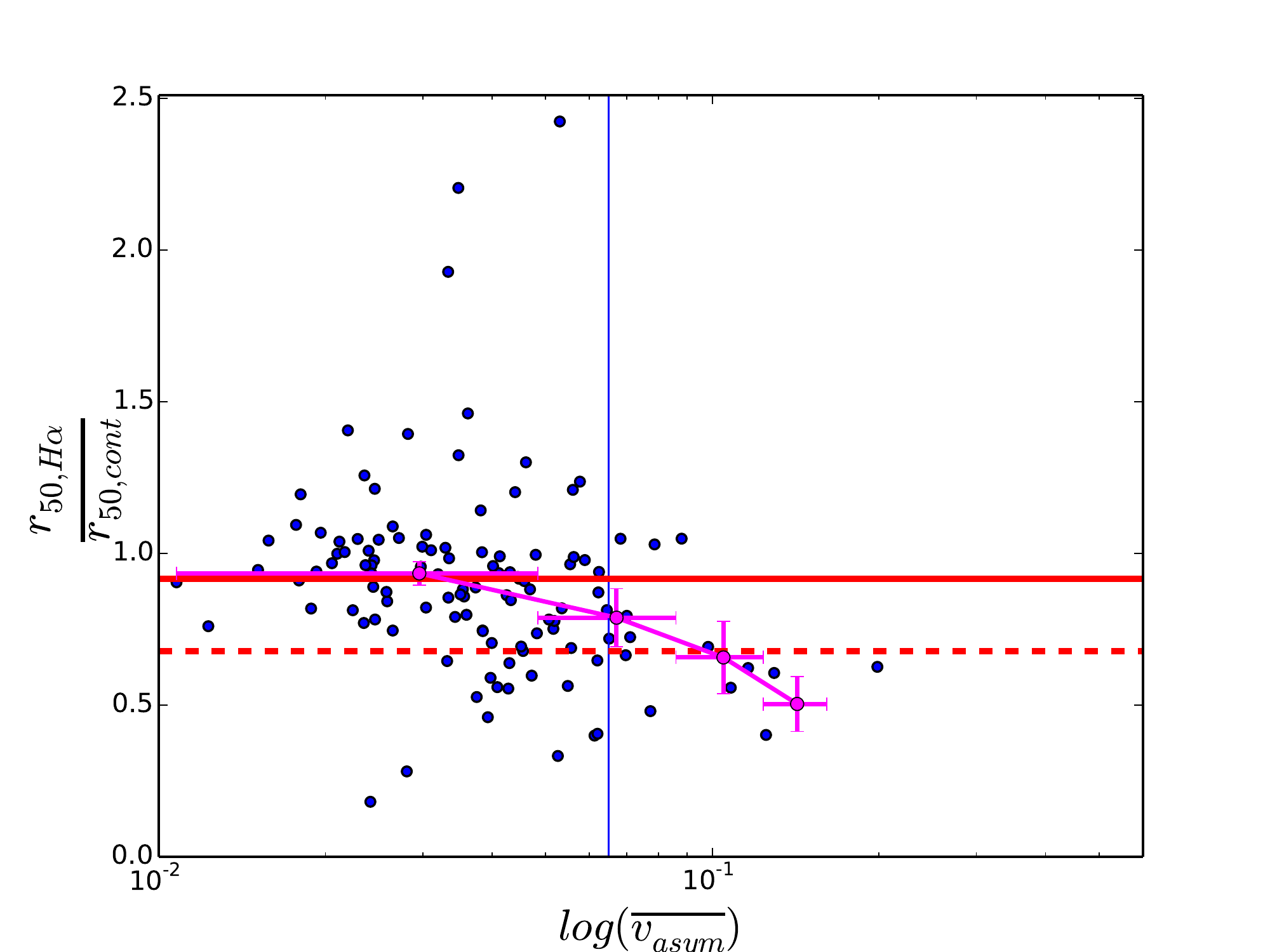}
\caption{{We show} $log(\overline{v_{asym}}$) against $\frac{r_{50,H\alpha}}{r_{50,cont}}$ for all galaxies (top) and galaxies with $log(M_{*})>10.0$ (bottom). The vertical blue line indicates the {kinematic} asymmetry cutoff $\overline{v_{asym}}>0.065$. Medians for normal (solid) and {kinematically} asymmetric (dashed) galaxies are shown in red. We see a 0.081$\pm$0.03 offset in median $\frac{r_{50,H\alpha}}{r_{50,cont}}$ for all galaxies, and 0.27$\pm$0.07 offset for high mass galaxies, as well as an overall inverse correlation between $\overline{v_{asym}}$ and $\frac{r_{50,H\alpha}}{r_{50,cont}}$ in both cases. In the bottom panel, binned medians of $\overline{v_{asym}}$ as a function of $\frac{r_{50,H\alpha}}{r_{50,cont}}$ are shown in magenta, further demonstrating the downward trend. Vertical error bars show the error on the median, and horizontal bars show the extend of the bins in $log(\overline{v_{asym}}$).}
\label{fig:r50}
\end{figure}

Fig. \ref{fig:r50} shows $\overline{v_{asym}}$ against the ratio of H$\alpha$ half light radius to $r$-band continuum half light ($\frac{r_{50,H\alpha}}{r_{50,cont}}$), both for all galaxies in our sample (top panel) and for only those galaxies with $log(M_{*})>10.0$ (bottom panel). The $\frac{r_{50,H\alpha}}{r_{50,cont}}$ ratio is calculated from the dust-corrected SAMI Galaxy Survey H$\alpha$ map and $r$-band continuum map for each galaxy. A curve of growth is calculated for both maps, defining the proportion of emission contained within a given radius. The $r_{50}$ radius is the radius containing 50$\%$ of the emission. Consequently, $\frac{r_{50,H\alpha}}{r_{50,cont}}$ is an indicator of the scale of star formation in a galaxy, compared to the stellar light (Schaefer et al., submitted). 

We see an offset of 0.08$\pm$0.03 in the median $\frac{r_{50,H\alpha}}{r_{50,cont}}$ for {kinematically} asymmetric and normal galaxies, and a Spearman's rank correlation test shows that there is an inverse correlation between $\overline{v_{asym}}$ and $\frac{r_{50,H\alpha}}{r_{50,cont}}$, with $\rho=-0.19$ and p-value$=7.6\times10^{-4}$, meaning that {kinematically} asymmetric galaxies have increased concentration of H$\alpha$ emission, compared to normal galaxies, and thus have increased central star formation.

It is possible that stochastic bursts of star formation in low mass galaxies may lead to more scatter in $\frac{r_{50,H\alpha}}{r_{50,cont}}$. If we restrict our sample to galaxies with $log(M_{*})>10.0$ (Fig. \ref{fig:r50}, bottom panel), the offset in median $\frac{r_{50,H\alpha}}{r_{50,cont}}$ increases to 0.24$\pm$0.07, and the Spearman's rank correlation test yields an even stronger result, with $\rho=-0.39$ and p-value$=1.9\times10^{-5}$. The reduction in p-value is particularly notable, given that the stellar mass cutoff restricts the sample to 143 galaxies. The full sample, excluding AGNs, contains 320 galaxies. To further show the trend, binned medians of $\frac{r_{50,H\alpha}}{r_{50,cont}}$ as a function of $\overline{v_{asym}}$ are shown in magenta, demonstrating the inverse correlation between $\overline{v_{asym}}$ and $\frac{r_{50,H\alpha}}{r_{50,cont}}$. A similar relationship is seen when the sample is restricted to galaxies with $log(M_{*})<9.0$ ($\rho=-0.19$,$p=0.05$), but is not significant for mid-mass galaxies $9.0<log(M_{*})<10.0$ ($\rho=-0.088$,$p=0.38$).

This indicates that whilst there is no significant increase in the amount of global star formation for {kinematically} asymmetric galaxies, there is a link between asymmetry and concentration of H$\alpha$ emission, or central star formation. We note that AGN will have biased $\frac{r_{50,H\alpha}}{r_{50,cont}}$ measurements, so we excluded them from the above analysis. AGN are identified from placement on a BPT diagram \citep{baldwin1981classification} derived from comparison of the [NII]/H$\alpha$ and [OIII]/H$\beta$ ratios. That is, the line ratios from the spectra of the central 2'' diameter of the AGN spectra sit above both the Kauffmann \citep{kauffmann2003host} and Kewley \citep{kewley2001theoretical} lines on the BPT diagram {We briefly discuss AGN asymmetry in Section ~\ref{sec:agn}}.

{A detailed study of $\frac{r_{50,H\alpha}}{r_{50,cont}}$, SFR and stellar mass will be the subject of future work by the SAMI Galaxy Survey team (Schaefer at al., submitted.)}

\begin{center}
\begin{table}
  \begin{tabular}{ b{2cm}  l  b{2cm}  l }
    \hline
\\
    Data Sample & $\overline{log(SSFR)}_{p}$ &$\overline{log(SSFR)}_{n}$&$\overline{log(SSFR)}$ \\
&&& offset \\ \hline
    SAMI Galaxy Survey SFR & -9.97$\pm$0.10 & -10.10$\pm$0.066 &0.12$\pm$0.12 \\
   GAMA Survey SFR &-9.72$\pm0.17$ & -10.05$\pm$0.10& 0.33$\pm$0.20 \\
SDSS Survey SFR &-10.04$\pm$0.12 &-10.19$\pm$0.070& 0.15$\pm$0.15 \\
    \hline
  \end{tabular}
\caption[Table caption text]{Median $log(SSFR)$ for {kinematically} asymmetric and normal galaxies (given as $\overline{log(SSFR)}_{p}$ and $\overline{log(SSFR)}_{n}$, respectively), and offsets between the two when using the SAMI Galaxy Survey, GAMA Survey and SDSS Survey SFRs, respectively. The offsets for the SAMI Galaxy Survey and SDSS values are not significant, due to the overlap in errors on the median SSFRs for asymmetric and normal galaxies.} 
\label{table:ssfr_offsets}
\end{table}
\end{center}

\subsection{Kinematic asymmetry and equivalent width of H$\alpha$ emission}
To further investigate the increase in concentration of star formation {(i.e. degree to which star formation is centralised)} due to {kinematic} asymmetry, we consider directly the relationship between central and extended star formation for {kinematically} asymmetric and normal galaxies. We use the equivalent width (EW) of H$\alpha$, which is approximately proportional to the SSFR. The equivalent width is the ratio of current star formation (derived from the H$\alpha$ emission) to star formation in the past (from the $r$-band continuum emission). We considered the EW within half an effective radius (henceforth inner EW), compared with the EW outside half an effective radius to the edge of the IFU (henceforth outer EW), for {kinematically} asymmetric and normal galaxies. We did not fix the outer radius, as we found that effective radius was not related to kinemetric asymmetry, so there would be no systematic effects.

\begin{figure}
\centering
\includegraphics[width=9cm]{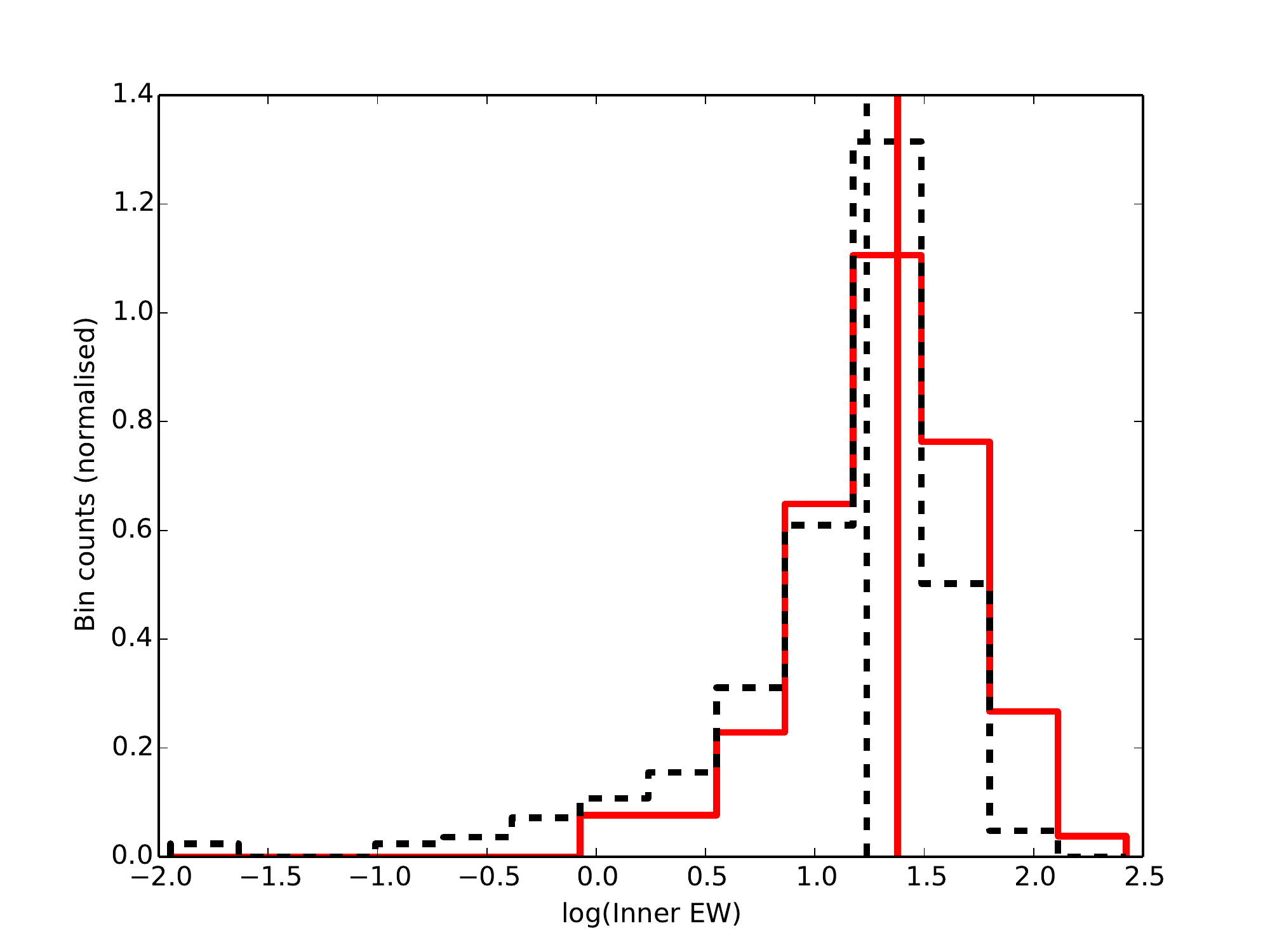}
\includegraphics[width=9cm]
{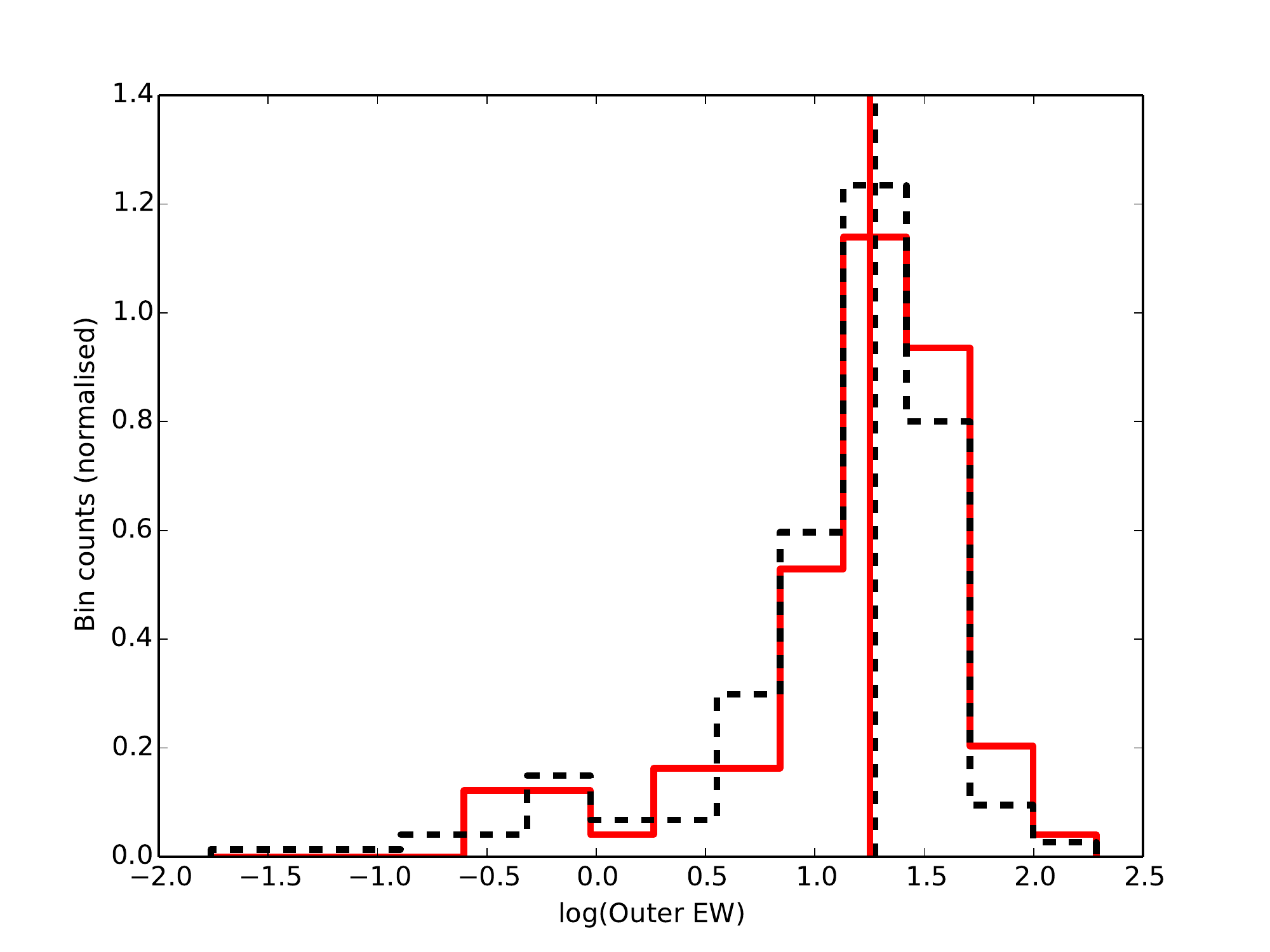}
\caption{{We show} normalised histograms of the equivalent width within half an effective radius (top) and outside this range (bottom) for {kinematically} asymmetric (red) and normal (black, dashed) galaxies. The median equivalent widths in each case are shown for asymmetric (red, solid) and normal (black, dashed) galaxies. The medians are the same for the log(outer EW), and there is a marginally significant offset for the log(inner EW). A two-sample Kolmogorov-Smirnov test of the distributions of log(inner EW) yields a p-value of 0.0090, indicating that the distributions of log(inner EW) for {kinematically} asymmetric and normal galaxies are statistically dissimilar, whereas this is not the case for log(outer EW).}
\label{fig:ew}
\end{figure}

Fig. ~\ref{fig:ew} shows histograms of the inner (top) and outer (bottom) EWs for {kinematically} asymmetric (red) and normal (black) galaxies, as well as the medians of the distributions in all cases. The median log(EW) for asymmetric and normal galaxies is the same for the outer EW, and there is a marginally significant offset of 0.14$\pm0.07$ dex in the inner EW, with the {kinematically} asymmetric and normal medians being 1.38$\pm0.05$ and 1.24$\pm0.04$, respectively (in log space). This result is similar in significance to the GAMA Survey results from SFR. A two-sample Kolmogorov-Smirnov test of the outer equivalent widths yields a p-value of 0.66 of the null hypothesis, whereas the same test for the inner equivalent widths yields a p-value of 0.0090. This indicates that only the distributions of inner equivalent widths of asymmetric and normal galaxies are statistically dissimilar. Table ~\ref{table:ew_offsets} gives the offsets in the medians for the inner and total EW, respectively. 

\begin{center}
\begin{table}
  \begin{tabular}{ b{5cm}  l  b{3cm}  l  l }
    \hline
    EW range & $\overline{log(EW)}$ offset \\ \hline
    Inner EW & 0.14$\pm$0.07 dex \\
  Outer EW & -0.02$\pm$0.16 dex\\
    \hline
  \end{tabular}
\caption[Table caption text]{Offsets in the median log(inner EW) and log(outer EW) for {kinematically} asymmetric and normal galaxies. There is no offset in log(outer EW), but there is an offset in log(inner EW), confirming that star formation is enhanced in the central region of asymmetric galaxies, relative to normal galaxies.}
\label{table:ew_offsets}
\end{table}
\end{center}

\subsection{Possible causes and implications of concentration of star formation}

There may be several causes for enhanced concentration of star formation in asymmetric galaxies. \citet{ellison2013galaxy} used close pairs and post-merger remnants in the SDSS survey to demonstrate that central star formation is greater for post-merger galaxies than for pre-merger galaxies, by comparing SFR enhancement within and outside fibre measurements.  This is qualitatively consistent with our results that {kinematic} asymmetry (a fraction of which is likely to be caused by interactions) is linked to star formation. We note that the central SSFR enhancement from asymmetry in \citet{ellison2013galaxy} is larger than our value. This may be explained by their selection of galaxies in pairs, in contrast to our broader sample. A more quantitative comparison may be possible with simulated data that will more effectively isolate specific causes of asymmetry and allow for analysis of their star formation.

\citet{barrera2015central} agree with our findings that there is an increase in central, but not extended star formation in interacting galaxies. They find a $2-3\times$ increase which, as in \citet{ellison2013galaxy}, is much larger than that found in this work. Their larger increase may also be explained their sample selection. They selected galaxies in interacting pairs, which excludes those that are asymmetric as a result of stochastic processes, or are post-merger remnants. Our results qualitatively agree with the theoretical findings of \citet{moreno2015mapping}, in which simulations of major mergers show an increase in central star formation, offset by suppression in the outskirts, i.e. a redistribution of gas. Further, their results indicating that lower-mass galaxies exhibit greater response to the interaction may be linked to our findings of a strong inverse correlation between stellar mass and kinematic asymmetry. This will be the focus of future work.

\citet{kaviraj2014importance} also shows a positive link between star formation and morphological disturbance that they assume to be caused by minor mergers. Their enhancement values are higher than ours, with ratios of SSFR for asymmetric and normal galaxies ranging from $\sim2$ to $\sim6$, for various populations. Given that they identify asymmetric galaxies through a visual classification, to which our kinemetric method has been shown to be a good match, and use star formation rates from the SDSS survey, the source of the discrepancy is not clear. Further, their asymmetric fractions of $\sim10-20\%$ are broadly consistent with our results.

Given the results of \citet{barrera2015central} and \citet{moreno2015mapping}, the increased central star formation in our sample points to minor mergers as a good candidate for the source of {kinematic} asymmetry in a significant fraction of cases. Further, we expect minor mergers to cause, in general, lower levels of kinemetric asymmetry than major mergers, which may contribute to the difference between our distribution and the SINS Survey results in Fig~\ref{fig:sins_points}. In addition, the radial inflows of gas associated with minor mergers lead to the accumulation of gas in central region of the host. In some cases, nearly half of all the gas initially distributed throughout the disk forms a dense region extending several hundred parsecs in the nucleus of the galaxy \citep{hernquist1995excitation}. This may lead to nuclear starbursts, contributing to the link we see between kinematic asymmetry and central star formation. 

Further, simulations show that tidal interactions can trigger repetitive central starbursts in spiral galaxies \citep{bekki2011transformation}, growing the bulge and transforming them into gas-poor S0s. Such interactions may disturb the kinematics of a galaxy, and lead to fluctuating star formation, causing either increased or decreased measurements of SSFR, depending on when in the course of the interactions the SSFR was calculated. 

\section{Kinematic asymmetry of AGN}
\label{sec:agn}

We do not find any difference in kinematic asymmetry in AGN, compared with the general sample. The median $\overline{v_{asym}}$ for the general population (excluding AGN) is $0.043\pm0.0053$, and for AGN it is $0.048\pm0.012$. Further, a two-sample Kolmogorov-Smirnov test on the $\overline{v_{asym}}$ values of AGN and non-AGN galaxies gives a probability of 0.71 of the null hypothesis, indicating that there is not a difference in distribution.

We do note that the AGN in this sample are relatively low-luminosity AGN, drawn from a mass-limited sample of regular galaxies. For a specific study of the spatially resolved kinematics of AGN, see McElroy et al. (in prep).

\section{Conclusions}
\label{sec:conclusions}
We have defined a sample of 360 galaxies from $\sim$ 450 available in the SAMI Galaxy Survey at the start of this work.

Using a method based on kinemetry, we calculated two coefficients for each galaxy, $\overline{v_{asym}}$ and $\overline{\sigma_{asym}}$, which represent the median kinematic asymmetry in the velocity and velocity dispersion, respectively. We found that these coefficients, particularly $\overline{v_{asym}}$, consistently distinguished between galaxies identified visually as normal or asymmetric. That is, our method and a visual classification agree for 90$\%$ of {visually} asymmetric galaxies and 95$\%$ of normal galaxies. The fraction of the sample classified as {kinematically} asymmetric using kinemetry was 23$\% \pm 7\%$.

A direct comparison of our results with classification methods using the Gini, $M_{20}$ and CAS $A$ coefficients {finds that galaxies classified as morphologically disturbed are all also kinematically asymmetric. We also find a significant population of galaxies with kinematic asymmetry that are not classified as disturbed using he Gini, $M_{20}$ and CAS $A$ coefficients. In contrast, a visual by-eye classification showed good agreement with the kinematic classification.}

There is a strong inverse correlation between mass and {kinematic} asymmetry, with $\rho=-0.30$ and p-value$=8.05\times10^{-8}$. Further, the proportion of galaxies classified as {kinematically} asymmetric falls as a function of stellar mass. This may be due to the higher gas fraction in low-mass galaxies leading to relatively larger gravitational instabilities, or to the influence of environment. {This result agrees with those from previous studies that have used different approaches to analyse the kinematic disturbance of low mass galaxies.}

Using SFRs from the SAMI Galaxy Survey data, we do not find a significant global offset in star formation between {kinematically} asymmetric and normal galaxies. However, we find that asymmetric galaxies have SFRs enhanced in their centres. We found an inverse correlation between $\overline{v_{asym}}$ and the ratio between $\frac{r_{50,H\alpha}}{r_{50,cont}}$, with $\rho=-0.19$ and p-value$=6.2\times10^{-4}$. If only high mass [log$(M_{*})>10.0$] galaxies are considered, the relationship becomes stronger, with $\rho=-0.37$ and p-value$=6.0\times10^{-6}$. This indicates that kinemetric asymmetry is linked with central star formation, as has also been seen in \citet{ellison2013galaxy} and others. A two-sample Kolmogorov-Smirnov test was able to distinguish the distributions of equivalent widths for asymmetric and normal galaxies for inner equivalent width only.

In future work, we shall investigate the extent to which {the physical} origin of {kinematic} asymmetry {e.g. mergers, turbulence and tidal tails} influences the results of kinemetry. Simulations will play an important role in furthering our understanding of the relationship between various kinds and magnitudes of asymmetry and the output of our kinemetric classification. We will also be able to make comparisons to high-redshift results, both through the use of simulations and artificial `redshifting' of SAMI Galaxy Survey velocity fields. The effects of environment on disturbance will also be investigated, particularly in light of the observed mass-$\overline{v_{asym}}$ relationship. The final size of the SAMI Galaxy Survey, 3400 galaxies, will also allow us to make more statistically robust statements in the future. 

\section*{ACKNOWLEDGMENTS}
The SAMI Galaxy Survey is based on observations made at the Anglo-Australian Telescope. The Sydney-AAO Multi-object Integral field spectrograph (SAMI) was developed jointly by the University of Sydney and the Australian Astronomical Observatory. The SAMI input catalogue is based on data taken from the Sloan Digital Sky Survey, the GAMA Survey and the VST ATLAS Survey. The SAMI Galaxy Survey is funded by the Australian Research Council Centre of Excellence for All-sky Astrophysics (CAASTRO), through project number CE110001020, and other participating institutions. The SAMI Galaxy Survey website is http://sami-survey.org/.

GAMA is a joint European-Australasian project based around a spectroscopic campaign using the Anglo-Australian Telescope. The GAMA input catalogue is based on data taken from the Sloan Digital Sky Survey and the UKIRT Infrared Deep Sky Survey. Complementary imaging of the GAMA regions is being obtained by a number of independent survey programs including GALEX MIS, VST KiDS, VISTA VIKING, WISE, Herschel- ATLAS, GMRT    ASKAP providing UV to radio coverage. GAMA is funded by the STFC (UK), the ARC (Australia), the AAO, and the participating institutions. The GAMA website is http://www.gama-survey.org/.

SMC acknowledges the support of an Australian Research Council Future Fellowship (FT100100457).

L.C. acknowledges financial support from the Australian Research Council (DP130100664). SB acknowledges the funding support from the Australian Research Council through a Future Fellowship (FT140101166)

M.S.O. acknowledges the funding support from the Australian Research Council through a Future Fellowship Fellowship (FT140100255).

\bibliography{paper_I}
\bibliographystyle{mn2e}
Galaxy properties and kinematic classifications
\appendix
\onecolumn
\section{Morphological and Kinematic Classifications, GAMA Catalogue information}
\begin{center}

\begin{longtable}{@{\extracolsep{\fill}}c c c c c c c c c c c@{}}
\hline

GAMA Survey & Vis. & Kin. & $\overline{v_{asym}}$ & $\overline{v_{asym}}$ & $\overline{\sigma_{asym}}$ & $\overline{\sigma_{asym}}$ &  log($M_{*}$) & Colour& log(SFR) & $\frac{r_{50,H\alpha}}{r_{50,cont}}$ \\
ID & Class. & Class. & & err. & & err. & & ($u-r$)& & \\
\hline
GAMA15218 & 	normal	 & 	normal	 & 0.031 & 0.005 & 0.057 & 0.016 & 9.11 & 1.16 & 0.08 & 0.96 \\
GAMA15510 & 	normal	 & 	normal	 & 0.062 & 0.005 & 0.184 & 0.039 & 10.12 & 1.84 & 0.92 & 0.87 \\
GAMA15561 & 	perturbed	 & 	perturbed	 & 0.111 & 0.019 & 0.448 & 0.338 & 8.19 & 1.16 & -0.71 & 0.88 \\
GAMA16026 & 	normal	 & 	normal	 & 0.065 & 0.013 & 0.147 & 0.021 & 10.21 & 1.83 & 1.25 & 0.81 \\
GAMA16294 & 	normal	 & 	normal	 & 0.034 & 0.002 & 0.084 & 0.011 & 8.95 & 1.13 & -0.36 & 1.04 \\
GAMA28860 & 	normal	 & 	normal	 & 0.037 & 0.003 & 0.037 & 0.012 & 9.43 & 1.28 & 0.61 & 0.98 \\
GAMA30377 & 	normal	 & 	normal	 & 0.028 & 0.039 & 0.041 & 0.033 & 8.24 & 1.12 & -0.35 & 1.13 \\
GAMA31452 & 	normal	 & 	normal	 & 0.053 & 0.004 & 0.071 & 0.015 & 9.43 & 1.17 & 0.49 & 0.89 \\
GAMA31509 & 	normal	 & 	normal	 & 0.047 & 0.005 & 0.046 & 0.011 & 8.38 & 1.09 & {-999.0} & 0.96 \\
GAMA31620 & 	normal	 & 	normal	 & 0.024 & 0.004 & 0.095 & 0.035 & 10.53 & 2.07 & -0.11 & 1.26 \\
GAMA32362 & 	normal	 & 	normal	 & 0.038 & 0.003 & 0.049 & 0.016 & 10.48 & 1.77 & 0.9 & 1.0 \\
GAMA36894 & 	normal	 & 	normal	 & 0.021 & 0.006 & 0.049 & 0.013 & 8.75 & 1.13 & -0.2 & 0.79 \\
GAMA37050 & 	normal	 & 	normal	 & 0.031 & 0.007 & 0.089 & 0.021 & 9.15 & 1.41 & -0.23 & 1.11 \\
GAMA39108 & 	perturbed	 & 	perturbed	 & 0.12 & 0.066 & 0.721 & 0.498 & 8.27 & 1.09 & -0.65 & 0.97 \\
GAMA41144 & 	normal	 & 	normal	 & 0.021 & 0.0041 & 0.051 & 0.005 & 10.37 & 1.66 & 1.31 & 1.04 \\
GAMA41164 & 	normal	 & 	normal	 & 0.059 & 0.007 & 0.093 & 0.019 & 8.34 & 0.95 & -0.46 & 1.13 \\
GAMA47342 & 	normal	 & 	normal	 & 0.03 & 0.003 & 0.032 & 0.019 & 10.05 & 1.61 & {-999.0} & 1.02 \\
GAMA47500 & 	normal	 & 	normal	 & 0.044 & 0.013 & 0.12 & 0.019 & 9.45 & 1.35 & 0.59 & 0.97 \\
GAMA47535 & 	normal	 & 	perturbed	 & 0.068 & 0.015 & 0.12 & 0.064 & 8.49 & 1.06 & {-999.0} & 0.89 \\
GAMA49857 & 	perturbed	 & 	perturbed	 & 0.071 & 0.0042 & 0.10 & 0.015 & 9.14 & 1.12 & 0.58 & 0.83 \\
GAMA53771 & 	perturbed	 & 	perturbed	 & 0.111 & 0.077 & 1.313 & 1.134 & 8.49 & 1.23 & {-999.0} & 0.82 \\
GAMA53809 & 	normal	 & 	normal	 & 0.034 & 0.005 & 0.086 & 0.014 & 9.02 & 1.23 & 0.26 & 0.83 \\
GAMA56064 & 	normal	 & 	normal	 & 0.048 & 0.012 & 0.115 & 0.071 & 10.51 & 1.88 & -0.12 & 1.4 \\
GAMA56140 & 	normal	 & 	normal	 & 0.024 & 0.011 & 0.144 & 0.033 & 11.28 & 2.29 & 1.47 & 0.93 \\
GAMA56183 & 	perturbed	 & 	perturbed	 & 0.091 & 0.0052 & 0.12 & 0.013 & 9.44 & 1.27 & 0.41 & 0.83 \\
GAMA62412 & 	normal	 & 	normal	 & 0.025 & 0.003 & 0.042 & 0.008 & 9.28 & 1.76 & -0.42 & 0.88 \\
GAMA62593 & 	perturbed	 & 	normal	 & 0.036 & 0.002 & 0.126 & 0.028 & 9.68 & 2.05 & 0.44 & 0.8 \\
GAMA62718 & 	perturbed	 & 	perturbed	 & 0.141 & 0.026 & 0.334 & 0.311 & 9.56 & 1.34 & 1.28 & 0.73 \\
GAMA64087 & 	normal	 & 	normal	 & 0.048 & 0.021 & 0.086 & 0.014 & 10.35 & 1.91 & 1.47 & 0.74 \\
GAMA65278 & 	normal	 & 	normal	 & 0.043 & 0.034 & 0.096 & 0.022 & 9.09 & 1.31 & -0.34 & 1.02 \\
GAMA65406 & 	normal	 & 	normal	 & 0.040 & 0.007 & 0.068 & 0.007 & 11.02 & 2.32 & 0.56 & 0.86 \\
GAMA65410 & 	normal	 & 	perturbed	 & 0.071 & 0.046 & 0.162 & 0.016 & 10.12 & 1.98 & 0.66 & 0.66 \\
GAMA77445 & 	normal	 & 	normal	 & 0.061 & 0.013 & 0.211 & 0.105 & 8.6 & 1.18 & {-999.0} & 1.03 \\
GAMA77754 & 	normal	 & 	normal	 & 0.022 & 0.004 & 0.116 & 0.016 & 10.49 & 1.51 & 1.66 & 0.81 \\
GAMA78531 & 	perturbed	 & 	perturbed	 & 0.068 & 0.039 & 0.111 & 0.021 & 10.65 & 2.11 & 0.81 & 1.05 \\
GAMA78667 & 	normal	 & 	normal	 & 0.03 & 0.003 & 0.09 & 0.018 & 10.18 & 1.36 & {-999.0} & 0.96 \\
GAMA79635 & 	normal	 & 	normal	 & 0.03 & 0.002 & 0.041 & 0.039 & 10.44 & 1.61 & {-999.0} & 1.06 \\
GAMA79693 & 	normal	 & 	normal	 & 0.034 & 0.021 & 0.163 & 0.041 & 10.04 & 2.16 & -0.3 & 0.79 \\
GAMA79710 & 	perturbed	 & 	perturbed	 & 0.07 & 0.064 & 0.266 & 0.043 & 9.0 & 1.53 & 0.22 & 0.86 \\
GAMA79712 & 	perturbed	 & 	perturbed	 & 0.167 & 0.117 & 0.984 & 0.723 & 8.04 & 1.52 & 0.21 & 0.86 \\
GAMA79771 & 	normal	 & 	normal	 & 0.033 & 0.004 & 0.038 & 0.005 & 8.81 & 1.09 & -0.55 & 1.23 \\
GAMA79850 & 	normal	 & 	normal	 & 0.052 & 0.021 & 0.098 & 0.007 & 9.81 & 2.07 & 0.37 & 0.66 \\
GAMA91924 & 	normal	 & 	perturbed	 & 0.066 & 0.008 & 0.439 & 0.084 & 10.65 & 1.67 & 1.15 & 0.98 \\
GAMA91926 & 	normal	 & 	normal	 & 0.024 & 0.003 & 0.181 & 0.027 & 10.09 & 2.25 & 0.16 & 0.78 \\
GAMA91963 & 	normal	 & 	normal	 & 0.061 & 0.025 & 0.447 & 0.084 & 11.04 & 2.26 & 0.55 & 0.4 \\
GAMA92770 & 	normal	 & 	normal	 & 0.023 & 0.004 & 0.069 & 0.008 & 9.87 & 1.91 & -0.77 & 1.3 \\
GAMA106376 & normal	 & 	normal	 & 0.041 & 0.017 & 0.097 & 0.026 & 10.17 & 1.21 & 1.43 & 0.94 \\
GAMA106389 & 	normal	 & 	normal	 & 0.012 & 0.002 & 0.075 & 0.024 & 10.22 & 2.02 & 0.77 & 0.76 \\
GAMA106717 & 	normal	 & 	normal	 & 0.018 & 0.005 & 0.057 & 0.007 & 10.13 & 1.29 & 1.21 & 1.19 \\
GAMA137789 & 	normal	 & 	normal	 & 0.036 & 0.029 & 0.131 & 0.028 & 8.41 & 1.23 & {-999.0} & 1.02 \\
GAMA137847 & 	perturbed	 & 	perturbed	 & 0.118 & 0.012 & 0.397 & 0.085 & 8.9 & 1.15 & 0.54 & 0.67 \\
GAMA138066 & 	normal	 & 	normal	 & 0.033 & 0.004 & 0.062 & 0.008 & 9.43 & 2.26 & 0.041 & 0.88 \\
GAMA138094 & 	normal	 & 	normal	 & 0.04 & 0.041 & 0.137 & 0.074 & 8.8 & 1.63 & -1.01 & 0.71 \\
GAMA144239 & 	normal	 & 	normal	 & 0.045 & 0.004 & 0.309 & 0.053 & 10.05 & 1.68 & 1.05 & 0.92 \\
GAMA144243 & 	perturbed	 & 	perturbed	 & 0.106 & 0.007 & 0.179 & 0.037 & 8.97 & 1.18 & {-999.0} & 0.83 \\
GAMA144320 & 	normal	 & 	perturbed	 & 0.07 & 0.03 & 0.153 & 0.027 & 10.28 & 1.77 & 1.01 & 0.79 \\

\newpage
\hline
GAMA Survey & Vis. & Kin. & $\overline{v_{asym}}$ & $\overline{v_{asym}}$ & $\overline{\sigma_{asym}}$ & $\overline{\sigma_{asym}}$ &  log($M_{*}$) & Colour& log(SFR) & $\frac{r_{50,H\alpha}}{r_{50,cont}}$ \\
ID & Class. & Class. & & err. & & err. & & ($u-r$)& & \\
\hline
GAMA144402 & 	normal	 & 	normal	 & 0.031 & 0.002 & 0.055 & 0.013 & 10.23 & 1.47 & {-999.0} & 1.01 \\
GAMA144465 & 	perturbed	 & 	perturbed	 & 0.159 & 0.071 & 0.376 & 0.057 & 8.66 & 1.31 & -1.08 & 0.58 \\
GAMA144480 & 	normal	 & 	normal	 & 0.057 & 0.028 & 0.189 & 0.149 & 8.27 & 1.14 & -0.6 & 0.82 \\
GAMA144498 & 	perturbed	 & 	perturbed	 & 0.082 & 0.013 & 0.202 & 0.12 & 9.89 & 2.1 & 0.72 & 0.52 \\
GAMA144682 & 	perturbed	 & 	perturbed	 & 0.533 & 0.075 & 0.779 & 0.349 & 8.97 & 1.19 & {-999.0} & 0.61 \\
GAMA178481 & 	normal	 & 	normal	 & 0.052 & 0.005 & 0.101 & 0.037 & 9.06 & 1.29 & -0.46 & 1.12 \\
GAMA178578 & 	normal	 & 	normal	 & 0.044 & 0.004 & 0.099 & 0.023 & 8.45 & 1.19 & {-999.0} & 0.9 \\
GAMA185252 & 	normal	 & 	normal	 & 0.05 & 0.008 & 0.128 & 0.027 & 8.55 & 1.23 & {-999.0} & 0.94 \\
GAMA185291 & 	normal	 & 	normal	 & 0.046 & 0.043 & 0.097 & 0.017 & 8.83 & 1.45 & -0.12 & 0.76 \\
GAMA185510 & 	normal	 & 	normal	 & 0.046 & 0.005 & 0.074 & 0.018 & 9.38 & 1.57 & {-999.0} & 0.73 \\
GAMA185532 & 	perturbed	 & 	normal	 & 0.052 & 0.004 & 0.096 & 0.02 & 9.14 & 1.34 & -0.32 & 0.97 \\
GAMA198503 & 	normal	 & 	normal	 & 0.032 & 0.026 & 0.06 & 0.038 & 8.61 & 1.41 & -0.16 & 0.81 \\
GAMA203037 & 	perturbed	 & 	perturbed	 & 0.116 & 0.091 & 0.731 & 0.282 & 8.23 & 1.22 & -1.78 & 1.11 \\
GAMA203114 & 	normal	 & 	normal	 & 0.03 & 0.008 & 0.09 & 0.019 & 10.91 & 2.51 & 0.18 & 0.34 \\
GAMA203140 & 	normal	 & 	normal	 & 0.058 & 0.007 & 0.075 & 0.021 & 10.95 & 2.47 & -0.9 & {-999.0} \\
GAMA203148 & 	normal	 & 	normal	 & 0.044 & 0.033 & 0.138 & 0.096 & 9.24 & 1.27 & 0.28 & 1.08 \\
GAMA204799 & 	normal	 & 	perturbed	 & 0.087 & 0.006 & 0.113 & 0.019 & 10.4 & 1.88 & 1.51 & {-999.0} \\
GAMA204906 & 	perturbed	 & 	perturbed	 & 0.183 & 0.124 & 1.407 & 1.026 & 10.21 & 2.02 & -0.18 & 1.51 \\
GAMA209181 & 	normal	 & 	normal	 & 0.026 & 0.003 & 0.071 & 0.012 & 10.27 & 1.24 & 1.27 & 1.09 \\
GAMA209319 & 	normal	 & 	normal	 & 0.041 & 0.024 & 0.104 & 0.048 & 8.39 & 0.96 & 0.33 & 0.85 \\
GAMA209414 & 	normal	 & 	normal	 & 0.051 & 0.006 & 0.075 & 0.009 & 8.99 & 1.23 & -0.31 & 0.85 \\
GAMA209698 & 	perturbed	 & 	perturbed	 & 0.086 & 0.016 & 0.615 & 0.113 & 10.35 & 2.04 & 1.52 & 0.48 \\
GAMA209701 & 	normal	 & 	normal	 & 0.024 & 0.002 & 0.123 & 0.041 & 10.81 & 2.19 & 0.022 & 1.16 \\
GAMA209708 & 	normal	 & 	normal	 & 0.061 & 0.016 & 0.194 & 0.014 & 8.61 & 1.26 & -0.53 & 0.99 \\
GAMA209743 & 	normal	 & 	normal	 & 0.024 & 0.003 & 0.091 & 0.011 & 10.19 & 1.56 & 1.03 & 0.89 \\
GAMA209807 & 	normal	 & 	normal	 & 0.052 & 0.007 & 0.191 & 0.042 & 10.83 & 1.89 & 1.40 & 0.75 \\
GAMA215053 & 	normal	 & 	normal	 & 0.047 & 0.005 & 0.09 & 0.014 & 10.07 & 2.25 & -0.62 & 0.61 \\
GAMA215292 & 	normal	 & 	normal	 & 0.046 & 0.002 & 0.05 & 0.005 & 10.11 & 1.62 & 0.82 & 0.91 \\
GAMA215335 & 	perturbed	 & 	perturbed	 & 0.129 & 0.105 & 0.89 & 0.32 & 10.12 & 1.81 & 1.18 & 0.61 \\
GAMA215698 & 	perturbed	 & 	perturbed	 & 0.457 & 0.137 & 1.12 & 0.576 & 8.19 & 1.47 & -1.35 & 0.65 \\
GAMA216843 & 	normal	 & 	normal	 & 0.031 & 0.006 & 0.042 & 0.016 & 9.19 & 1.24 & 0.52 & 0.73 \\
GAMA218713 & 	perturbed	 & 	perturbed	 & 0.116 & 0.182 & 0.427 & 0.21 & 10.05 & 2.08 & 0.64 & 0.62 \\
GAMA220320 & 	normal	 & 	normal	 & 0.039 & 0.005 & 0.097 & 0.008 & 8.97 & 1.67 & -0.52 & 0.35 \\
GAMA220371 & 	normal	 & 	normal	 & 0.058 & 0.008 & 0.057 & 0.012 & 9.56 & 1.64 & -0.71 & 1.09 \\
GAMA220383 & 	normal	 & 	normal	 & 0.029 & 0.007 & 0.088 & 0.011 & 8.51 & 1.77 & -9.0 & 1.511 \\
GAMA227278 & 	perturbed	 & 	perturbed	 & 0.124 & 0.02 & 0.63 & 0.229 & 10.12 & 2.32 & -0.66 & 0.61 \\
GAMA227351 & 	normal	 & 	normal	 & 0.041 & 0.004 & 0.049 & 0.005 & 9.47 & 1.71 & 0.23 & 0.72 \\
GAMA227371 & 	normal	 & 	normal	 & 0.045 & 0.007 & 0.111 & 0.076 & 8.53 & 1.13 & -0.53 & 0.88 \\
GAMA227407 & 	normal	 & 	normal	 & 0.036 & 0.029 & 0.104 & 0.064 & 8.44 & 1.44 & -1.21 & 1.08 \\
GAMA227428 & 	normal	 & 	perturbed	 & 0.073 & 0.022 & 0.116 & 0.021 & 9.99 & 1.6 & 0.72 & 0.92 \\
GAMA227572 & 	perturbed	 & 	perturbed	 & 0.089 & 0.009 & 0.346 & 0.068 & 9.87 & 1.47 & 1.53 & 0.75 \\
GAMA227614 & 	perturbed	 & 	perturbed	 & 0.098 & 0.032 & 0.381 & 0.033 & 10.01 & 2.02 & 1.2 & 0.69 \\
GAMA227961 & 	normal	 & 	normal	 & 0.026 & 0.002 & 0.116 & 0.02 & 10.37 & 2.29 & 0.71 & 0.84 \\
GAMA227970 & 	normal	 & 	normal	 & 0.024 & 0.003 & 0.119 & 0.021 & 10.15 & 1.32 & 1.24 & 0.98 \\
GAMA228066 & 	normal	 & 	normal	 & 0.036 & 0.003 & 0.13 & 0.009 & 10.37 & 2.06 & 0.05 & 0.86 \\
GAMA228086 & 	perturbed	 & 	normal	 & 0.057 & 0.006 & 0.118 & 0.065 & 9.12 & 1.14 & -0.013 & 0.98 \\
GAMA228428 & 	normal	 & 	normal	 & 0.046 & 0.002 & 0.05 & 0.011 & 8.85 & 0.77 & 0.68 & 0.81 \\
GAMA228432 & 	normal	 & 	normal	 & 0.058 & 0.006 & 0.072 & 0.019 & 9.33 & 1.29 & 0.99 & 0.77 \\
GAMA230714 & 	normal	 & 	normal	 & 0.024 & 0.004 & 0.028 & 0.006 & 10.21 & 1.66 & 1.05 & 1.01 \\
GAMA230776 & 	perturbed	 & 	perturbed	 & 0.807 & 0.197 & 1.414 & 0.507 & 11.71 & 2.64 & -0.43 & {-999.0} \\
GAMA238080 & 	normal	 & 	perturbed	 & 0.065 & 0.035 & 0.401 & 0.201 & 8.26 & 1.12 & -0.76 & 0.88 \\
GAMA238085 & 	perturbed	 & 	perturbed	 & 0.18 & 0.042 & 0.456 & 0.151 & 9.11 & 1.22 & -0.25 & 1.05 \\
GAMA238125 & 	normal	 & 	normal	 & 0.037 & 0.004 & 0.045 & 0.009 & 9.56 & 1.57 & 0.74 & 0.76 \\
GAMA238164 & 	perturbed	 & 	perturbed	 & 0.077 & 0.007 & 0.103 & 0.02 & 8.98 & 1.15 & -0.21 & 0.88 \\
GAMA238203 & 	normal	 & 	normal	 & 0.029 & 0.007 & 0.073 & 0.011 & 10.14 & 2.44 & -1.17 & {-999.0} \\
GAMA238204 & 	normal	 & 	normal	 & 0.029 & 0.008 & 0.095 & 0.021 & 10.69 & 1.87 & -0.57 & 0.88 \\
GAMA238216 & 	normal	 & 	normal	 & 0.046 & 0.03 & 0.143 & 0.023 & 10.17 & 1.81 & 0.71 & 0.68 \\
GAMA238221 & 	normal	 & 	normal	 & 0.024 & 0.003 & 0.077 & 0.016 & 10.19 & 1.6 & 0.64 & 0.96 \\
GAMA238276 & 	normal	 & 	normal	 & 0.056 & 0.006 & 0.072 & 0.013 & 10.56 & 2.31 & {-999.0} & 1.21 \\
\newpage
\hline
GAMA Survey & Vis. & Kin. & $\overline{v_{asym}}$ & $\overline{v_{asym}}$ & $\overline{\sigma_{asym}}$ & $\overline{\sigma_{asym}}$ &  log($M_{*}$) & Colour& log(SFR) & $\frac{r_{50,H\alpha}}{r_{50,cont}}$ \\
ID & Class. & Class. & & err. & & err. & & ($u-r$)& & \\
\hline

GAMA238282 & 	normal	 & 	normal	 & 0.039 & 0.005 & 0.081 & 0.014 & 10.59 & 2.14 & -1.0 & 0.74 \\
GAMA238328 & 	normal	 & 	normal	 & 0.031 & 0.05 & 0.183 & 0.169 & 8.76 & 1.23 & 0.06 & 0.81 \\
GAMA238351 & 	normal	 & 	normal	 & 0.039 & 0.004 & 0.074 & 0.014 & 9.93 & 2.26 & 0.81 & 0.31 \\
GAMA238358 & 	normal	 & 	normal	 & 0.043 & 0.004 & 0.099 & 0.075 & 10.86 & 1.73 & {-999.0} & 0.86 \\
GAMA238360 & 	normal	 & 	normal	 & 0.035 & 0.005 & 0.145 & 0.029 & 10.19 & 2.04 & 0.56 & 0.88 \\
GAMA238395 & 	normal	 & 	normal	 & 0.024 & 0.007 & 0.048 & 0.004 & 9.82 & 1.33 & 1.04 & 0.96 \\
GAMA238496 & 	normal	 & 	normal	 & 0.043 & 0.008 & 0.136 & 0.024 & 10.2 & 2.21 & 0.61 & 0.55 \\
GAMA238925 & 	normal	 & 	normal	 & 0.018 & 0.009 & 0.166 & 0.03 & 10.49 & 1.70 & 1.22 & 0.91 \\
GAMA239172 & 	perturbed	 & 	perturbed	 & 0.14 & 0.137 & 0.348 & 0.215 & 8.16 & 1.10 & -0.51 & 1.27 \\
GAMA239249 & 	normal	 & 	normal	 & 0.019 & 0.025 & 0.037 & 0.005 & 9.34 & 1.61 & 0.13 & 0.81 \\
GAMA239292 & 	normal	 & 	normal	 & 0.043 & 0.022 & 0.102 & 0.027 & 10.03 & 1.93 & -0.011 & 0.94 \\
GAMA239376 & 	normal	 & 	normal	 & 0.039 & 0.004 & 0.04 & 0.009 & 9.61 & 1.46 & 0.39 & 0.83 \\
GAMA250192 & 	normal	 & 	normal	 & 0.051 & 0.006 & 0.085 & 0.02 & 10.71 & 2.39 & -1.06 & 0.78 \\
GAMA250277 & 	normal	 & 	normal	 & 0.044 & 0.083 & 0.231 & 0.127 & 9.99 & 1.35 & 1.23 & 0.53 \\
GAMA272831 & 	normal	 & 	normal	 & 0.064 & 0.012 & 0.208 & 0.033 & 11.12 & 2.36 & 0.89 & 0.94 \\
GAMA278643 & 	normal	 & 	normal	 & 0.051 & 0.004 & 0.022 & 0.006 & 9.16 & 1.48 & -0.52 & 1.01 \\
GAMA278684 & 	normal	 & 	normal	 & 0.013 & 0.018 & 0.056 & 0.056 & 8.07 & 1.27 & -0.83 & 0.63 \\
GAMA278702 & 	normal	 & 	perturbed	 & 0.088 & 0.048 & 0.822 & 0.307 & 8.27 & 1.16 & -0.070 & 0.81 \\
GAMA278741 & 	normal	 & 	normal	 & 0.038 & 0.003 & 0.062 & 0.006 & 9.25 & 1.35 & 0.15 & 0.91 \\
GAMA278760 & 	perturbed	 & 	normal	 & 0.051 & 0.005 & 0.077 & 0.010 & 9.85 & 1.29 & 1.51 & 0.62 \\
GAMA278787 & 	normal	 & 	normal	 & 0.061 & 0.042 & 0.127 & 0.046 & 9.16 & 1.39 & -0.46 & 0.98 \\
GAMA278846 & 	normal	 & 	normal	 & 0.055 & 0.029 & 0.195 & 0.118 & 9.52 & 1.57 & 0.33 & 0.74 \\
GAMA278909 & 	normal	 & 	normal	 & 0.039 & 0.006 & 0.097 & 0.017 & 9.24 & 1.43 & -0.19 & 1.11 \\
GAMA278995 & 	perturbed	 & 	perturbed	 & 0.211 & 0.075 & 0.445 & 0.054 & 7.84 & 1.14 & 0.98 & 0.65 \\
GAMA279818 & 	normal	 & 	normal	 & 0.04 & 0.010 & 0.082 & 0.022 & 9.4 & 1.28 & 0.47 & 1.01 \\
GAMA279886 & 	normal	 & 	normal	 & 0.058 & 0.005 & 0.074 & 0.022 & 8.85 & 1.97 & -2.97 & {-999.0} \\
GAMA279891 & 	normal	 & 	normal	 & 0.053 & 0.031 & 0.164 & 0.132 & 8.02 & 0.97 & -0.79 & 1.11 \\
GAMA279917 & 	normal	 & 	normal	 & 0.054 & 0.004 & 0.074 & 0.008 & 9.19 & 1.22 & 0.57 & 0.77 \\
GAMA279943 & 	perturbed	 & 	perturbed	 & 0.239 & 0.063 & 1.525 & 1.222 & 8.43 & 1.24 & -0.48 & 0.80 \\
GAMA288992 & 	normal	 & 	normal	 & 0.028 & 0.006 & 0.086 & 0.008 & 10.49 & 2.5 & -1.4 & 0.28 \\
GAMA289116 & 	normal	 & 	normal	 & 0.031 & 0.005 & 0.096 & 0.018 & 8.74 & 1.38 & -0.58 & 0.93 \\
GAMA289200 & 	normal	 & 	normal	 & 0.052 & 0.004 & 0.085 & 0.026 & 8.34 & 1.44 & -0.97 & 0.76 \\
GAMA296685 & 	perturbed	 & 	perturbed	 & 0.121 & 0.032 & 0.371 & 0.087 & 9.33 & 1.44 & 0.06 & 0.78 \\
GAMA296798 & 	perturbed	 & 	perturbed	 & 0.585 & 0.137 & 1.675 & 0.787 & 8.36 & 1.21 & -0.60 & 0.54 \\
GAMA296829 & 	normal	 & 	normal	 & 0.052 & 0.016 & 0.101 & 0.015 & 10.19 & 1.83 & 0.19 & 0.78 \\
GAMA296847 & 	normal	 & 	normal	 & 0.025 & 0.003 & 0.042 & 0.024 & 9.16 & 1.4 & -0.86 & 0.95 \\
GAMA296848 & 	normal	 & 	normal	 & 0.042 & 0.011 & 0.173 & 0.077 & 8.35 & 1.12 & -0.8 & 0.86 \\
GAMA296934 & 	normal	 & 	normal	 & 0.04 & 0.04 & 0.141 & 0.058 & 10.21 & 1.61 & 1.65 & 0.71 \\
GAMA297557 & 	perturbed	 & 	perturbed	 & 0.084 & 0.041 & 0.319 & 0.041 & 8.42 & 1.4 & -0.59 & 0.84 \\
GAMA297667 & 	normal	 & 	normal	 & 0.032 & 0.003 & 0.041 & 0.007 & 10.25 & 1.77 & 0.94 & 0.93 \\
GAMA297705 & 	normal	 & 	normal	 & 0.037 & 0.003 & 0.059 & 0.035 & 8.69 & 1.46 & -0.95 & 0.95 \\
GAMA300350 & 	normal	 & 	perturbed	 & 0.089 & 0.089 & 0.756 & 0.45 & 8.32 & 1.32 & -1.33 & 1.12 \\
GAMA300372 & 	normal	 & 	perturbed	 & 0.071 & 0.058 & 0.212 & 0.107 & 9.15 & 1.39 & 0.11 & 0.92 \\
GAMA300477 & 	normal	 & 	normal	 & 0.054 & 0.023 & 0.075 & 0.036 & 9.23 & 1.31 & -0.08 & 1.18 \\
GAMA301098 & 	perturbed	 & 	perturbed	 & 0.295 & 0.208 & 0.415 & 0.41 & 7.67 & 0.97 & -1.60 & 0.81 \\
GAMA301799 & 	normal	 & 	normal	 & 0.043 & 0.025 & 0.094 & 0.029 & 9.87 & 1.79 & 0.34 & 0.85 \\
GAMA302846 & 	normal	 & 	normal	 & 0.033 & 0.0032 & 0.054 & 0.005 & 9.85 & 1.69 & -0.21 & 1.12 \\
GAMA302994 & 	normal	 & 	normal	 & 0.042 & 0.006 & 0.093 & 0.024 & 8.79 & 1.44 & -0.3 & 0.74 \\
GAMA318936 & 	normal	 & 	normal	 & 0.032 & 0.008 & 0.079 & 0.015 & 8.76 & 1.02 & 0.020 & 0.98 \\
GAMA319018 & 	perturbed	 & 	perturbed	 & 0.088 & 0.008 & 0.178 & 0.024 & 9.98 & 1.48 & 0.84 & 0.76 \\
GAMA319049 & 	normal	 & 	normal	 & 0.047 & 0.008 & 0.08 & 0.015 & 10.06 & 1.91 & 0.29 & 0.88 \\
GAMA319057 & 	normal	 & 	normal	 & 0.037 & 0.009 & 0.11 & 0.016 & 10.15 & 1.51 & 1.34 & 0.89 \\
GAMA319059 & 	normal	 & 	normal	 & 0.062 & 0.008 & 0.093 & 0.023 & 10.05 & 2.17 & -1.13 & 0.41 \\
GAMA319067 & 	normal	 & 	normal	 & 0.035 & 0.006 & 0.097 & 0.018 & 10.95 & 2.21 & -0.50 & 1.32 \\
GAMA319070 & 	normal	 & 	normal	 & 0.033 & 0.005 & 0.098 & 0.018 & 10.65 & 2.37 & -0.54 & 1.93 \\
GAMA319139 & 	normal	 & 	normal	 & 0.035 & 0.006 & 0.065 & 0.007 & 8.56 & 1.08 & -0.28 & 1.03 \\
GAMA319150 & 	perturbed	 & 	perturbed	 & 0.216 & 0.071 & 1.834 & 0.48 & 8.5 & 1.21 & -0.41 & 0.84 \\
GAMA319157 & 	normal	 & 	normal	 & 0.035 & 0.006 & 0.074 & 0.015 & 10.21 & 2.27 & -1.24 & 2.20 \\
GAMA319197 & 	normal	 & 	normal	 & 0.056 & 0.008 & 0.078 & 0.013 & 10.4 & 2.38 & 1.00 & 0.89 \\
\newpage
\hline
GAMA Survey & Vis. & Kin. & $\overline{v_{asym}}$ & $\overline{v_{asym}}$ & $\overline{\sigma_{asym}}$ & $\overline{\sigma_{asym}}$ &  log($M_{*}$) & Colour& log(SFR) & $\frac{r_{50,H\alpha}}{r_{50,cont}}$ \\
ID & Class. & Class. & & err. & & err. & & ($u-r$)& & \\
\hline

GAMA319243 & 	perturbed	 & 	perturbed	 & 0.208 & 0.12 & 0.549 & 0.286 & 8.53 & 1.43 & -0.34 & 0.88 \\
GAMA319272 & 	normal	 & 	normal	 & 0.032 & 0.007 & 0.336 & 0.229 & 8.51 & 1.17 & -0.25 & 1.04 \\
GAMA319381 & 	normal	 & 	normal	 & 0.011 & 0.003 & 0.091 & 0.011 & 10.29 & 1.57 & 0.94 & 0.90 \\
GAMA319385 & 	normal	 & 	perturbed	 & 0.077 & 0.063 & 0.23 & 0.103 & 10.14 & 2.11 & 0.86 & 0.48 \\
GAMA319400 & 	normal	 & 	normal	 & 0.02 & 0.0010 & 0.121 & 0.028 & 10.46 & 1.88 & 0.53 & 1.07 \\
GAMA319453 & 	normal	 & 	normal	 & 0.019 & 0.0040 & 0.14 & 0.02 & 10.44 & 2.04 & 0.52 & 0.94 \\
GAMA320068 & 	normal	 & 	normal	 & 0.044 & 0.037 & 0.052 & 0.027 & 9.2 & 1.34 & 0.17 & 0.93 \\
GAMA322910 & 	normal	 & 	normal	 & 0.058 & 0.019 & 0.201 & 0.062 & 9.74 & 1.47 & 0.75 & 0.93 \\
GAMA323504 & 	normal	 & 	normal	 & 0.04 & 0.008 & 0.067 & 0.025 & 10.99 & 2.06 & 1.10 & 0.59 \\
GAMA323505 & 	normal	 & 	normal	 & 0.023 & 0.0060 & 0.114 & 0.017 & 9.93 & 1.71 & 0.18 & 0.87 \\
GAMA323577 & 	normal	 & 	normal	 & 0.048 & 0.029 & 0.266 & 0.045 & 9.4 & 1.49 & -0.57 & 1.10 \\
GAMA324323 & 	normal	 & 	normal	 & 0.032 & 0.031 & 0.112 & 0.039 & 9.72 & 1.51 & 0.74 & 0.71 \\
GAMA324351 & 	normal	 & 	normal	 & 0.038 & 0.005 & 0.071 & 0.014 & 10.45 & 2.38 & -0.24 & 0.53 \\
GAMA325390 & 	normal	 & 	normal	 & 0.032 & 0.0040 & 0.078 & 0.02 & 8.33 & 1.57 & -1.74 & 0.72 \\
GAMA345682 & 	normal	 & 	normal	 & 0.061 & 0.0060 & 0.052 & 0.013 & 9.28 & 1.42 & -0.09 & 0.84 \\
GAMA345820 & 	normal	 & 	normal	 & 0.06 & 0.007 & 0.089 & 0.011 & 10.17 & 2.06 & -0.29 & 0.38 \\
GAMA346046 & 	normal	 & 	normal	 & 0.058 & 0.0070 & 0.088 & 0.024 & 10.31 & 2.37 & -9.0 & 1.24 \\
GAMA346718 & 	perturbed	 & 	perturbed	 & 0.084 & 0.050 & 0.176 & 0.088 & 9.20 & 0.82 & 0.92 & 1.07 \\
GAMA346793 & 	normal	 & 	normal	 & 0.024 & 0.003 & 0.053 & 0.007 & 10.31 & 1.56 & 1.13 & 0.93 \\
GAMA346839 & 	normal	 & 	normal	 & 0.061 & 0.0050 & 0.095 & 0.020 & 10.39 & 2.29 & -0.82 & 0.47 \\
GAMA346890 & 	normal	 & 	normal	 & 0.025 & 0.0050 & 0.097 & 0.015 & 10.45 & 2.25 & -1.41 & 1.04 \\
GAMA346892 & 	normal	 & 	normal	 & 0.025 & 0.0040 & 0.068 & 0.014 & 10.29 & 1.55 & 1.07 & 0.78 \\
GAMA346894 & 	normal	 & 	normal	 & 0.041 & 0.0060 & 0.071 & 0.02 & 10.49 & 2.54 & -0.79 & 0.56 \\
GAMA348115 & 	normal	 & 	normal	 & 0.054 & 0.0070 & 0.091 & 0.017 & 11.19 & 2.45 & 0.45 & {-999.0} \\
GAMA348116 & 	normal	 & 	normal	 & 0.019 & 0.002 & 0.108 & 0.030 & 10.63 & 2.30 & 0.63 & 0.82 \\
GAMA373173 & 	normal	 & 	normal	 & 0.024 & 0.006 & 0.082 & 0.016 & 11.06 & 2.6 & -0.020 & 0.18 \\
GAMA373202 & 	normal	 & 	normal	 & 0.057 & 0.005 & 0.139 & 0.015 & 9.27 & 1.81 & -0.45 & 1.04 \\
GAMA373284 & 	perturbed	 & 	perturbed	 & 0.071 & 0.013 & 0.132 & 0.016 & 9.89 & 1.95 & 0.62 & 0.89 \\
GAMA375402 & 	perturbed	 & 	perturbed	 & 0.159 & 0.009 & 0.2 & 0.115 & 8.44 & 1.27 & -0.43 & 0.91 \\
GAMA375531 & 	perturbed	 & 	perturbed	 & 0.082 & 0.009 & 0.13 & 0.013 & 9.19 & 0.77 & -9.0 & 0.74 \\
GAMA376001 & 	normal	 & 	normal	 & 0.016 & 0.03 & 0.137 & 0.118 & 10.31 & 1.97 & -0.070 & 1.04 \\
GAMA376121 & 	normal	 & 	normal	 & 0.031 & 0.006 & 0.051 & 0.005 & 11.08 & 2.18 & 0.23 & 1.25 \\
GAMA376185 & 	normal	 & 	normal	 & 0.027 & 0.017 & 0.077 & 0.095 & 9.02 & 1.18 & -0.23 & 1.22 \\
GAMA377962 & 	normal	 & 	normal	 & 0.019 & 0.052 & 0.047 & 0.076 & 9.05 & 1.33 & 0.09 & 0.76 \\
GAMA381159 & 	normal	 & 	normal	 & 0.025 & 0.009 & 0.065 & 0.012 & 9.99 & 2.03 & 0.48 & 0.86 \\
GAMA381207 & 	normal	 & 	normal	 & 0.036 & 0.005 & 0.073 & 0.014 & 10.56 & 2.43 & -9.0 & 1.46 \\
GAMA381215 & 	normal	 & 	normal	 & 0.033 & 0.008 & 0.141 & 0.027 & 10.4 & 2.46 & 1.01 & 0.65 \\
GAMA381225 & 	normal	 & 	normal	 & 0.045 & 0.004 & 0.073 & 0.012 & 10.18 & 1.65 & 0.95 & 0.92 \\
GAMA381229 & 	normal	 & 	normal	 & 0.053 & 0.006 & 0.096 & 0.014 & 10.47 & 2.33 & -0.89 & 2.42 \\
GAMA382152 & 	perturbed	 & 	perturbed	 & 0.088 & 0.006 & 0.138 & 0.051 & 10.19 & 1.67 & 0.49 & 1.05 \\
GAMA382158 & 	normal	 & 	normal	 & 0.056 & 0.004 & 0.085 & 0.011 & 10.49 & 2.37 & -0.96 & 0.99 \\
GAMA383259 & 	perturbed	 & 	perturbed	 & 0.071 & 0.008 & 0.375 & 0.062 & 10.77 & 1.73 & 1.81 & 0.72 \\
GAMA383283 & 	normal	 & 	normal	 & 0.016 & 0.001 & 0.067 & 0.012 & 9.2 & 1.70 & -0.44 & 1.01 \\
GAMA383318 & 	normal	 & 	normal	 & 0.041 & 0.004 & 0.206 & 0.022 & 9.88 & 1.38 & 0.97 & 0.78 \\
GAMA386268 & 	normal	 & 	normal	 & 0.053 & 0.005 & 0.092 & 0.013 & 11.0 & 2.48 & -0.24 & 0.33 \\
GAMA388451 & 	normal	 & 	normal	 & 0.051 & 0.004 & 0.072 & 0.010 & 8.46 & 1.26 & -0.74 & 0.93 \\
GAMA388476 & 	normal	 & 	normal	 & 0.022 & 0.019 & 0.156 & 0.064 & 10.47 & 2.18 & -0.38 & 1.05 \\
GAMA388552 & 	normal	 & 	normal	 & 0.042 & 0.007 & 0.084 & 0.009 & 11.01 & 2.46 & 0.03 & 0.28 \\
GAMA388603 & 	normal	 & 	normal	 & 0.029 & 0.002 & 0.043 & 0.004 & 9.82 & 1.50 & 0.39 & 0.98 \\
GAMA417392 & 	normal	 & 	normal	 & 0.04 & 0.006 & 0.081 & 0.006 & 8.84 & 1.16 & -0.19 & 1.08 \\
GAMA417424 & 	normal	 & 	normal	 & 0.034 & 0.003 & 0.039 & 0.008 & 9.33 & 1.41 & -0.020 & 0.98 \\
GAMA417678 & 	perturbed	 & 	perturbed	 & 0.065 & 0.032 & 0.097 & 0.03 & 10.05 & 1.91 & 1.71 & 0.72 \\
GAMA419632 & 	normal	 & 	normal	 & 0.051 & 0.046 & 0.038 & 0.065 & 8.91 & 1.62 & -0.98 & 1.01 \\
GAMA422320 & 	normal	 & 	normal	 & 0.027 & 0.006 & 0.058 & 0.009 & 9.45 & 1.22 & 0.11 & 0.96 \\
GAMA422355 & 	normal	 & 	normal	 & 0.05 & 0.007 & 0.074 & 0.013 & 9.28 & 1.34 & -0.56 & 1.06 \\
GAMA422359 & 	normal	 & 	normal	 & 0.045 & 0.025 & 0.128 & 0.011 & 10.1 & 1.75 & 1.04 & 0.69 \\
GAMA422366 & 	normal	 & 	normal	 & 0.035 & 0.003 & 0.038 & 0.007 & 9.69 & 1.39 & 0.31 & 1.00 \\
GAMA422683 & 	perturbed	 & 	perturbed	 & 0.121 & 0.073 & 1.239 & 0.457 & 8.47 & 1.30 & -0.13 & 0.86 \\
GAMA422721 & 	normal	 & 	normal	 & 0.046 & 0.035 & 0.253 & 0.115 & 8.54 & 1.30 & -1.04 & 0.70 \\
\newpage
\hline
GAMA Survey & Vis. & Kin. & $\overline{v_{asym}}$ & $\overline{v_{asym}}$ & $\overline{\sigma_{asym}}$ & $\overline{\sigma_{asym}}$ &  log($M_{*}$) & Colour& log(SFR) & $\frac{r_{50,H\alpha}}{r_{50,cont}}$ \\
ID & Class. & Class. & & err. & & err. & & ($u-r$)& & \\
\hline

GAMA422888 & 	normal	 & 	normal	 & 0.052 & 0.04 & 0.061 & 0.032 & 8.36 & 1.36 & -0.92 & {-999.0} \\
GAMA422907 & 	perturbed	 & 	perturbed	 & 0.121 & 0.007 & 0.27 & 0.054 & 9.14 & 1.23 & 0.62 & 1.05 \\
GAMA422921 & 	normal	 & 	normal	 & 0.055 & 0.035 & 0.191 & 0.057 & 8.14 & 1.04 & -1.12 & 0.99 \\
GAMA422933 & 	normal	 & 	normal	 & 0.026 & 0.007 & 0.071 & 0.005 & 10.12 & 1.98 & 0.95 & 0.75 \\
GAMA485504 & 	normal	 & 	normal	 & 0.021 & 0.014 & 0.076 & 0.028 & 10.24 & 1.62 & 0.53 & 1.00 \\
GAMA485690 & 	perturbed	 & 	perturbed	 & 0.199 & 0.010 & 0.306 & 0.047 & 10.14 & 2.06 & 1.18 & 0.63 \\
GAMA485885 & 	normal	 & 	normal	 & 0.036 & 0.004 & 0.047 & 0.016 & 10.28 & 1.71 & 1.03 & 0.8 \\
GAMA485924 & 	normal	 & 	normal	 & 0.022 & 0.002 & 0.064 & 0.015 & 10.44 & 1.72 & 0.41 & 1.41 \\
GAMA486872 & 	normal	 & 	normal	 & 0.026 & 0.004 & 0.054 & 0.015 & 10.46 & 1.76 & 0.62 & 1.13 \\
GAMA486957 & 	normal	 & 	normal	 & 0.018 & 0.005 & 0.058 & 0.021 & 10.86 & 1.69 & 1.43 & 1.09 \\
GAMA487010 & 	normal	 & 	normal	 & 0.055 & 0.03 & 0.283 & 0.043 & 8.98 & 1.30 & {-999.0} & 0.97 \\
GAMA487027 & 	perturbed	 & 	perturbed	 & 0.073 & 0.003 & 0.15 & 0.016 & 9.98 & 1.48 & 1.66 & 0.74 \\
GAMA492384 & 	normal	 & 	normal	 & 0.055 & 0.039 & 0.197 & 0.015 & 10.48 & 1.87 & 1.41 & 0.56 \\
GAMA493621 & 	normal	 & 	normal	 & 0.036 & 0.006 & 0.060 & 0.009 & 8.98 & 1.17 & -0.27 & 1.27 \\
GAMA493811 & 	normal	 & 	normal	 & 0.064 & 0.053 & 0.089 & 0.211 & 8.71 & 1.24 & {-999.0} & 0.72 \\
GAMA493825 & 	perturbed	 & 	perturbed	 & 0.159 & 0.081 & 1.014 & 0.367 & 8.27 & 1.20 & -0.75 & 1.00 \\
GAMA496966 & 	normal	 & 	normal	 & 0.043 & 0.006 & 0.062 & 0.005 & 10.37 & 1.91 & 0.84 & 0.85 \\
GAMA504713 & 	normal	 & 	normal	 & 0.033 & 0.004 & 0.046 & 0.013 & 10.46 & 1.43 & 1.22 & 0.98 \\
GAMA504882 & 	normal	 & 	normal	 & 0.027 & 0.031 & 0.073 & 0.059 & 10.15 & 1.84 & -0.05 & 1.05 \\
GAMA504922 & 	normal	 & 	normal	 & 0.038 & 0.010 & 0.066 & 0.044 & 10.03 & 1.96 & 0.17 & 0.75 \\
GAMA505979 & 	normal	 & 	normal	 & 0.024 & 0.003 & 0.064 & 0.009 & 9.74 & 1.37 & 0.56 & 0.95 \\
GAMA508414 & 	perturbed	 & 	perturbed	 & 0.093 & 0.016 & 0.139 & 0.015 & 9.63 & 1.45 & 0.84 & 0.69 \\
GAMA508421 & 	normal	 & 	normal	 & 0.035 & 0.005 & 0.059 & 0.017 & 10.42 & 1.93 & 1.11 & 0.86 \\
GAMA508480 & 	normal	 & 	normal	 & 0.011 & 0.048 & 0.046 & 0.158 & 9.67 & 1.78 & 0.27 & 0.46 \\
GAMA508481 & 	normal	 & 	normal	 & 0.033 & 0.017 & 0.078 & 0.027 & 10.0 & 1.69 & {-999.0} & 1.05 \\
GAMA508682 & 	perturbed	 & 	perturbed	 & 0.108 & 0.081 & 0.736 & 0.703 & 7.87 & 0.99 & 0.05 & 0.86 \\
GAMA509557 & 	perturbed	 & 	perturbed	 & 0.113 & 0.095 & 0.685 & 0.387 & 8.62 & 1.27 & 0.45 & 0.74 \\
GAMA509576 & 	perturbed	 & 	perturbed	 & 0.157 & 0.011 & 0.425 & 0.189 & 7.93 & 0.83 & -0.72 & 1.3 \\
GAMA509670 & 	perturbed	 & 	normal	 & 0.056 & 0.008 & 0.171 & 0.059 & 8.75 & 1.06 & -0.27 & 1.28 \\
GAMA509727 & 	perturbed	 & 	perturbed	 & 0.081 & 0.055 & 0.28 & 0.138 & 8.95 & 1.31 & -0.12 & 1.01 \\
GAMA511789 & 	normal	 & 	normal	 & 0.027 & 0.002 & 0.097 & 0.011 & 8.81 & 0.99 & 0.44 & 0.87 \\
GAMA511863 & 	normal	 & 	normal	 & 0.044 & 0.015 & 0.093 & 0.023 & 9.37 & 2.03 & -0.26 & 0.48 \\
GAMA511867 & 	normal	 & 	normal	 & 0.019 & 0.002 & 0.082 & 0.013 & 10.68 & 1.63 & 1.33 & {-999.0} \\
GAMA514260 & 	perturbed	 & 	perturbed	 & 0.076 & 0.006 & 0.238 & 0.015 & 8.9 & 1.56 & 0.26 & 0.73 \\
GAMA517070 & 	normal	 & 	normal	 & 0.023 & 0.003 & 0.054 & 0.022 & 10.17 & 1.49 & 0.54 & 1.05 \\
GAMA517164 & 	normal	 & 	normal	 & 0.028 & 0.005 & 0.042 & 0.023 & 10.45 & 1.92 & -0.32 & 1.39 \\
GAMA517302 & 	perturbed	 & 	perturbed	 & 0.125 & 0.072 & 0.234 & 0.18 & 10.26 & 2.05 & 1.08 & 0.4 \\
GAMA517594 & 	normal	 & 	normal	 & 0.049 & 0.007 & 0.093 & 0.018 & 8.95 & 1.32 & 0.38 & 0.94 \\
GAMA522127 & 	perturbed	 & 	perturbed	 & 0.22 & 0.020 & 1.864 & 0.501 & 8.36 & 1.12 & -0.80 & 1.00 \\
GAMA522166 & 	normal	 & 	normal	 & 0.05 & 0.009 & 0.175 & 0.206 & 8.83 & 1.09 & 0.29 & 0.91 \\
GAMA534654 & 	normal	 & 	normal	 & 0.033 & 0.016 & 0.075 & 0.009 & 10.32 & 1.66 & 0.80 & 0.85 \\
GAMA534655 & 	perturbed	 & 	perturbed	 & 0.237 & 0.106 & 1.161 & 0.37 & 11.13 & 2.58 & 0.010 & 0.46 \\
GAMA534710 & 	normal	 & 	normal	 & 0.035 & 0.017 & 0.132 & 0.011 & 9.14 & 1.54 & 0.42 & 0.77 \\
GAMA534753 & 	normal	 & 	normal	 & 0.035 & 0.004 & 0.117 & 0.015 & 9.56 & 1.59 & 1.76 & 0.5 \\
GAMA534759 & 	perturbed	 & 	perturbed	 & 0.145 & 0.061 & 1.025 & 0.993 & 9.49 & 1.60 & -0.49 & 0.93 \\
GAMA536625 & 	normal	 & 	normal	 & 0.062 & 0.008 & 0.091 & 0.053 & 10.27 & 1.87 & 0.41 & 0.94 \\
GAMA536626 & 	normal	 & 	normal	 & 0.056 & 0.005 & 0.079 & 0.007 & 8.91 & 1.04 & 0.32 & 0.94 \\
GAMA537163 & 	perturbed	 & 	perturbed	 & 0.086 & 0.018 & 0.189 & 0.145 & 8.11 & 1.04 & -0.37 & 1.02 \\
GAMA537171 & 	normal	 & 	normal	 & 0.036 & 0.024 & 0.12 & 0.028 & 9.36 & 1.63 & -0.17 & 0.98 \\
GAMA537187 & 	normal	 & 	normal	 & 0.038 & 0.008 & 0.045 & 0.005 & 9.24 & 1.32 & 0.14 & 0.85 \\
GAMA537367 & 	normal	 & 	normal	 & 0.058 & 0.147 & 0.703 & 0.36 & 9.14 & 1.41 & 0.44 & 0.81 \\
GAMA537417 & 	normal	 & 	normal	 & 0.039 & 0.005 & 0.125 & 0.031 & 8.91 & 1.17 & -0.59 & 1.13 \\
GAMA543752 & 	perturbed	 & 	perturbed	 & 0.094 & 0.081 & 0.349 & 0.04 & 8.89 & 1.12 & -0.040 & 1.13 \\
GAMA543763 & 	normal	 & 	normal	 & 0.028 & 0.003 & 0.114 & 0.012 & 8.45 & 1.11 & {-999.0} & 1.08 \\
GAMA543769 & 	perturbed	 & 	normal	 & 0.06 & 0.005 & 0.145 & 0.033 & 8.37 & 1.05 & {-999.0} & 0.89 \\
GAMA543812 & 	normal	 & 	normal	 & 0.037 & 0.005 & 0.057 & 0.018 & 9.19 & 1.11 & -0.011 & 1.13 \\
GAMA543859 & 	normal	 & 	normal	 & 0.015 & 0.002 & 0.045 & 0.005 & 10.74 & 2.14 & 1.09 & 0.95 \\
GAMA551192 & 	perturbed	 & 	perturbed	 & 0.232 & 0.075 & 1.029 & 0.751 & 8.76 & 1.38 & -0.30 & 0.75 \\
GAMA551202 & 	normal	 & 	normal	 & 0.032 & 0.002 & 0.083 & 0.017 & 9.91 & 1.68 & 1.36 & 0.62 \\
\newpage
\hline
GAMA Survey & Vis. & Kin. & $\overline{v_{asym}}$ & $\overline{v_{asym}}$ & $\overline{\sigma_{asym}}$ & $\overline{\sigma_{asym}}$ &  log($M_{*}$) & Colour& log(SFR) & $\frac{r_{50,H\alpha}}{r_{50,cont}}$ \\
ID & Class. & Class. & & err. & & err. & & ($u-r$)& & \\
\hline

GAMA558861 & 	normal	 & 	normal	 & 0.035 & 0.006 & 0.047 & 0.006 & 8.52 & 1.44 & -0.85 & 1.11 \\
GAMA558887 & 	normal	 & 	normal	 & 0.037 & 0.025 & 0.069 & 0.045 & 8.80 & 1.23 & -0.04 & 0.81 \\
GAMA561856 & 	normal	 & 	normal	 & 0.044 & 0.003 & 0.073 & 0.023 & 10.06 & 1.63 & 0.16 & 1.20 \\
GAMA567545 & 	normal	 & 	normal	 & 0.038 & 0.004 & 0.098 & 0.032 & 10.93 & 2.08 & 0.91 & 1.14 \\
GAMA567624 & 	normal	 & 	normal	 & 0.024 & 0.003 & 0.030 & 0.008 & 9.31 & 1.33 & 0.11 & 0.92 \\
GAMA567676 & 	perturbed	 & 	perturbed	 & 0.159 & 0.02 & 0.503 & 0.077 & 8.52 & 1.23 & -0.22 & 1.37 \\
GAMA567678 & 	normal	 & 	perturbed	 & 0.079 & 0.017 & 0.075 & 0.016 & 9.47 & 2.17 & -0.25 & 0.77 \\
GAMA567736 & 	perturbed	 & 	perturbed	 & 0.223 & 0.015 & 0.45 & 0.102 & 8.47 & 0.64 & 0.32 & 0.99 \\
GAMA567750 & 	perturbed	 & 	perturbed	 & 0.391 & 0.092 & 6.925 & 4.161 & 8.22 & 1.18 & -0.52 & {-999.0} \\
GAMA567876 & 	perturbed	 & 	perturbed	 & 0.107 & 0.074 & 0.76 & 0.755 & 8.27 & 1.33 & -1.05 & {-999.0} \\
GAMA567983 & 	normal	 & 	normal	 & 0.042 & 0.008 & 0.169 & 0.033 & 8.43 & 1.13 & {-999.0} & 1.00 \\
GAMA570206 & 	perturbed	 & 	perturbed	 & 0.131 & 0.023 & 0.518 & 0.061 & 10.58 & 2.25 & -0.012 & 0.56 \\
GAMA570227 & 	normal	 & 	normal	 & 0.043 & 0.004 & 0.107 & 0.03 & 10.7 & 2.31 & 0.59 & 0.64 \\
GAMA574200 & 	normal	 & 	normal	 & 0.035 & 0.005 & 0.086 & 0.021 & 9.35 & 1.25 & 0.73 & 0.97 \\
GAMA583443 & 	normal	 & 	normal	 & 0.026 & 0.063 & 0.057 & 0.017 & 8.86 & 1.25 & -0.27 & 1.04 \\
GAMA592401 & 	normal	 & 	normal	 & 0.049 & 0.018 & 0.023 & 0.18 & 8.26 & 1.11 & {-999.0} & {-999.0} \\
GAMA592421 & 	normal	 & 	normal	 & 0.033 & 0.005 & 0.113 & 0.019 & 10.9 & 1.91 & 1.36 & 1.02 \\
GAMA592466 & 	normal	 & 	normal	 & 0.038 & 0.009 & 0.21 & 0.018 & 8.31 & 0.95 & -0.38 & 0.89 \\
GAMA592542 & 	perturbed	 & 	perturbed	 & 0.242 & 0.073 & 0.875 & 0.353 & 8.35 & 1.19 & {-999.0} & 1.00 \\
GAMA592621 & 	perturbed	 & 	normal	 & 0.021 & 0.002 & 0.214 & 0.023 & 10.23 & 1.28 & 1.72 & 0.97 \\
GAMA592835 & 	normal	 & 	normal	 & 0.059 & 0.004 & 0.069 & 0.003 & 10.36 & 1.68 & 1.21 & 0.98 \\
GAMA592863 & 	normal	 & 	normal	 & 0.043 & 0.004 & 0.108 & 0.03 & 9.55 & 1.46 & 0.21 & 0.87 \\
GAMA593645 & 	normal	 & 	normal	 & 0.056 & 0.012 & 0.258 & 0.024 & 8.53 & 1.20 & -0.23 & 0.79 \\
GAMA593680 & 	normal	 & 	normal	 & 0.026 & 0.013 & 0.053 & 0.011 & 10.42 & 2.14 & 0.98 & 0.87 \\
GAMA594906 & 	normal	 & 	normal	 & 0.04 & 0.011 & 0.124 & 0.014 & 9.7 & 1.34 & 1.05 & 0.87 \\
GAMA594986 & 	normal	 & 	normal	 & 0.023 & 0.019 & 0.088 & 0.007 & 10.12 & 2.18 & 0.55 & 0.77 \\
GAMA594990 & 	perturbed	 & 	normal	 & 0.039 & 0.051 & 0.17 & 0.048 & 10.37 & 2.16 & 0.58 & 0.46 \\
GAMA595027 & 	normal	 & 	normal	 & 0.027 & 0.004 & 0.042 & 0.005 & 9.86 & 1.63 & 1.10 & 0.85 \\
GAMA595060 & 	normal	 & 	normal	 & 0.025 & 0.002 & 0.068 & 0.012 & 10.38 & 1.72 & -0.31 & 1.21 \\
GAMA599582 & 	perturbed	 & 	perturbed	 & 0.079 & 0.006 & 0.185 & 0.076 & 10.7 & 1.91 & 0.86 & 1.03 \\
GAMA599761 & 	normal	 & 	normal	 & 0.022 & 0.005 & 0.075 & 0.013 & 10.93 & 2.10 & -0.1 & {-999.0} \\
GAMA599839 & 	normal	 & 	normal	 & 0.03 & 0.007 & 0.077 & 0.072 & 9.68 & 1.88 & 0.38 & 0.83 \\
GAMA599873 & 	normal	 & 	normal	 & 0.035 & 0.022 & 0.173 & 0.11 & 8.89 & 1.29 & 0.020 & 0.88 \\
GAMA599877 & 	normal	 & 	normal	 & 0.056 & 0.052 & 0.191 & 0.02 & 10.3 & 2.23 & 0.050 & 0.69 \\
GAMA600014 & 	perturbed	 & 	perturbed	 & 0.095 & 0.006 & 0.223 & 0.029 & 8.98 & 1.11 & 0.24 & 0.96 \\
GAMA600026 & 	normal	 & 	normal	 & 0.048 & 0.006 & 0.03 & 0.029 & 10.17 & 1.46 & 1.2 & 1.00 \\
GAMA600030 & 	normal	 & 	normal	 & 0.046 & 0.008 & 0.153 & 0.027 & 10.25 & 2.12 & 1.41 & 1.30 \\
GAMA617945 & 	normal	 & 	normal	 & 0.036 & 0.03 & 0.165 & 0.105 & 8.38 & 1.09 & {-999.0} & 1.00 \\
GAMA618071 & 	normal	 & 	normal	 & 0.037 & 0.005 & 0.108 & 0.014 & 8.9 & 1.17 & 0.07 & 1.16 \\
GAMA618108 & 	normal	 & 	normal	 & 0.043 & 0.007 & 0.185 & 0.015 & 10.45 & 2.47 & -0.37 & 0.70 \\
GAMA618116 & 	normal	 & 	normal	 & 0.022 & 0.003 & 0.081 & 0.006 & 10.25 & 1.38 & 1.03 & 1.00 \\
GAMA618152 & 	normal	 & 	normal	 & 0.051 & 0.029 & 0.328 & 0.16 & 10.03 & 1.77 & 0.89 & 0.78 \\
GAMA618220 & 	normal	 & 	normal	 & 0.046 & 0.003 & 0.067 & 0.037 & 10.62 & 1.98 & 0.10 & 1.15 \\
GAMA618906 & 	normal	 & 	normal	 & 0.053 & 0.006 & 0.134 & 0.023 & 10.61 & 2.17 & 0.83 & 0.82 \\
GAMA618935 & 	normal	 & 	normal	 & 0.023 & 0.006 & 0.077 & 0.013 & 9.8 & 1.48 & 0.95 & 0.73 \\
GAMA618952 & 	normal	 & 	normal	 & 0.04 & 0.0050 & 0.205 & 0.034 & 10.79 & 2.38 & 0.030 & 0.97 \\
GAMA618992 & 	perturbed	 & 	perturbed	 & 0.108 & 0.018 & 0.24 & 0.017 & 10.76 & 2.06 & 1.60 & 0.56 \\
GAMA618993 & 	normal	 & 	normal	 & 0.21 & 0.07 & 0.25 & 0.063 & 10.82 & 1.98 & 2.39 & 0.53 \\
GAMA618993 & 	normal	 & 	normal	 & 0.063 & 0.013 & 0.265 & 0.074 & 10.82 & 1.98 & 2.39 & 0.53 \\
GAMA619046 & 	normal	 & 	normal	 & 0.037 & 0.0050 & 0.037 & 0.006 & 9.12 & 1.42 & -0.19 & {-999.0} \\
GAMA619095 & 	normal	 & 	normal	 & 0.024 & 0.0030 & 0.078 & 0.011 & 10.46 & 1.75 & 1.44 & 0.96 \\
GAMA619097 & 	normal	 & 	normal	 & 0.04 & 0.004 & 0.123 & 0.018 & 9.99 & 1.89 & 0.94 & 0.78 \\
GAMA619098 & 	normal	 & 	normal	 & 0.035 & 0.005 & 0.044 & 0.008 & 9.35 & 1.25 & -0.00 & 1.06 \\
GAMA619105 & 	normal	 & 	normal	 & 0.03 & 0.003 & 0.051 & 0.006 & 9.76 & 1.43 & 0.40 & 0.85 \\
GAMA620034 & 	normal	 & 	normal	 & 0.041 & 0.002 & 0.038 & 0.013 & 10.22 & 1.71 & 0.35 & 0.99 \\
GAMA620087 & 	normal	 & 	normal	 & 0.041 & 0.085 & 0.066 & 0.014 & 9.20 & 1.11 & 0.38 & 0.97 \\
GAMA620098 & 	normal	 & 	normal	 & 0.042 & 0.004 & 0.156 & 0.024 & 8.97 & 1.08 & 0.22 & 0.78 \\
GAMA622394 & 	perturbed	 & 	perturbed	 & 0.095 & 0.032 & 0.243 & 0.165 & 9.21 & 1.27 & -0.12 & 1.46 \\
GAMA622434 & 	normal	 & 	normal	 & 0.045 & 0.008 & 0.074 & 0.041 & 10.74 & 2.17 & -0.12 & {-999.0} \\
\newpage
\hline
GAMA Survey & Vis. & Kin. & $\overline{v_{asym}}$ & $\overline{v_{asym}}$ & $\overline{\sigma_{asym}}$ & $\overline{\sigma_{asym}}$ &  log($M_{*}$) & Colour& log(SFR) & $\frac{r_{50,H\alpha}}{r_{50,cont}}$ \\
ID & Class. & Class. & & err. & & err. & & ($u-r$)& & \\
\hline

GAMA622534 & 	normal	 & 	normal	 & 0.027 & 0.007 & 0.042 & 0.009 & 9.12 & 1.38 & -0.02 & 1.10 \\
GAMA622694 & 	normal	 & 	normal	 & 0.03 & 0.007 & 0.081 & 0.033 & 10.75 & 1.87 & 1.26 & 0.82 \\
GAMA622744 & 	normal	 & 	normal	 & 0.059 & 0.007 & 0.155 & 0.026 & 8.99 & 1.15 & 0.85 & 0.67 \\
GAMA622770 & 	normal	 & 	normal	 & 0.062 & 0.049 & 0.097 & 0.052 & 10.02 & 2.07 & 1.12 & 0.65 \\
GAMA623620 & 	normal	 & 	normal	 & 0.055 & 0.005 & 0.07 & 0.005 & 10.22 & 1.49 & 1.18 & 0.96 \\
GAMA623641 & 	normal	 & 	normal	 & 0.065 & 0.005 & 0.114 & 0.061 & 9.32 & 1.62 & -0.81 & 1.00 \\
GAMA623679 & 	normal	 & 	normal	 & 0.04 & 0.003 & 0.071 & 0.008 & 10.19 & 1.78 & 0.29 & 0.96 \\
GAMA623712 & 	perturbed	 & 	perturbed	 & 0.09 & 0.03 & 0.223 & 0.096 & 9.2 & 1.53 & -0.55 & 0.76 \\
GAMA623722 & 	normal	 & 	normal	 & 0.036 & 0.017 & 0.125 & 0.034 & 8.9 & 1.73 & -0.96 & 0.85 \\
GAMA623726 & 	perturbed	 & 	perturbed	 & 0.179 & 0.012 & 0.165 & 0.073 & 8.32 & 1.23 & -1.31 & 0.89 \\
\hline
\caption{Visual classification (performed by members of the SAMI Galaxy Survey team), kinemetric classification, median asymmetry in velocity dispersion and velocity, with errors, GAMA Survey catalogue values for stellar mass \citep{taylor2011galaxy} and colour \citep{hill2011galaxy} and SAMI Galaxy Survey values for SFR and $\frac{r_{50,H\alpha}}{r_{50,cont}}$ (Schaefer et al., submitted) for all galaxies in the sample. Where there is data missing in the SAMI Galaxy Survey values, {-999.0} is the default value. The asymmetry cutoff derived in this work is $\overline{v_{asym}}>0.065$, with galaxies above the cutoff having kinemetric classification `asymmetric' and those below it being `normal'.}
\end{longtable}

\end{center}
\end{document}